\documentclass[letterpaper,11pt]{article}

\def\thefootnote{\arabic{footnote}}

\usepackage{epsfig}
\usepackage{graphics}
    \renewcommand{\theequation}{\arabic{section}.\arabic{equation}}
\DeclareMathAlphabet   {\mathsc}{OT1}{cmr}{m}{sc}

\def\[{\left [}
\def\]{\right ]}
\def\({\left (}
\def\){\right )}
\newcommand{\lang}{\left\langle}
\newcommand{\rang}{\right\rangle}
\newcommand{\lbr}{\left\{}
\newcommand{\rbr}{\right\}}
\newcommand{\beq}{\begin{equation}}
\newcommand{\eeq}{\end{equation}}
\newcommand{\bea}{\begin{eqnarray}}
\newcommand{\eea}{\end{eqnarray}}
\newcommand{\oline}[1]{\overline{#1}}

\newcommand{\wtd}[1]{\widetilde{#1}}
\newcommand{\Lag}{\mathcal{L}}

\newcommand{\slashed}[1]{\not{\hspace{-.05in}#1}}

\newcommand{\GeV}      {~\mathrm{GeV}}
\newcommand{\TeV}      {~\mathrm{TeV}}

\newcommand{\SM}       {\mathsc{sm}}
\newcommand{\EW}       {\mathsc{ew}}

\newcommand{\UV}       {\mathsc{uv}}

\newcommand{\PL}       {\mathsc{pl}}

\newcommand{\GUT}      {\mathsc{gut}}
\newcommand{\STR}      {\mathsc{str}}
\newcommand{\SUSY}     {\mathsc{susy}}

\newcommand{\lowest}{|_{\theta =\bar{\theta}=0}}

\newcommand{\diff}{\mbox{d}}
\newcommand{\order}{\mathcal{O}}
\newcommand{\re}{{\rm Re}}

\newcommand{\gappeq}{\mathrel{\rlap {\raise.5ex\hbox{$>$}}
{\lower.5ex\hbox{$\sim$}}}}
\newcommand{\lappeq}{\mathrel{\rlap{\raise.5ex\hbox{$<$}}
{\lower.5ex\hbox{$\sim$}}}}

\newcommand{\DS}{(\Delta S)^2}
\newcommand{\DSaa}{(\Delta S_{AA})^2}
\newcommand{\DSab}{(\Delta S_{AB})^2}

\newcommand{\met}{\not{\hspace{-.05in}{E_T}}}

\begin{document}

\begin{titlepage}
\begin{center}

\vskip .1in {\large \bf Studying Gaugino Mass Unification at the
LHC}

\vskip .4in Baris~Altunkaynak,$^{(1)}$ Phillip~Grajek,$^{(2)}$
Michael~Holmes,$^{(1)}$ \\ Gordon Kane,$^{(2)}$
and Brent~D.~Nelson$^{(1)}$ \vskip .1in

(1) {\em Department of Physics, Northeastern University, Boston, MA
02115} \vskip .1in

(2) {\em Michigan Center for Theoretical Physics, Randall Lab.,\\
University of Michigan, Ann Arbor, MI 48109}\vskip .1in


\end{center}

\begin{abstract}
    We begin a systematic study of how gaugino mass unification can
    be probed at the CERN Large Hadron Collider (LHC) in a
    quasi-model independent manner. As a first step in that
    direction we focus our attention on the theoretically
    well-motivated mirage pattern of gaugino masses, a one-parameter
    family of models of which universal (high scale) gaugino masses are a
    limiting case. We improve on previous methods to define an
    analytic expression for the metric on signature space and use it
    to study one-parameter deviations from universality in the
    gaugino sector, randomizing over other soft
    supersymmetry-breaking parameters. We put forward three
    ensembles of observables targeted at the physics of the gaugino
    sector, allowing for a determination of this non-universality
    parameter without reconstructing individual mass eigenvalues or
    the soft supersymmetry-breaking gaugino masses themselves. In
    this controlled environment we find that approximately
    80\% of the supersymmetric parameter space would
    give rise to a model for which our method will detect
    non-universality in the gaugino mass sector at the 10\% level
    with $\order(10\,{\rm fb^{-1}})$ of integrated luminosity. We
    discuss strategies for improving the method and for adding
    more realism in dealing with the actual experimental
    circumstances of the LHC.
\end{abstract}
\end{titlepage}
\newpage

\renewcommand{\thepage}{\arabic{page}}
\setcounter{page}{1}
\def\thefootnote{\arabic{footnote}}
\setcounter{footnote}{0}

\section{Introduction} \label{intro} 

As the Large Hadron Collider (LHC) era fast approaches, the
theoretical community is increasingly focused on how the new
discoveries made there will be interpreted. The first step, most
obviously, is to establish the presence of physics beyond the
Standard Model. This will be done using search strategies that are
by now well-established, though many interesting ``what-if''
scenarios continue to be proposed and investigated~\cite{whatif}. We
continue to believe that supersymmetry (SUSY) is the best-motivated
extension to the Standard Model for physics at the LHC energy scale.
Furthermore, if supersymmetry is indeed relevant at the electroweak
scale there are many reasons to expect that its presence will be
established early on in the LHC program~\cite{Kane:2002ap}. Indeed,
even some properties of the spectrum, such as the masses and spins
of low-lying new states, may be crudely known even after relatively
little integrated
luminosity~\cite{Kane:2008kw,Hubisz:2008gg,Yamamoto:2007it}. In this
paper we begin a research program into what comes next: how to
connect the multiple LHC observations to organizing principles in
some (high-energy) effective Lagrangian of underlying physics.

This secondary problem can be further divided into two sub-problems.
The first has come to be called the ``inversion'' problem. Briefly
stated, the inversion problem is the recognition that even in very
restrictive model frameworks it is quite likely that more than one
set of model parameters will give predictions for LHC observations
that are in good agreement with the experimental
data~\cite{ArkaniHamed:2005px}. Much recent work has focused on how
to address this
issue~\cite{Kane:2006hd,Berger:2007yu,Berger:2007ut,Altunkaynak:2008ry,Kane:2008gb},
and we will borrow much of the philosophy and many of the useful
techniques from this recent literature. But our focus here is on
what we might call the second sub-problem: how to turn the ensemble
of distinct LHC signatures into a determination of certain broad
properties of the underlying Lagrangian at low energies. Clearly the
most direct attack on this second sub-problem is to perform a global
fit to the parameters of a particular
model~\cite{Binetruy:2003cy,fits}, modulo the degeneracy issue just
described above. Not surprisingly, therefore, the work we will
describe in this paper will make significant use of likelihood fits.
But our ultimate goal is to fit to certain broad properties of the
{\em underlying physics itself} -- and not simply to a particular
model of that physics.

We will refine this rather vague-sounding goal in a moment. But it
is helpful to first consider an example of what we mean by the
phrase ``broad properties of the underlying physics.'' Consider a
high energy theorist interested in connecting the (supersymmetric)
physics at the LHC to physics at an even higher energy scale, such
as some underlying string theory. What sort of information would be
of most use to him or her in this pursuit? Would it be a precise
measurement of the gluino mass, or of the mass splitting in the top
squark sector, or some other such measurement? Obtaining such
information is (at least in principle) possible at the LHC, but far
more valuable would be knowledge of the size of the supersymmetric
$\mu$-parameter or whether $\tan\beta$ is very small. Such
information is far more difficult to obtain at the
LHC~\cite{Brhlik:1998gu} but is more correlated with moduli
stabilization and/or how the $\mu$-parameter is generated in string
models~\cite{Nath:2002nb}. For example, this knowledge may tell us
whether the $\mu$-parameter is fundamental in the superpotential or
generated via the K\"ahler potential as in the Giudice-Masiero
mechanism~\cite{Giudice:1988yz}. This, in turn, is far more powerful
in discriminating between potential string constructions than the
gluino mass itself -- no matter how accurately it is determined. We
might refer to the genesis of the $\mu$-parameter as a ``broad
property of the underlying physics.''

If all such key broad properties of the underlying physics were
enumerated, it is our view that one of the most important such
properties would be the question of gaugino mass universality. That
is, the notion that at the energy scale at which supersymmetry
breaking is transmitted to the observable sector, the gauginos of
the minimal supersymmetric Standard Model (MSSM) all acquired soft
masses of the same magnitude. This issue is intimately related to
another, perhaps equally important issue: the wave-function of the
lightest supersymmetric particle, typically the lightest neutral
gaugino. Few properties of the superpartner spectrum have more
far-reaching implications for low-energy phenomenology, the nature
of supersymmetry breaking, and the structure of the underlying
physics Lagrangian~\cite{Binetruy:2005ez}. If the theorist could be
told only one ``result'' from the LHC data the answer to the simple
question ``Is there evidence for gaugino mass universality?'' might
well be it. But these soft parameters are not themselves directly
measurable at the LHC~\cite{Kneur:1998gy}.\footnote{Even a
measurement of the physical gluino mass is not a direct measurement
of the associated $SU(3)$ soft mass $M_3$. Quantum corrections to
the gluino bare mass can be sizable and their theoretical
computation involves a large set of other MSSM soft
parameters~\cite{Martin:1993yx,Krasnikov:1994sb} -- which are also
not directly measurable!} One might consider performing a fit to
some particular theory, such as minimal supergravity (mSUGRA), in
which universal gaugino masses are assumed~\cite{Arnowitt:1992aq} --
or perhaps to certain models with fixed, non-universal gaugino mass
ratios~\cite{Bhattacharya:2007dr,Bhattacharya:2008qu}. But we are
not so much interested in whether mSUGRA -- or any other particular
theory for which gaugino mass universality is a feature -- is a good
fit to the data. Rather, we wish to know whether gaugino mass
universality is a property of the underlying physics {\em
independent of all other properties of the model.} From this example
both the ambitiousness and the difficulty inherent in our task is
clear.

We have therefore decided to begin our attack by considering a
concrete parametrization of non-universalities in soft gaugino
masses. Many such frameworks present themselves, but we will choose
a parametrization that has the virtue of also having a strong
theoretical motivation from string theory. In recent work by Choi
and Nilles~\cite{Choi:2007ka} soft supersymmetry-breaking gaugino
mass patterns were explored in a variety of string-motivated
contexts. In particular, the so-called ``mirage pattern'' of gaugino
masses provides an interesting case study in gaugino mass
non-universality. Yet as mentioned above, these soft supersymmetry
breaking parameters are not themselves directly measurable. Linking
the soft parameters to the underlying Lagrangian is important, but
without the crucial step of linking the parameters to the data
itself it will be impossible to reconstruct the underlying physics
from the LHC observations.

The mirage paradigm gets its name from the fact that should the
mirage pattern of gaugino masses be used as the low-energy boundary
condition of the (one-loop) renormalization group equations then
there will exist some high energy scale at which all three gaugino
masses are identical. This unification has nothing to do with grand
unification of gauge groups, however, and the gauge couplings will
in general {\em not} unify at this particular energy scale -- hence
the name ``mirage.'' The set of all such low-energy boundary
conditions that satisfy the mirage condition defines a one-parameter
family of models. This parameter can be taken to be the mirage
unification scale itself, or some other parameter, such as the ratio
between various contributions to the gaugino soft masses. We note
that the minimal supergravity paradigm of soft supersymmetry
breaking is itself a member of this family of models since it is
{\em defined} by the property that gaugino masses are universal at
the scale $M_{\GUT} \simeq 2 \times 10^{16} \GeV$. Indeed, in the
parametrization we adopt from~\cite{Choi:2007ka}, the gaugino mass
ratios at the electroweak scale take the form
\begin{equation}
M_1\,:\,M_2\,:\,M_3 \, \simeq \, (1+0.66\alpha)\, : \, (2+0.2\alpha)
\, : \, (6-1.8\alpha)\, , \label{mirage_ratios} \end{equation}
where the case $\alpha=0$ is precisely the unified mSUGRA limit.
Note that when we speak of testing gaugino mass universality,
therefore, we do not imagine a common gaugino soft mass at the
low-energy scale. Instead, the ``universality'' paradigm implies the
ratios
\begin{equation}
M_1\,:\,M_2\,:\,M_3 \simeq 1\, : \, 2 \, : \, 6\, .
\label{msugra_ratios} \end{equation}
The goal of this work is to ask whether it is possible to determine
that the $\alpha$ parameter of~(\ref{mirage_ratios}) is different
from zero -- and if so, how.

The theoretical details behind the ratios of~(\ref{mirage_ratios})
will be the topic of Section~\ref{theory} in this paper. These
details are largely irrelevant for the analysis that follows in
Sections~\ref{method} and~\ref{results}, but may nevertheless be of
interest to many readers. For those who are only interested in the
methodology we will pursue and the results, this section can be
omitted. At the end of Section~\ref{theory} we will present two
benchmark scenarios that arise from concrete realizations of the
mirage pattern of gaugino masses in certain classes of string
models. As this is a paper about the interface of theory and
experiment at the LHC -- and not about string phenomenology per se
-- we will leave the theoretical description of these models to the
Appendix. In Section~\ref{method} we discuss how we will go about
attempting to measure the value of the parameter $\alpha$
in~(\ref{mirage_ratios}) and describe the process that led us to an
ensemble of specific LHC~observables targeted for precisely this
purpose. In Section~\ref{results} this list of signatures is tested
on a large collection of MSSM models, as well as on our two special
benchmarks from Section~\ref{theory}. We will see that the signature
lists constructed using the method of Section~\ref{method} do an
excellent job of detecting the presence of non-universality in the
gaugino soft masses over a very wide array of supersymmetric spectra
hierarchies and mass ranges. Non-universality on the order of
30-50\% should become apparent within the first 10~${\rm fb}^{-1}$
of analyzed data for most supersymmetric models consistent with
current experimental constraints. Detecting non-universality at the
10\% level would require an increase in data by roughly a factor of
two. Nevertheless, depending on the details of the superpartner
spectrum, some cases will require far more data to truly measure the
presence of non-universality. Of course all of these statements must
here be understood in the context of the very particular assumptions
of this study. Some thoughts on how the process can be taken further
in the direction of increased realism are discussed in the
concluding section.

Before moving to the body of the paper, however, we would like to
take a moment to emphasize a few broad features of the theoretical
motivation behind the parametrization in~(\ref{mirage_ratios}). In
the limit of very large values for the parameter $\alpha$ the ratios
among the gaugino masses approach those of the anomaly-mediated
supersymmetry breaking (AMSB)
paradigm~\cite{Giudice:1998xp,Randall:1998uk}. In fact, the mirage
pattern is most naturally realized in scenarios in which a common
contribution to all gaugino masses is balanced against an equally
sizable contribution proportional to the beta-function coefficients
of the three Standard Model gauge groups. Such an outcome arises in
string-motivated contexts, such as KKLT-type moduli stabilization in
D-brane models~\cite{Kachru:2003aw,Grana:2005jc} and K\"ahler
stabilization in heterotic string models~\cite{Gaillard:2007jr}.
These string-derived manifestations can also be extended easily to
include the presence of gauge mediation, in which the mirage pattern
is maintained in the gaugino
sector~\cite{Everett:2008qy,Everett:2008ey}. Importantly, however,
it can arise in {\em non}-stringy models, such as deflected anomaly
mediation~\cite{Katz:1999uw,Rattazzi:1999qg}. We note that in none
of these cases is the pure-AMSB limit likely to be obtained, so our
focus here will be on small to moderate values of the parameter
$\alpha$ in~(\ref{mirage_ratios}).\footnote{In any event, the
phenomenology of the AMSB scenario is sufficiently distinct from the
models we will consider that distinguishing between them should not
be difficult~\cite{Gherghetta:1999sw}.} We will further refine these
observations in Section~\ref{theory} before turning our attention to
the measurement of the parameter $\alpha$ at the~LHC.


\section{Theoretical Motivation and Background} \label{theory} 

In this section we wish to understand the origin of the mass ratios
in~(\ref{mirage_ratios}) from first principles. We will treat the
mirage mass pattern here in complete generality, without any
reference to its possible origin from string-theoretic
considerations. This short section concludes with two specific sets
of soft parameters, both of which represent models with the mirage
gaugino mass pattern (though the physics behind the rest of their
soft supersymmetry breaking parameters are quite different). In the
Appendix we will recast the discussion of this section in terms of
the degrees of freedom present in low-energy effective Lagrangians
from string model building. There we will also present the string
theory origin of the two benchmark models that appear in
Table~\ref{tbl:inputs} at the end of this section.


Let us begin by imagining a situation in which there are two
contributions to the soft supersymmetry breaking gaugino masses. We
assume that these contributions arise at some effective high-energy
scale at which supersymmetry breaking is transmitted from some
hidden sector to the observable sector. Let us refer to this scale
as simply the ultraviolet scale $\Lambda_{\UV}$. It is traditional
in phenomenological treatments to take this scale to be the GUT
scale at which gauge couplings unify, but in string constructions
one might choose a different (possibly higher scale) at which the
supergravity approximation for the effective Lagrangian becomes
valid. We will further assume that one contribution to gaugino
masses is universal in nature while the other contribution is
proportional to the beta-function coefficient of the Standard Model
gauge group. More specifically, consider the universal piece to be
given by
\begin{equation} M^1_a \(\Lambda_{\UV}\) = M_u\, , \label{piece1} \end{equation}
where $a=1,2,3$ labels the Standard Model gauge group factors ${\cal
G}_a$ and $M_u$ represents some mass scale in the theory. The second
piece is the so-called anomaly mediated piece, which arises from
loop diagrams involving the auxiliary scalar field of
supergravity~\cite{Gaillard:1999yb,Bagger:1999rd}. It will take the
form
\begin{equation} M^2_a \(\Lambda_{\UV}\) = g_a^2 \(\Lambda_{\UV}\)
\frac{b_a}{16\pi^2} M_g \, , \label{piece2} \end{equation}
where the $b_a$ are the beta-function coefficients for the Standard
Model gauge groups. In our conventions these are given by
\begin{equation}
b_a =  -(3 C_a - \sum_i C_a^i) ,  \label{ba}
\end{equation}
where $C_a$, $C_a^i$ are the quadratic Casimir operators for the
gauge group ${\cal G}_a$, respectively, in the adjoint
representation and in the representation of the matter fields
$\Phi^i$ charged under that group.\footnote{The convention chosen
in~(\ref{ba}) is opposite of the one chosen in~\cite{Kane:2002qp}.}
For the MSSM these are
\begin{equation} \lbr b_1, b_2, b_3\rbr = \lbr \frac{33}{5}, 1, -3
\rbr \, . \label{baSM} \end{equation}
Note that if we take $\Lambda_{\UV} = \Lambda_{\GUT}$ then we have
\begin{equation} g_1^2\(\Lambda_{\UV}\) = g_2^2\(\Lambda_{\UV}\) =
g_3^2 \(\Lambda_{\UV}\) = g_{\GUT}^2 \simeq \frac{1}{2}\, .
\label{gGUT} \end{equation}
The mass scale $M_g$ is common to all three gauge groups; the
subscript is meant to indicate that the contribution
in~(\ref{piece2}) is related to the gravitino mass. The full gaugino
mass at the high energy boundary condition scale is therefore
\begin{equation} M_a \(\Lambda_{\UV}\) = M_a^1 \(\Lambda_{\UV}\) + M_a^2 \(\Lambda_{\UV}\) =
M_u + g_a^2\(\Lambda_{\UV}\) \frac{b_a}{16\pi^2} M_g \, .
\label{Mafull} \end{equation}

Now imagine evolving the boundary conditions in~(\ref{Mafull}) to
some low-energy scale $\Lambda_{\EW}$ via the (one-loop)
renormalization group equations (RGEs). For the anomaly-generated
piece of~(\ref{piece2}) we need only replace the gauge coupling with
the value at the appropriate scale
\begin{equation} M^2_a \(\Lambda_{\EW}\) = g_a^2 \(\Lambda_{\EW}\)
\frac{b_a}{16\pi^2} M_g \, , \label{piece2b} \end{equation}
while for the universal piece we can use the fact that $M_a/g_a^2$
is a constant for the one-loop RGEs. After some manipulation this
yields
\begin{equation} M^1_a \(\Lambda_{\EW}\) = M_u \[1 -
g_a^2\(\Lambda_{\EW}\) \frac{b_a}{8\pi^2}
\ln\(\frac{\Lambda_{\UV}}{\Lambda_{\EW}}\)\] \, . \label{piece1b}
\end{equation}
Combining~(\ref{piece1b}) and~(\ref{piece2b}) gives the low scale
expression
\begin{equation} M_a \(\Lambda_{\EW}\) = M_u \lbr 1 -
g_a^2\(\Lambda_{\EW}\) \frac{b_a}{8\pi^2}
\ln\(\frac{\Lambda_{\UV}}{\Lambda_{\EW}}\)
\[1-\frac{1}{2}\frac{M_g}{M_u\ln\(\frac{\Lambda_{\UV}}{\Lambda_{\EW}}\)}\]\rbr \,
. \label{Malowfull} \end{equation}
For gaugino masses to be unified {\em at the low scale}
$\Lambda_{\EW}$ then the quantity in the square brackets
in~(\ref{Malowfull}) must be engineered to vanish. This can be
achieved with a judicious choice of the values $M_u$ and $M_g$ for a
particular high-energy input scale $\Lambda_{\UV}$. Put differently,
for a given $\Lambda_{\UV}$ (such as the GUT scale) and a given
overall scale $M_u$, there is a one-parameter family of models
defined by the choice $M_g$.

It is possible, however, to find a more convenient parametrization
of the family of gaugino mass patterns defined by~(\ref{Malowfull}).
Consider defining the parameter $\alpha$ by
\begin{equation} \alpha =
\frac{M_g}{M_u\ln\(\Lambda_{\UV}/\Lambda_{\EW}\)} \, ,
\label{alpha_example} \end{equation}
so that~(\ref{Malowfull}) becomes
\begin{equation} M_a \(\Lambda_{\EW}\) = M_u \[ 1 - \(1-\frac{\alpha}{2}\)
g_a^2\(\Lambda_{\EW}\) \frac{b_a}{8\pi^2}
\ln\(\frac{\Lambda_{\UV}}{\Lambda_{\EW}}\)\] \, \label{Malowfull2}
\end{equation}
and the requirement of universality at the scale $\Lambda_{\EW}$ now
implies $\alpha=2$. Normalizing the three gaugino masses by
$M_1\(\Lambda_{\EW}\)|_{\alpha=0}$ and evaluating the gauge
couplings at a scale $\Lambda_{\EW} = 1000 \GeV$ we obtain the
mirage ratios
\begin{equation}
M_1\,:\,M_2\,:\,M_3 \, = \, (1.0+0.66\alpha)\, : \,
(1.93+0.19\alpha) \, : \, (5.87-1.76\alpha)\, ,
\label{mirage_ratios2} \end{equation}
for $\Lambda_{\UV} = \Lambda_{\GUT}$, in good agreement with the
expression in~(\ref{mirage_ratios}).

Let us generalize the parametrization in~(\ref{alpha_example}) once
more. Instead of defining the parameter in terms of the starting and
stoping points in the RG evolution of the gaugino mass parameters,
we will fix them in terms of mass scales in the theory itself. Thus
we follow the convention of Choi et al.~\cite{Choi:2005uz} and
define
\begin{equation} \alpha \equiv
\frac{M_g}{M_u\ln\(M_{\PL}/M_g\)} \, , \label{alpha}
\end{equation}
where $M_{\PL}$ is the reduced Planck mass $M_{\PL} = 2.4 \times
10^{18} \GeV$. Our parametrization is now divorced from the boundary
condition scales of the RG flow and can be fixed in advance. The
choice of mass parameters in the logarithm of~(\ref{alpha}) may seem
arbitrary -- and at this point it is indeed completely arbitrary --
but they have been chosen so as to make better contact with string
constructions, such as those which we present in the Appendix.
Inserting~(\ref{alpha}) into~(\ref{Malowfull}) yields
\begin{eqnarray} M_a \(\Lambda_{\EW}\) &=& M_u \lbr 1 -
g_a^2\(\Lambda_{\EW}\) \frac{b_a}{8\pi^2}\[
\ln\(\frac{\Lambda_{\UV}}{\Lambda_{\EW}}\) - \frac{\alpha}{2}
\ln\(\frac{M_{\PL}}{M_g}\)\]\rbr \nonumber \\
 &=& M_u \lbr 1 -
g_a^2\(\Lambda_{\EW}\) \frac{b_a}{8\pi^2}\[
\ln\(\frac{\Lambda_{\UV}\(M_g/M_{\PL}\)^{\alpha/2}}{\Lambda_{\EW}}\)
\]\rbr \, . \label{Malowfull3}
\end{eqnarray}
Comparing this expression with~(\ref{piece1b}) it is clear if gauge
couplings unify at a scale $\Lambda_{\UV} = \Lambda_{\GUT}$, then we
should expect the soft supersymmetry breaking gaugino masses to
unify at an effective scale given by
\begin{equation} \Lambda_{\rm mir} = \Lambda_{\GUT}
\(\frac{M_g}{M_{\PL}}\)^{\alpha/2}\, . \label{mirage_scale}
\end{equation}
We see that our parametrization in terms of $\alpha$ is indeed
equivalent to a parametrization in terms of the effective
unification scale, as suggested in the introduction.

The value of $\alpha$ as defined in~(\ref{alpha_example})
or~(\ref{alpha}) can be crudely thought of as the ratio of the
anomaly contribution to the universal contribution to gaugino
masses. Indeed, the limit $\alpha \to 0$ is the limit of the minimal
supergravity paradigm, while $\alpha \to \infty$ is the AMSB limit.
But as~(\ref{Mafull}) makes clear, these two contributions will be
of comparable size only if $M_g$ is at least an order of magnitude
larger than $M_u$. We could therefore have chosen a parametrization
based on the ratio $r = M_g/M_u$, with interesting values being in
the range $r \simeq \order(10-100)$. But such a parametrization
spoils the simple relation with the mirage unification
scale~(\ref{mirage_scale}). Furthermore, the introduction of the
factor $\ln(M_{\PL}/M_g)$ in~(\ref{alpha}) provides the needed large
factor, taking a value of $\ln(M_{\PL}/M_g) \simeq 35$ for $M_g
\simeq 1$ TeV. To obtain the mirage pattern it is therefore
necessary for the underlying theory to generate some large number $c
\simeq \ln(M_{\PL}/M_g) \simeq 30$. Specific examples of how this is
achieved in explicit string-based models are given in the Appendix
to this paper.

\begin{table}[t]
{\begin{center}
\begin{tabular}{|l|c|c|c|l|c|c|}\cline{1-7}
Parameter & Point A & Point B & & Parameter & Point A & Point B \\
\cline{1-7} $\alpha$ & 0.3 & 1.0  &
& $m_{Q_3}^{2}$ & $(1507)^2$ & $(430.9)^2$ \\
$M_{g}$ & 1.5 TeV & 16.3 TeV  &
& $m_{U_3}^{2}$ & $(1504)^2$ & $(610.3)^2$ \\
$M_{1}$ & 198.7 & 851.6 &
& $m_{D_3}^{2}$ &  $(1505)^2$ & $(352.2)^2$ \\
$M_{2}$ & 172.1 & 553.3 & & $m_{\tilde{c}_{R}}$,
$m_{L_3}^{2}$ & $(1503)^2$ & $(381.6)^2$ \\
$M_{3}$ & 154.6 & 339.1 & & $m_{E_3}^{2}$ & $(1502)^2$ & $(407.9)^2$
\\ \cline{1-7}
$A_t$ & 193.0 & 1309 & & $m_{Q_{1,2}}^{2}$ & $(1508)^2$ &
$(208.4)^2$ \\
$A_b$ & 205.3 & 1084 &
& $m_{U_{1,2}}^{2}$ & $(1506)^2$ & $(302.7)^2$ \\
$A_{\tau}$ & 188.4 & 1248 & & $m_{D_{1,2}}^{2}$ & $(1505)^2$ &
$(347.0)^2$ \\
$m_{H_u}^{2}$ & $(1500)^2$ & $(752.0)^2$ & &
$m_{L_{1,2}}^{2}$ & $(1503)^2$ & $(379.8)^2$ \\
$m_{H_d}^{2}$ & $(1503)^2$ & $(388.7)^2$ & &
$m_{E_{1,2}}^{2}$ & $(1502)^2$ & $(404.5)^2$ \\
\cline{1-7}
\end{tabular}
\end{center}}
{\caption{\label{tbl:inputs}\footnotesize{\bf Soft Term Inputs}.
Initial values of supersymmetry breaking soft terms in GeV at the
initial scale given by $\Lambda_{\UV} = 2 \times 10^{16} \GeV$. Both
points are taken to have $\mu > 0$ and $\tan\beta=10$. The actual
value of $\tan\beta$ is fixed in the electroweak symmetry-breaking
conditions.}}
\end{table}


In Table~\ref{tbl:inputs} we have collected the necessary soft
supersymmetry-breaking parameters to completely specify two
benchmark points for further analysis in what follows. The details
behind these two models are described in the Appendix. Here we will
simply indicate that point~A represents a heterotic string model
with K\"ahler stabilization of the dilaton which was studied in
detail in~\cite{Kane:2002qp}. This particular example has a value of
$\alpha = 0.3$. Point~B is an example from a class of Type~IIB
string compactifications with fluxes which was studied
in~\cite{Choi:2005uz}. This second example has a value $\alpha =
1.0$. Both are examples of the mirage pattern of gaugino masses,
having mirage unification scales of $\Lambda_{\rm mir} = 2.0 \times
10^{14} \GeV$ and $\Lambda_{\rm mir} = 1.5 \times 10^{9} \GeV$,
respectively. Note that these soft supersymmetry breaking terms are
taken to be specified at the GUT energy scale of $\Lambda_{\rm GUT}
= 2.0 \times 10^{16} \GeV$ and must be evolved to electroweak scale
energies through the renormalization group equations.

\section{Determining $\alpha$: Methodology} \label{method} 

\subsection{Setting Up the Problem}
\label{sec:setup}

As mentioned in the introduction, the ultimate goal of this avenue
of study is to determine whether or not soft supersymmetry breaking
gaugino masses obey some sort of universality condition independent
of all other facts about the supersymmetric model. Such a goal
cannot be met in a single paper so we have begun by asking a simpler
question: {\em assuming} the world is defined by the MSSM with
gaugino masses obeying the relation~(\ref{mirage_ratios}), how well
can we determine the value of the parameter~$\alpha$. At the very
least we would like to be able to establish that $\alpha \neq 0$
with a relatively small amount of integrated luminosity. The first
step in such an incremental approach is to demonstrate that some set
of ``targeted observables''~\cite{Binetruy:2003cy} (we will call
them ``signatures'' in what follows) is sensitive to small changes
in the value of the parameter $\alpha$ in a world where all other
parameters which define the SUSY model are kept fixed. In subsequent
work we intend to relax this strong constraint and treat the issue
of gaugino mass universality more generally. Despite the lack of
realism we feel this is a logical point of departure -- very much in
the spirit of the ``slopes'' of the Snowmass Points and
Slopes~\cite{Allanach:2002nj} and other such benchmark studies.
Thus, where the Snowmass benchmarks talk of slopes, we will here
speak of ``model lines'' in which all parameters are kept fixed but
the value of $\alpha$ is varied in a controlled manner.

To construct a model line we must specify the supersymmetric model
in all aspects other than the gaugino sector. The MSSM is completely
specified by~105 distinct parameters,
%
%
but only a small subset are in any way relevant for the
determination of LHC~collider observables~\cite{Brhlik:1998gu}. We
will therefore choose a simplified set of 17~parameters as in the
two benchmark models of Table~\ref{tbl:inputs}
\beq \lbr \begin{array}{c} \tan\beta,\,\,m^2_{H_u},\,\,m^2_{H_d} \\
M_3,\,\,A_t,\,\,A_b,\,\,A_{\tau} \\
m_{Q_{1,2}},\,m_{U_{1,2}},\,m_{D_{1,2}},\,m_{L_{1,2}},\,m_{E_{1,2}} \\
m_{Q_3},\,m_{U_3},\,m_{D_3},\,m_{L_3},\,m_{E_3} \end{array} \rbr \,
. \label{paramset} \eeq
The parameters in~(\ref{paramset}) are understood to be taken at the
electroweak scale (specifically $\Lambda_{\EW} = 1000\GeV$) so no
renormalization group evolution is required. The gluino soft mass
$M_3$ will set the overall scale for the gaugino mass sector. The
other two gaugino masses $M_1$ and $M_2$ are then determined
relative to $M_3$ via~(\ref{mirage_ratios2}). A model line will take
the inputs of~(\ref{paramset}) and then construct a family of
theories by varying the parameter $\alpha$ from $\alpha=0$ (the
mSUGRA limit) to some non-zero value in even increments.

For each point along the model line we pass the model parameters to
{\tt PYTHIA 6.4}~\cite{Sjostrand:2006za} for spectrum calculation
and event generation. Events are then sent to the {\tt
PGS4}~\cite{PGS} package to simulate the detector response.
Additional details of the analysis will be presented in later
sections. The end result of our procedure is a set of observable
quantities that have been designed and (at least crudely) optimized
so as to be effective at separating $\alpha = 0$ from other points
along the model line in the least amount of integrated luminosity
possible. In Section~\ref{sec:separate} we describe the manner in
which we perform this separation between models. The signature
lists, and the analysis behind their construction, is presented in
Section~\ref{sec:sigs}. In Section~\ref{results} we will demonstrate
the effectiveness of these signature lists on a large sample of
randomly generated model lines and provide some deeper insight on
why the whole procedure works by examining our benchmarks in greater
detail.

\subsection{Distinguishability}
\label{sec:separate}

The technique we will employ to distinguish between candidate
theories using LHC~observables was suggested
in~\cite{Binetruy:2003cy} and subsequently refined
in~\cite{ArkaniHamed:2005px}. The basic premise is to construct a
variable similar to a traditional chi-square statistic
\beq (\Delta S_{AB})^2 = \frac{1}{n} \, \sum_i \[ \frac{S_i^A -
S_i^B}{\delta S_i^{AB}}\]^2, \label{DeltaS} \eeq
where $S$ is some observable quantity (or signature). The index
$i=1,\dots,n$ labels these signatures, with $n$ being the total
number of signatures considered. The labels $A$ and $B$ indicate two
distinct theories which give rise to the signature sets $S_i^A$ and
$S_i^B$, respectively. Finally, the error term $\delta S_i^{AB}$ is
an appropriately-constructed measure of the uncertainty of the term
in the numerator, i.e. the difference between the signatures.
%
In this work we will always define a signature $S$ as an observation
interpreted as a count (or number) and denote it with capital $N$.
One example is the number of same-sign, same-flavor lepton pairs in
a certain amount of integrated luminosity. Another example is taking
the invariant mass of all such pairs and forming a histogram of the
results, then integrating from some minimum value to some maximum
value to obtain a number. In principle there can be an infinite
number of signatures defined in this manner. In practice
experimentalists will consider a finite number and many such
signatures are redundant.

We can identify any signature $N_i$ with an effective cross section
$\bar{\sigma}_i$ via the relation
\beq \bar{\sigma}_i = N_i / L\, ,  \label{barsigma} \eeq
where $L$ is the integrated luminosity. We refer to this as an {\em
effective} cross-section as it is defined by the counting signature
$N_i$ which contains in its definition such things as the geometric
cuts that are performed on the data, the detector efficiencies, and
so forth. Furthermore these effective cross sections, whether
inferred from actual data or simulated data, are subject to
statistical fluctuations. As we increase the integrated luminosity
we expect that this effective cross section $\bar{\sigma}_i$ (as
inferred from the data) converges to an ``exact'' cross section
$\sigma_i$ given by
\beq \label{xsection} \sigma_i = \lim_{L \rightarrow \infty}
\bar{\sigma}_i\, . \eeq
These exact cross sections are (at least in principle) calculable
predictions of a particular theory, making them the more natural
quantities to use when trying to distinguish between theories. The
transformation in~(\ref{barsigma}) allows for a comparison of two
signatures with differing amounts of integrated luminosity. This
will prove useful in cases where the experimental data is presented
after a limited amount of integrated luminosity $L_A$, but the
simulation being compared to the data involves a much higher
integrated luminosity $L_B$. Using these notions we can re-express
our chi-square variable $\DSab$ in terms of the cross sections
\beq (\Delta S_{AB})^2 = \frac{1}{n} \, \sum_i \[
\frac{\bar{\sigma}_i^A - \bar{\sigma}_i^B}{\delta
\bar{\sigma}_i^{AB}}\]^2\, . \eeq
We will assume that the errors associated with the signatures $N_i$
are purely statistical in nature and that the integrated
luminosities $L_A$ and $L_B$ are precisely known, so that
\beq \delta \bar{\sigma}_i^{AB} = \sqrt{ (\delta \bar{\sigma}_i^A)^2
+ (\delta \bar{\sigma}_i^B)^2 } = \sqrt{ \bar{\sigma}_i^A / L_A  +
\bar{\sigma}_i^B / L_B } \, , \label{dsigma} \eeq
and therefore $(\Delta S_{AB})^2$ is given by
\beq (\Delta S_{AB})^2 = \frac{1}{n} \, \sum_i \[
\frac{\bar{\sigma}_i^A - \bar{\sigma}_i^B}{\sqrt{ \bar{\sigma}_i^A /
L_A  + \bar{\sigma}_i^B / L_B }}\]^2, \label{DeltaS2} \eeq
where each cross section includes the (common) Standard Model
background, i.e. $\bar{\sigma}_i = \bar{\sigma}^{\SUSY}_i +
\bar{\sigma}^{\SM}$.

The variable $\DSab$ forms a measure of the distance between any two
theories in the space of signatures defined by the $S_i$. We can use
this metric on signature space to answer the following question: how
far apart should two sets of signatures $S_i^A$ and $S_i^B$ be
before we conclude that theories $A$ and $B$ are truly distinct? The
original criterion used in~\cite{ArkaniHamed:2005px} was as follows.
Imagine taking any supersymmetric theory and performing a collider
simulation. Now choose a new random number seed and repeat the
simulation. Due to random fluctuations we expect that even the same
set of input parameters, after simulation and event reconstruction,
will produce a slightly different set of signatures. That is, we
expect $\DSaa \neq 0$ since it involves the {\em effective}
cross-sections as extracted from the simulated data. Now repeat the
simulation a large number of times, each with a different random
number seed. Use~(\ref{DeltaS2}) to compute the distance of each new
simulation with the original simulation in signature space. The set
of all $\DSaa$ values so constructed will form a distribution. Find
the value of $\DSaa\big|_{95}$ in this distribution which represents
the 95th percentile of the distribution. This might be taken as a
measure of the uncertainty in ``distance'' measurements associated
with statistical fluctuations.

This procedure for defining distinguishability is unwieldy in a
number of respects. Determining the threshold for separating models
by $\DSab > \DSaa\big|_{95}$ is computationally intensive as it
requires many repeated simulations of the same model (as well as the
Standard Model background). More importantly, the ``brute force''
determination of $\DSaa\big|_{95}$ is particular to model~$A$ as
well as the list of signatures used in~(\ref{DeltaS2}). Each change
in either the model parameters or the signature mix demands a new
determination of the threshold for distinguishability. We will
therefore propose a new criterion that has the benefit of being
analytically calculable with a form that is universal to any pair of
models and any set of signatures.

To do that let us reconsider the quantum fluctuations. At a finite
integrated luminosity $L$ we can describe the outcome of a counting
experiment as a Poisson distribution approximated by a normal
distribution (this is a good approximation for approximately
10~counts or more), which can be expressed as
\begin{equation} N_i = L \sigma_i + \sqrt{L \sigma_i} \, Z \, .
\label{Ni} \end{equation}
Here $Z$ is a standard random variable, {\em i.e.} a random variable
having a normal distribution centered at 0 with a standard deviation
of 1. Note that by introducing statistical fluctuations via the
variable $Z$ we can replace $\bar{\sigma}_i$ in~(\ref{Ni}) with the
exact cross section. Equation~(\ref{Ni}) then merely states the well
known fact that the distribution in measured values $N_i$ should
form a normal distribution about the value $L \sigma_i$. To combine
two such distributions $N_1$ and $N_2$ we may write
\bea \label{signature} N_{\rm tot} &=& \Big(
L\sigma_1+\sqrt{L\sigma_1} Z_1 \Big)
+ \Big( L\sigma_2 + \sqrt{L\sigma_2} Z_2 \Big) \nonumber \\
&\equiv& L \sigma^{T} + \sqrt{L \sigma^{T}} Z, \eea
where $Z$ is a new standard random variable and $\sigma^T$ is the
total cross-section. For example, $\sigma_1$ might be the
contribution to a particular final state arising from Standard Model
processes while $\sigma_2$ might be the contribution arising from
production of supersymmetric particles.

With the above in mind we can re-visit the
definition~(\ref{DeltaS2}) and obtain an analytic approximation for
the distribution in $\DSab$ values by using random variables to
represent the signatures. The measured cross sections can be related
to the exact cross sections via
\begin{equation} \bar{\sigma}_i^A = N_i^A / L_A = \sigma_i^A + \sqrt{\sigma_i^A
/ L_A} \, Z_A \, ,\label{sigrelate} \end{equation}
with a similar expression for the model $B$.
Substituting~(\ref{sigrelate}) into~(\ref{DeltaS2}) gives
\bea
\label{ds2f} \DSab &=& \frac{1}{n}  \, \sum_i \frac{ \[ \sigma_i^A -
\sigma_i^B + \sqrt{\frac{\sigma_i^A}{L_A} + \frac{\sigma_i^B}{L_B}}
Z \]^2 }{ \frac{\sigma_i^A}{L_A} + \frac{\sigma_i^B}{L_B} + \sqrt{
\frac{1}{L_A^2} \frac{\sigma_i^A}{L^A}
+ \frac{1}{L_B^2} \frac{\sigma_i^B}{L^B} } Z' }  \nonumber \\
&\approx& \frac{1}{n}  \, \sum_i
\[ \frac{\sigma_i^A - \sigma_i^B}{ \sqrt{\frac{\sigma_i^A}{L_A}
+ \frac{\sigma_i^B}{L_B}} } + Z \]^2 \, , \eea
where we have combined $Z_A$ and $Z_B$ into the random variables $Z$
and $Z'$ and have assumed that $L_A$ and $L_B$ are sufficiently
large to be able to neglect the term proportional to $Z'$. In this
limit we immediately see that $\DSab$ is itself a random variable
with a probability distribution for the quantity $\DSab$ given by
\begin{equation}
P(\Delta S^2) = n \, \chi_{n,\lambda}^2(n \Delta S^2)\, ,
\label{prob}
\end{equation}
where $\chi_{n,\lambda}^2$ is the non-central chi-squared
distribution for $n$ degrees of freedom.\footnote{If we had chosen
to define the separation variable~(\ref{DeltaS}) without the factor
of $1/n$ we would have found that the distribution of $\DSab$ values
was exactly given by the non-central chi-square distribution. The
two are related by a simple change of variables.} The non-centrality
parameter $\lambda$ is given by
\beq \lambda = \sum_i \frac{(\sigma_i^A - \sigma_i^B)^2}{\sigma_i^A
/ L_A  + \sigma_i^B / L_B }\, , \label{lambdadef} \eeq
and now the $\sigma_i$ represent {\em exact} cross sections.
This is actually the result we expect since the original $\DSab$
in~(\ref{DeltaS2}) is essentially a chi-square like function. Note
that since the $\sigma_i$ in the distribution of~(\ref{ds2f}) are
exact, we have the anticipated result that fluctuations of the
quantity $\DSaa$ should be given by the central chi-square
distribution $\chi^2_n(0)$. We note, however, that the derivation
of~(\ref{ds2f}) implicitly assumed that the signatures $S_i$ which
we consider are uncorrelated -- or more precisely that the {\em
fluctuations} in these signatures are uncorrelated. We will have
more to say about signature correlations in Section~\ref{sec:sigs}
below. We now have a measure of separation in signature space that
is related to well known functions in probability
theory.\footnote{In fact, the non-central chi-square distribution is
related to the regularized confluent geometric functions
%
%
.}


\begin{figure}[tb]
\begin{center}
\includegraphics[scale=0.5,angle=0]{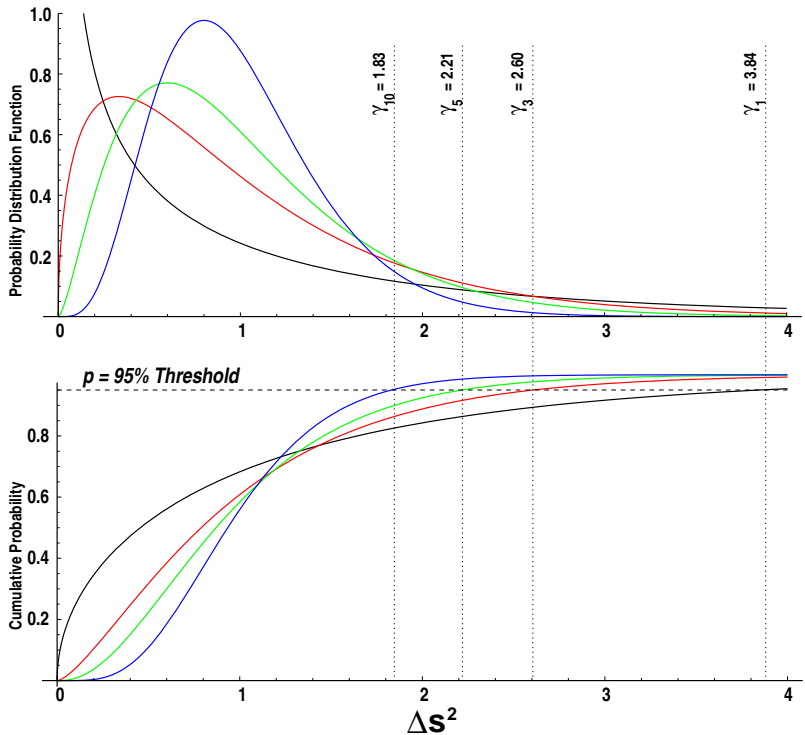}
\caption{\footnotesize \textbf{Plot of distribution in $\DSaa$
values.} The top panel plots the probability distribution
function~(\ref{prob}) for $\lambda = 0$ and $n = 1,\,3,\,5$ and 10.
The lower panel plots the cumulative distribution function -- the
absolute probability for obtaining that value of $\DS$. The 95\%
percent threshold is indicated by the horizontal lines, and the
corresponding values of $\DS\big|_{\rm 95th}$ are indicated by the
marked values of $\gamma_n(0.95)$.} \label{fig:centralchi}
\end{center}
\end{figure}

Armed with this technology, let us return to the issue of
distinguishing a model from itself. From~(\ref{ds2f}),
(\ref{prob})~and~(\ref{lambdadef}) it is apparent that all the
physics behind the distribution of possible $\DSab$ values is
contained in the values of $\lambda$ and $n$. In particular the
distribution of possible $\DSaa$ values (a central chi-square
distribution) should depend {\em only} on the number $n$ of
signatures considered -- not on the model point nor on the nature of
those signatures. When comparing a model with itself we can
therefore dispense with the subscript and write $\DSaa = \DS$. We
plot the probability distribution $P\DS$ of~(\ref{prob}) for
$\lambda = 0$ and various values of $n$ in the top panel of
Figure~\ref{fig:centralchi}.
%
%
%
We have also plotted the cumulative distribution function for the
same $n$ values in the lower panel of Figure~\ref{fig:centralchi}.
To rule out the null hypothesis ({\em i.e.} the hypothesis that
models~$A$ and~$B$ are in fact the same model) to a level of
confidence $p$ requires demanding that $\DS$ is larger than the
$p$-th percentile value for the distribution~(\ref{prob}) for the
appropriate $n$ value. For example, if we use the criterion
from~\cite{ArkaniHamed:2005px} and require $\DSab > \DS\big|_{\rm
95th}$ then $p=0.95$. We have indicated this value for the
cumulative distribution function by the horizontal dashed line in
Figure~\ref{fig:centralchi}. In general we will denote this
particular value of $\DS\big|_{p}$ for each value of $n$ by the
symbol $\gamma_n(p)$. It can be found via the cumulative
distribution function as in Figure~\ref{fig:centralchi}, or by
numerically solving the equation
\begin{equation} \Gamma \(\frac{n}{2},\, \frac{n}{2} \, \gamma_n(p)\)
= \Gamma \(\frac{n}{2}\) (1-p)\, , \end{equation}
where $\Gamma(n)$ is Euler's gamma function and $\Gamma(n,m)$ is the
incomplete gamma function. A summary of these values for smaller $n$
values is given in Table~\ref{gammatable}. If we measure our $n$
signatures, extract the cross-sections, form $\DSab$ and the number
is greater than $\gamma_n(p)$ then we can say that the null
hypothesis can be ruled out at a level of confidence given by
$p\times100$\%. The value of this critical
$\DS\big|_{p}=\gamma_n(p)$ is a universal number determined {\em
only} by our choice of $p$ value and the number of signatures $n$
that we choose to consider.

\begin{table}[tb]
\centering
\begin{tabular}{c||c|c|c|c}
 & \multicolumn{4}{c}{Confidence Level $p$}\\ \hline
 n & 0.95 & 0.975 & 0.99 & 0.999  \\
 \hline \hline
 1 & 3.84 & 5.02 & 6.64 & 10.83 \\
 2 & 3.00 & 3.69 & 4.61 & 6.91 \\
 3 & 2.61 & 3.12 & 3.78 & 5.42 \\
 4 & 2.37 & 2.79 & 3.32 & 4.62 \\
 5 & 2.21 & 2.57 & 3.02 & 4.10 \\
 6 & 2.10 & 2.41 & 2.80 & 3.74 \\
 7 & 2.01 & 2.29 & 2.64 & 3.48 \\
 8 & 1.94 & 2.19 & 2.51 & 3.27 \\
 9 & 1.88 & 2.11 & 2.41 & 3.10 \\
 10 & 1.83 & 2.05 & 2.32 & 2.96 \\
\hline
\end{tabular}
{\caption{\label{gammatable}\footnotesize {\bf List of $\gamma_n(p)$
values for various values of the parameters $n$ and $p$}. The value
$\gamma_n(p)$ represents the position of the $p$-th percentile in
the distribution of $P\DS$ for any list of $n$ signatures. For
example, if we consider a list of 10~signatures, then the quantity
$\DSab$ formed by these ten measurements must be larger than~1.83 to
say that models~$A$ and~$B$ are distinct, with 95\% confidence. If
we demand 99\% confidence this threshold becomes~2.32.}}
\end{table}

\begin{figure}[tb]
\begin{center}
\includegraphics[scale=1.2,angle=0]{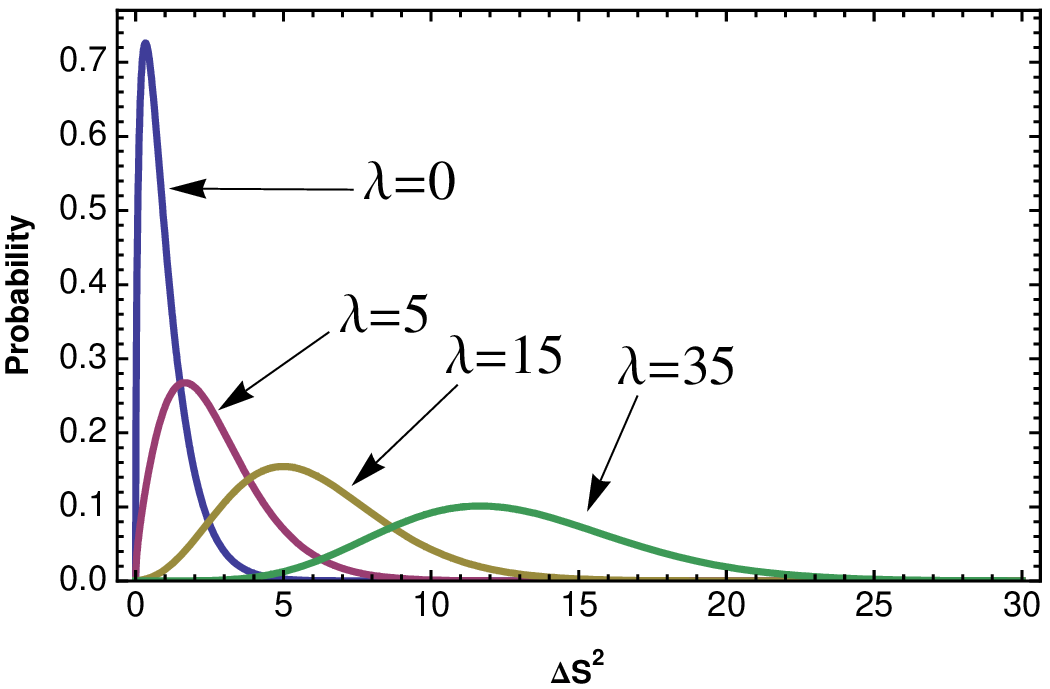}
\caption{\footnotesize \textbf{Plot of distribution in $\DSab$
values for $n=3$ and various $\lambda$.} The probability
distribution function~(\ref{prob}) for $\lambda = 0,\,5,\,15$ and~35
is plotted for the case of $n = 3$. The curves are normalized such
that the total area under each distribution remains unity. Note that
the peak in the distribution moves to larger values of $\DSab$ as
the non-centrality parameter $\lambda$ is increased.}
\label{fig:noncentralchi}
\end{center}
\end{figure}

If, however, our measurement gives $\DSab <\gamma_n(p)$ then we
cannot say the two models are distinct, at least not at the
confidence level $p$. But they may still be separate models and we
were simply unfortunate, with statistical fluctuations producing a
small value of $\DSab$. If we accumulate more data and measure
$\DSab$ again, we may find a different result. To quantify the
probability that two different models $A$ and $B$ will give a
particular value of $\DSab$ requires the use of the non-central
chi-square distribution in~(\ref{prob}). The degree of
non-centrality is given by the quantity $\lambda$
in~(\ref{lambdadef}). Clearly, the more distinct the predictions
$\sigma_i^A$ and $\sigma_i^B$ are from one another, the larger this
number will be. In Figure~\ref{fig:noncentralchi} we plot the
distribution for $\DSab$ for $n=3$ signatures and several values of
$\lambda$. As expected, the larger this parameter is, the more
likely we are to find large values of $\DSab$.

Let us assume for the moment that ``model $A$'' is the experimental
data, which corresponds to an integrated luminosity of $L^{\rm
exp}$. Our ``model $B$'' can then be a simulation with integrated
luminosity $L^{\rm sim} = q L^{\rm exp}$. We might imagine that $q$
can be arbitrarily large, limited only by computational
resources.\footnote{Among other benefits of a large value for $q$
would be the reduction in uncertainties arising from the simulation
side of the comparison, {\em i.e.} assuming that the simulation
perfectly captures both the physics and the detector response, the
remaining uncertainty would be that associated with the experimental
observation associated with $\sigma_i^A$.} We can then
rewrite~(\ref{lambdadef}) as
\begin{equation} \lambda = L^{\rm exp} \sum_i
\frac{ (\sigma_i^A - \sigma_i^B)^2 }{ \sigma_i^A + \frac{1}{q} \,
\sigma_i^B}\, . \label{lambda2} \end{equation}
From this expression it is clear that we can expect the value of
this parameter $\lambda$ to {\em increase} as experimental data is
collected. The larger the value of $L^{\rm exp}$ the less likely it
becomes to find a particularly small value of $\DSab$. This confirms
out basic intuition that given {\em any} observable (or set of
observables) for which the two models predict different values then
with sufficient integrated luminosity it should always be possible
to distinguish the models to arbitrary degree of confidence.

For any given value of $\lambda \neq 0$, the probability that a
measurement of $\DSab$ will fluctuate to a value so small that it is
not possible to separate two distinct models (to confidence level
$p$) is simply the fraction of the probability distribution
in~(\ref{prob}) that lies to the left of the value $\gamma_n(p)$. If
we wish to be at least 95\% certain that our measurements will
correctly recognize that two different models are indeed distinct we
must require
\begin{equation}
P = \int_{\gamma_n(p)}^{\infty} n \, \chi_{n,\lambda}^2(n \Delta
S^2_{AB})\, d(\Delta S^2_{AB}) = \int_{n\gamma_n(p)}^{\infty}
\chi_{n,\lambda}^2(y) \, dy \geq 0.95\, . \label{condition2}
\end{equation}
Since the value of the integral $P$ in~(\ref{condition2}) decreases
monotonically as $\lambda$ increases the value of this parameter
which makes~(\ref{condition2}) an equality is the minimum
non-centrality value $\lambda_{\rm min}(n,p)$ such that the two
models can be distinguished.

\begin{table}[tb]
\centering
\begin{tabular}{c||c|c|c|c}
 & \multicolumn{4}{c}{Confidence Level $p$}\\ \hline
 n & 0.95 & 0.975 & 0.99 & 0.999  \\
 \hline \hline
 1 & 12.99 & 17.65 & 24.03 & 40.71 \\
 2 & 15.44 & 20.55 & 27.41 & 44.99 \\
 3 & 17.17 & 22.60 & 29.83 & 48.10 \\
 4 & 18.57 & 24.27 & 31.79 & 50.66 \\
 5 & 19.78 & 25.71 & 33.50 & 52.88 \\
 6 & 20.86 & 26.99 & 35.02 & 54.88 \\
 7 & 21.84 & 28.16 & 36.41 & 56.71 \\
 8 & 22.74 & 29.25 & 37.69 & 58.40 \\
 9 & 23.59 & 30.26 & 38.89 & 59.99 \\
 10 & 24.39 & 31.21 & 40.02 & 61.48 \\
\hline
\end{tabular}
{\caption{\label{tgammatable}\footnotesize {\bf List of
$\lambda_{\rm min}(n,p)$ values for various values of the parameters
$n$ and $p$}. A distribution such as those in
Figure~\ref{fig:noncentralchi} with $\lambda = \lambda_{\rm
min}(n,p)$ will have precisely the fraction $p$ of its total area at
larger values of $\DSab$ than the corresponding critical value
$\gamma_n(p)$ from Table~\ref{gammatable}. A graphical example of
this statement is shown in Figure~\ref{fig:lambda_min}.}}
\end{table}

\begin{figure}[tb]
\begin{center}
\includegraphics[scale=1.0,angle=0]{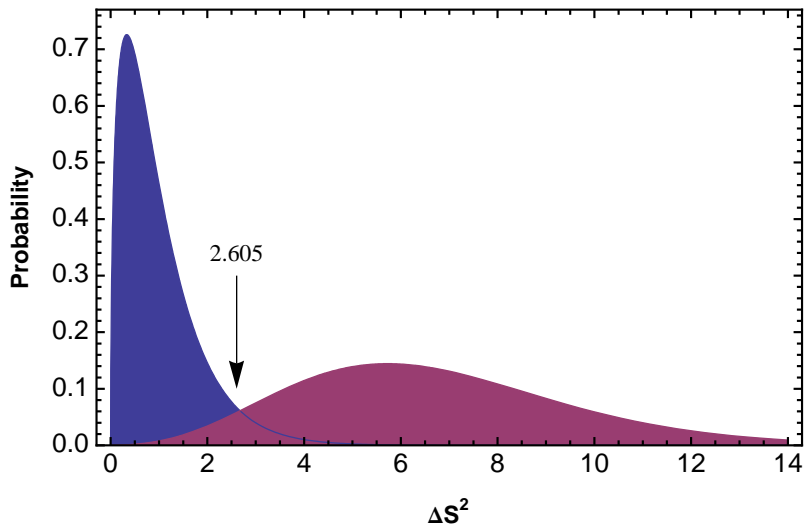}
\caption{\footnotesize \textbf{Determination of $\lambda_{\rm min}$
for the case $n=3$.} The plot shows an example of the distribution
of $\DSab$ for $n=3$. The curve on the left represent $\lambda=0$
case, i.e. values we will get when we compare a model to itself.
95\% of the possible outcomes of this comparison are below 2.61
which is shown on the plot. The curve on the right has
$\lambda=17.17$ and 95\% of the curve is beyond 2.61. As $\lambda$
increases, this curve moves further to the right and gets flatter.}
\label{fig:lambda_min}
\end{center}
\end{figure}

In other words for two distinct models~$A$ and~$B$, any combination
of $n$ experimental signatures such that $\lambda > \lambda_{\rm
min}(n,p=0.95)$ will be effective in demonstrating that the two
models are indeed different 95\% of the time, with a confidence
level of 95\%. We have successfully reduced the problem to an
exercise in pure mathematics, as these $\lambda_{\rm min}(n,p)$
values can be calculated analytically without regard to the physics
involved. A collection of values for small values of $n$ are given
in Table~\ref{tgammatable}. Note that as we increase $n$ the
necessary value $\lambda_{\rm min}$ increases, reflecting the fact
that as more observations are made we should expect that it will
become increasingly likely that at least one will show a large
deviation. Indeed, the quantity $\lambda$ can be thought of as a
measure of the overall distance from $\DSab = 0$ in the
$n$-dimensional signature space in units of the variances. As an
example, again consider the case where $n=3$. For this value of $n$
the corresponding $\gamma_3(0.95) = 2.61$ value can be found from
Table~\ref{gammatable}, while we can find $\lambda_{\rm min}(3,0.95)
= 17.17$ from Table~\ref{tgammatable}. We plot the
distributions~(\ref{prob}) for $\lbr n, \lambda \rbr = \lbr 3,0\rbr$
and $\lbr 3,17.17\rbr$ simultaneously in
Figure~\ref{fig:lambda_min}. By construction, the area of the
non-central distribution to the left of the indicated value of
$(\DSab) = 2.61$ will be precisely 5\% of the total area.

Having reached the end of our somewhat lengthy digression on
probability theory we now return to the physics issue at hand. The
requirement that $\lambda \geq \lambda_{\rm min}(n,p)$ can be
translated into a condition on the signature set and/or luminosity
via the definition in~(\ref{lambda2}). Let us make one final
notational definition
\begin{equation} R_{AB} = \sum_i (R_{AB})_i = \sum_i
\frac{ (\sigma_i^A - \sigma_i^B)^2 }{ \sigma_i^A + \frac{1}{q} \,
\sigma_i^B} \, \label{Rdef} \end{equation}
where $R_{AB}$ has the units of a cross section. Our condition for
95\% certainty that we will be able to separate two truly distinct
models at the 95\% confidence level becomes
\begin{equation}  L_{\rm exp}  \geq \frac{\lambda_{\rm min}(n,0.95)}{R_{AB}}\,
. \label{ourcond} \end{equation}
Given two models~$A$ and~$B$ and a selection of $n$ signatures both
$\lambda_{\rm min}(n,0.95)$ and $R_{AB}$ are completely determined.
Therefore the minimum amount of integrated luminosity needed to
separate the models experimentally will be given by
\begin{equation}  L_{\rm min}(p) = \frac{\lambda_{\rm
min}(n,p)}{R_{AB}}\, . \label{Lmin} \end{equation}
We will be using~(\ref{Lmin}) repeatedly throughout the rest of this
paper. A well-chosen set of signatures will be the set that makes
the resulting value of $L_{\rm min}$ determined from~(\ref{Lmin}) as
small as it can possibly be.

\subsection{Specific Signature Choice}
\label{sec:sigs}

Following the discussion in Section~\ref{sec:separate} we are in a
position to define the goal behind our signature selection more
precisely. We wish to select a set of $n$ signatures $S_i$ such that
the quantity $L_{\rm min}(p)$ as defined in~(\ref{Lmin}), for a
given value of $p$, is as small as it can possibly be over the
widest possible array of model pairs~$A$ and~$B$. We must also do
our best to ensure that the $n$~signatures we choose to consider are
reasonably uncorrelated with one another so that the statistical
treatment of the preceding section is applicable.
We will address the latter issue below, but let us first turn our
attention to the matter of optimizing the signature list.
%

We took as our starting point an extremely large initial set of
possible signatures. These included all the counting signatures and
most of the kinematic distributions used
in~\cite{ArkaniHamed:2005px}, all of the signatures
of~\cite{Feldman:2008hs}, several ``classic'' observables common in
the literature~\cite{Branson:2001ak} and several more which we
constructed ourselves. Removing redundant instances of the same
signature this yielded 46~independent counting signatures and
82~kinematic distributions represented by histograms, for 128
signatures in total. We might naively think that the best strategy
is to include {\em all} of these signatures in the analysis
(neglecting for now the issue of possible correlations among them).
In fact, if the goal is statistically separating two models, the
optimal strategy is generally to choose a rather small subset of the
total signatures. Let us understand why that is the case. To do so
we need a quantitative way of establish an absolute measure of the
``power'' of any given signature to separate two models~$A$ and~$B$.
This can be provided by considering the condition in~(\ref{Lmin}).
For any signature $S_i$ we can define an individual $(L_{\rm
min})_i$ by
\begin{equation}
(L_{\rm min})_i = \lambda_{\rm min}(1,p) \, \frac{\sigma_i^A +
\frac{1}{q} \, \sigma_i^B}{(\sigma_i^A - \sigma_i^B)^2} \,
,\label{Lmindef}
\end{equation}
where, for example, $\lambda_{\rm min}(1,0.95) = 12.99$. This
quantity is exactly the integrated luminosity required to separate
models $A$ and $B$, to confidence level $p$, by using the single
observable $S_i$. For a list of $N$ signatures it is possible to
construct $N$ such $(L_{\rm min})_i$ values and order them from
smallest value (most powerful) to largest value (least powerful). If
we take any subset $n$ of these, then the requisite $L_{\rm min}$
that results from considering all $n$ simultaneously is given by
\begin{equation}
L_{\rm min} = \frac{\lambda_{\rm min}(n,p)}{\lambda_{\rm min}(1,p)}
\, \left\{ (L_{\rm min})_1^{-1} + (L_{\rm min})_2^{-1} + \cdots +
(L_{\rm min})_n^{-1} \right\}^{-1}\,  . \label{resistor}
\end{equation}
Referring back to Table~\ref{tgammatable} we see that the ratio
$\lambda_{\rm min}(n,p) / \lambda_{\rm min}(1,p)$ grows with $n$.
This indicates that as we add signatures with ever diminishing
$(L_{\rm min})_i$ values we will eventually encounter a point of
negative returns, where the resulting overall $L_{\rm min}$ starts
to grow again.

\begin{figure}[tb]
\begin{center}
\includegraphics[scale=0.60,angle=0]{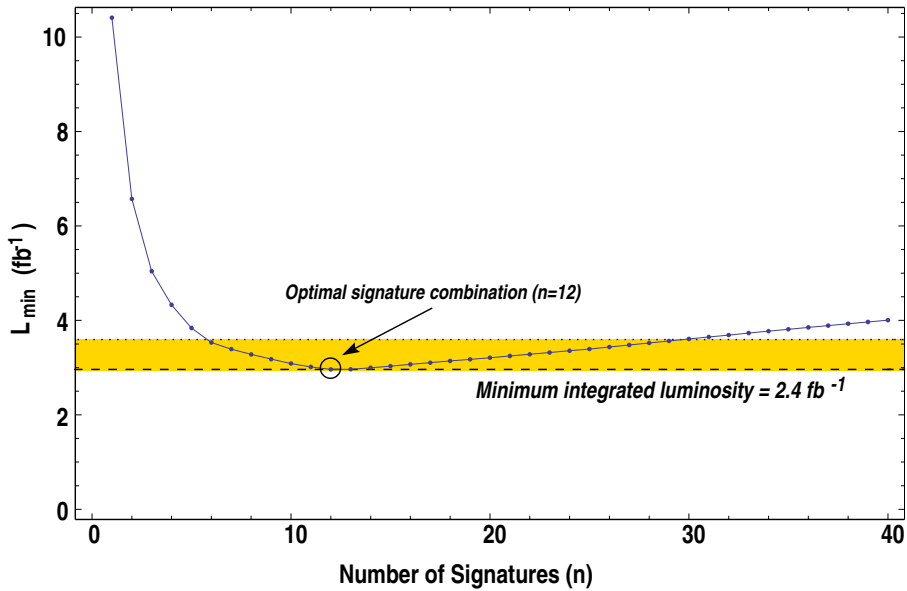}
\caption{\footnotesize \textbf{An example of finding an ``optimal''
signature list.} By sequentially ordering the calculated $(L_{\rm
min})_i$ values for any particular pair of models in ascending
order, it is always possible to find the optimal set of signatures
for that pair by applying~(\ref{resistor}). In this particular
example the minimum value of $L_{\rm min}$ is found after combining
just the first~12 signatures. After just the best six signatures we
are already within 20\% of the optimal value, as indicated by the
shaded band.} \label{fig:ri_plot}
\end{center}
\end{figure}

As more signatures are added, the threshold for adding the next
signature in the list gets steadily stronger. For a {\em particular
pair} of models, $A$~and~$B$, it is always possible to find the
optimal list of signatures from among a given grand set by ordering
the resulting $(L_{\rm min})_i$ values and adding them sequentially
until a minimum of $L_{\rm min}$ is observed. To do so, we note that
kinematic distributions must be converted into counts (and all
counts are then converted into effective cross sections). This
conversion requires specifying an integration range for each
histogram. The choice of this range can itself be optimized, by
considering each integration range as a separate signature and
choosing the values such that $(L_{\rm min})_i$ is minimized.

Figure~\ref{fig:ri_plot}, based on an actual pair of models from one
of our model lines, represents the outcome of just such an
optimization procedure. In this case a clearly optimal signature set
is given by the 12~signatures represented by the circled point,
which yields $L_{\rm min} = 2.4$ fb$^{-1}$. The situation in
Figure~\ref{fig:ri_plot} is typical of the many examples we studied:
the optimal signature set usually consisted of $\order(10)$
signatures. If we are willing to settle for a luminosity just 20\%
higher than this minimal value then we need only $\order(5)$
signatures, typically.\footnote{It is interesting to compare this to
the results of~\cite{ArkaniHamed:2005px} in which the effective
dimension of signature space was found to be also $\order(5)$ to
$\order(10)$.} This 20\% range is indicated by the shaded band in
Figure~\ref{fig:ri_plot}. Of course this ``optimal'' set of
signatures $\lbr S_i \rbr$ is only optimal for the specific pair of
models~$A$ and~$B$. We must repeat this optimization procedure on a
large collection of model pairs and form a suitable average of the
results in order to find a set of signatures $\lbr S_i \rbr$ that
best approximates the truly optimal set over the widest possible set
of model pairs $\lbr A, B \rbr$. The lists we will present at the
end of this section represent the results of just such a procedure.

But before we present them, we must now address the issue of
correlations. To be able to use the analytic results of our
statistical presentation of the problem in
Section~\ref{sec:separate} we must be careful to only choose
signatures from a list in which all the members are uncorrelated
with one another. This immediately suggests a dilemma: once a
signature is chosen, many others in the grand set will now be
excluded for being correlated with the first. This complicates the
process of optimization considerably -- the task now becomes to
perform the above optimization procedure over the largest possible
list of {\em uncorrelated} (or at least minimally correlated)
signatures. To find the correlation between any two signatures $S_i$
and $S_j$ it is sufficient to construct their correlation
coefficient $\rho_{ij}$, given by
\begin{equation} \rho_{ij} = \frac{{\rm cov}(i,j)}{{\rm var}(i) {\rm var}(j)}
= \lim_{N \rightarrow \infty}  \frac{ \frac{1}{N} \sum_k \left[
\bar{\sigma}^k_i - \sigma_i \right] \left[\bar{\sigma}_j^k -\sigma_j
\right] } {\sqrt{ \frac{1}{N} \sum_k \left[\bar{\sigma}_i^k -
\sigma_i \right]^2} \sqrt{ \frac{1}{N} \sum_k \left[\bar{\sigma}_j^k
- \sigma_j \right]^2}}\, , \label{correlation} \end{equation}
where the $\bar{\sigma}^k$ represent the individual results obtained
from each of the $N$ cross section measurements, labeled by the
index $k$.

In our analysis we estimated the entries in the $128 \times 128$
dimensional matrix of~(\ref{correlation}) in the following crude
manner. We began with a simple MSSM model specified by a parameter
set as in~(\ref{paramset}), with gaugino masses having the unified
ratios of~(\ref{msugra_ratios}). We simulated this model
$N=2000$~times, each time with a different random number seed. The
simulation involved generating 5~fb$^{-1}$ of events using {\tt
PYTHIA 6.4}, which were passed to the detector simulator {\tt PGS4}.
After simulating the detector response and object reconstruction the
default level-one triggers included in the {\tt PGS4} detector
simulation were applied. Further object-level cuts were then
performed, as summarized in Table~\ref{objcuts}. After these
object-specific cuts we then applied an event-level cut on the
surviving detector objects similar to those used
in~\cite{ArkaniHamed:2005px}. Specifically we required all events to
have missing transverse energy $\slashed{E}_T
> 150 \GeV$, transverse sphericity $S_T > 0.1$, and $H_T > 600 \GeV$ (400
GeV for events with 2 or more leptons) where $H_T = \slashed{E}_T +
\sum_{\rm Jets} p^{\rm jet}_T$. Once all cuts were applied the grand
list of 128~signatures was then computed for each run, and from
these signatures the covariance matrix in~(\ref{correlation}) was
constructed. All histograms and counting signatures were constructed
and analyzing using the {\tt ROOT}-based analysis package {\tt
Parvicursor}~\cite{Parvicursor}.

\begin{table}[tb]
\centering
\begin{tabular}{|c||c|c||}
\hline
 Object & Minimum $p_T$ & Minimum $|\eta|$\\
 \hline \hline
 Photon & 20 GeV & 2.0 \\
 Electron & 20 GeV & 2.0 \\
 Muon & 20 GeV & 2.0 \\
 Tau & 20 GeV & 2.4 \\
 Jet & 50 GeV & 3.0 \\
\hline
\end{tabular}
{\caption{\label{objcuts}\footnotesize {\bf Initial cuts to keep an
object in the event record}. After event reconstruction using the
package {\tt PGS4} we apply additional cuts to the individual
objects in the event record. Detector objects that fail to meet the
above criteria are removed from the event record and do not enter
our signature analysis. These cuts are applied to all analysis
described in this paper.}}
\end{table}

Not surprisingly, many of the signatures considered in our grand
list of 128~observables were highly correlated with one another. For
example, the distribution of transverse momenta for the hardest jet
in any event was correlated with the overall effective mass of the
jets in the events (defined as the scalar sum of all jet $p_T$
values: $M_{\rm eff} = \sum_{\rm Jets} p^{\rm jet}_T$). Both were
correlated with the distribution of $H_T$ values for the events, and
so forth. The consistency of our approach would then require that
only a subset of these signatures can be included. One way to
eliminate correlations is to partition the experimental data into
mutually-exclusive subsets through some topological criteria such as
the number of jets and/or leptons. For example, the distribution of
$H_T$ values in the set having any number of jets and zero leptons
will be uncorrelated with the same signature in the set having any
number of jets and at least one lepton. Our analysis indicated that
this partitioning strategy has its limitations, however. The
resolving power of any given signature tends to diminish as the set
it is applied to is made ever more exclusive. This is in part due to
the diminishing cross-section associated with the more exclusive
final state (recall that our metric for evaluating signatures is
proportional to the cross-section). It is also the case that the
statistical error associated with extracting these cross-section
values from the counts will grow as the number of events drops. We
were thus led to consider a very simple two-fold partitioning of the
data:
\begin{equation} \begin{array}{c} N_{\rm jets} \leq 4
\,\, {\rm versus}\,\, N_{\rm jets} \geq 5, \\
N_{\rm leptons} =0\,\, {\rm versus}\,\, N_{\rm leptons} \geq 1.
\end{array}
\label{partition} \end{equation}
This choice of data partitioning is reflected in the signature
tables at the end of this section.

Within each if the four subsets it is still necessary to perform a
correlation analysis and construct the matrix
in~(\ref{correlation}). Let us for the moment imagine that we are
willing to tolerate a correlation among signatures given by some
value $\epsilon$. Then the matrix of correlations
in~(\ref{correlation}) can be converted into a matrix $C_{ab}$ which
defines the uncorrelated signatures by assigning the values
\begin{equation} C_{ab} = \left\{
\begin{array}{rl}
 1 & \textrm{if } \rho_{ab} \leq \epsilon \\
 0 & \textrm{if } \rho_{ab} > \epsilon \, .
\end{array} \right.
\label{Cmatrix} \end{equation}
The matrix $C_{ab}$ is actually the adjacency matrix of a
graph\footnote{A graph is a set of vertices connected by edges. An
element of an adjacency matrix of a graph is 1 if there is an edge
between two vertices, 0 otherwise.} and the problem of finding all
the possible sets of uncorrelated signatures is equivalent to
finding all the complete subgraphs (or `clique') of that graph. A
complete graph is a graph which has an edge between each vertex. In
terms of our problem, this means a set of signatures having at most
a correlation at the level of $\epsilon$ between any two of them.
This is a well-known problem in combinatorics that becomes
exponentially more difficult to solve as the number of signatures
increases. For our purposes we will be working with relatively small
sets of signatures which were pre-selected on the basis of their
effectiveness for separating $\alpha = 0$ from non-zero values of
this parameter. Then from these sets we will proceed to build the
maximal subgraph for our choice of allowed correlation $\epsilon$.

\begin{table}[t]
{\begin{center} \begin{small}
\begin{tabular}{|l|l|c|c|} \cline{1-4}
 & Description & Min Value & Max Value \\  \cline{1-4}
1 & $M_{\rm eff}^{\rm any}$ = $\met$ + $\sum_{\rm all} p^{\rm all}_T$ [All events]
& 1250 GeV & End \\
\cline{1-4}
\end{tabular}
\end{small} \end{center}} {\caption{\label{tbl:sigA}\footnotesize{\bf
Signature List A}. The effective mass formed from the transverse
momenta of all objects in the event (including the missing
transverse energy) was the single most effective signature of the
128~signatures we investigated. Since this ``list'' is a single item
it was not necessary to partition the data in any way. For this
distribution we integrate from the minimum value of 1250~GeV to the
end of the distribution.}}
\end{table}

We constructed a large number of model families in the manner
described in Section~\ref{sec:setup}, each involving the range $-0.5
\leq \alpha \leq 1.0$ for the parameter $\alpha$ in steps of
$\Delta\alpha = 0.05$. For each point along these model lines we
generated 100,000~events using {\tt PYTHIA 6.4} and {\tt PGS4}. To
this we added an appropriately-weighted Standard Model background
sample consisting of 5 fb$^{-1}$ each of $t$/$\bar{t}$ and
$b$/$\bar{b}$ pair production, high-$p_T$ QCD dijet production,
single $W^{\pm}$ and $Z$-boson production, pair production of
electroweak gauge bosons ($W^+\,W^-$, $W^{\pm}\,Z$ and $Z\,Z$), and
Drell-Yan processes. To examine which of our 128~signatures would be
effective in measuring the value of the parameter $\alpha$ we fixed
``model~$A$'' to be the point on each of the model lines with
$\alpha = 0$ and then treated each point along the line with $\alpha
\neq 0$ as a candidate ``model~$B$.'' Clearly each model line we
investigated -- and each $\alpha$ value along that line -- gave
slightly different sets of maximally effective signatures. The lists
we will present in Tables~\ref{tbl:sigA},~\ref{tbl:sigB}
and~\ref{tbl:sigC} represent an ensemble average over these model
lines, restricted to a maximum correlation amount $\epsilon$ as
described above.

\begin{table}[t]
{\begin{center} \begin{small}
\begin{tabular}{|l|l|c|c|} \cline{1-4}
 & Description & Min Value & Max Value \\  \cline{1-4}
%
%
1 & $M_{\rm eff}^{\rm jets}$ [0 leptons, $\geq 5$ jets] & 1100 GeV & End \\
2 & $M_{\rm eff}^{\rm any}$ [0 leptons, $\leq 4$ jets]
& 1450 GeV & End \\
3 & $M_{\rm eff}^{\rm any}$ [$\geq 1$ leptons, $\leq 4$ jets]
& 1550 GeV & End \\
\cline{1-4}
%
%
4 & $p_T$(Hardest Lepton) [$\geq 1$ lepton, $\geq 5$ jets] & 150 GeV & End \\
\cline{1-4}
%
%
5 & $M_{\rm inv}^{\rm jets}$ [0 leptons, $\leq 4$ jets] & 0 GeV & 850 GeV \\
\cline{1-4}
\end{tabular}
\end{small} \end{center}} {\caption{\label{tbl:sigB}\footnotesize{\bf
Signature List B}. The collection of our most effective observables,
restricted to the case where the maximum correlation between any two
of these signatures is 10\%. Note that the jet-based effective mass
variables would normally be highly-correlated if we had not
partitioned the data according to~(\ref{partition}). For these
distributions we integrate from ``Min Value'' to ``Max Value''.}}
\end{table}

Let us begin with Table~\ref{tbl:sigA}, which gives the single most
effective signature at separating models with different values of
the parameter $\alpha$. It is the effective mass formed from all
objects in the event
\begin{equation} M_{\rm eff}^{\rm any} = \met + \sum_{\rm all} p^{\rm
all}_T\, , \label{sigA} \end{equation}
where we form the distribution from all events which pass our
initial cuts.  That this one signature would be the most powerful is
not a surprise given the way we have set up the problem. It is the
most inclusive possible signature one can imagine (apart from the
overall event rate itself) and therefore has the largest overall
cross-section. Furthermore, the variable in~(\ref{sigA}) is
sensitive to the mass differences between the gluino mass and the
lighter electroweak gauginos -- precisely the quantity that is
governed by the parameter $\alpha$. Yet as we will see in
Section~\ref{results} this one signature can often fail to be
effective at all in certain circumstances, resulting in a rather
large required $L_{\rm min}$ to be able to separate $\alpha =0$ from
non-vanishing cases. In addition it is built from precisely the
detector objects that suffer the most from experimental uncertainty.
This suggests a larger and more varied set of signatures would be
preferable.

We next consider the five signatures in Table~\ref{tbl:sigB}. These
signatures were chosen by taking our most effective observables and
restricting ourselves to that set for which $\epsilon$ = 10\%. We
again see the totally inclusive effective mass variable
of~(\ref{sigA}) as well as the more traditional effective mass
variable, $M_{\rm eff}^{\rm jets}$, defined via~(\ref{sigA}) but
with the scalar sum of $p_T$ values now running over the jets only.
We now include the $p_T$ of the hardest lepton in events with at
least one lepton and five or more jets, as well as the invariant
mass $M_{\rm inv}^{\rm jets}$ of the jets in events with zero
leptons and 4~or less jets. The various jet-based effective mass
variables would normally be highly correlated with one another if we
were not forming them from disjoint partitions of the overall data
set. The favoring of jet-based observables to those based on leptons
is again largely due to the fact that jet-based signatures will have
larger effective cross-sections for reasonable values of the SUSY
parameters in~(\ref{paramset}) than leptonic signatures. The best
signatures are those which track the narrowing gap between the
gluino mass and the electroweak gauginos and the narrowing gap
between the lightest chargino/second-lightest neutralino mass and
the LSP mass. In this case the first leptonic signature to appear --
the transverse momentum of the leading lepton in events with at
least one lepton -- is an example of just such a signature.

\begin{table}[t]
{\begin{center} \begin{small}
\begin{tabular}{|l|l|c|c|} \cline{1-4}
 & Description & Min Value & Max Value \\  \cline{1-4}
\multicolumn{4}{c}{Counting Signatures} \\ \cline{1-4}
1 & $N_{\ell}\,\,\quad$ [$\geq 1$ leptons, $\leq 4$ jets] & & \\
2 & $N_{\ell^+ \ell^-}$ [$M^{\ell^+\ell^-}_{\rm inv} = M_Z \pm 5 \GeV$] & & \\
3 & $N_B\,\,\quad$ [$\geq 2$ B-jets] &  &  \\ \cline{1-4}
\multicolumn{4}{c}{[0 leptons, $\leq 4$ jets]} \\ \cline{1-4}
4 & $M_{\rm eff}^{\rm any}$ & 1000 GeV & End \\
5 & $M_{\rm inv}^{\rm jets}$ & 750 GeV & End \\
6 & $\met$ & 500 GeV & End \\ \cline{1-4}
\multicolumn{4}{c}{[0 leptons, $\geq 5$ jets]} \\ \cline{1-4}
7 & $M_{\rm eff}^{\rm any}$ & 1250 GeV & 3500 GeV \\
8 & $r_{\rm jet}$ [3 jets $>$ 200 GeV] & 0.25 & 1.0 \\
9 & $p_T$(4th Hardest Jet) & 125 GeV & End \\
10 & $\met$/$M_{\rm eff}^{\rm any}$ & 0.0 & 0.25 \\
\cline{1-4}
\multicolumn{4}{c}{[$\geq 1$ leptons, $\geq 5$ jets]} \\ \cline{1-4}
11 & $\met$/$M_{\rm eff}^{\rm any}$ & 0.0 & 0.25 \\
12 & $p_T$(Hardest Lepton) & 150 GeV & End \\
13 & $p_T$(4th Hardest Jet) & 125 GeV & End \\
14 & $\met$ + $M_{\rm eff}^{\rm jets}$ & 1250 GeV & End \\
\cline{1-4}
\end{tabular}
\end{small} \end{center}} {\caption{\label{tbl:sigC}\footnotesize{\bf
Signature List C}. In this collection of signatures we have allowed
the maximum correlation between any two signatures to be as high as
30\%. Note that some of the signatures are normalized signatures,
(\#8, \#10 and \#11), while the first three are truly counting
signatures. A description of each of these observables is given in
the text. For all distributions we integrate from ``Min Value'' to
``Max Value''.}}
\end{table}

Finally, let us consider the larger ensemble of signatures in
Table~\ref{tbl:sigC}. In this final set we have relaxed our concern
over the issue of correlated signatures, allowing as much as 30\%
correlation between any two signatures in the list. This allows for
a larger number as well as a wider variety of observables to be
included. As we will see in Section~\ref{results} this can be very
important in some cases in which the supersymmetric model has
unusual properties, or in cases where the two $\alpha$ values being
considered give rise to different mass orderings (or hierarchies) in
the superpartner spectrum. In displaying the signatures in
Table~\ref{tbl:sigC} we find it convenient to group them according
to the partition of the data being considered. Note that the
counting signatures are taken over the entire data set.

The first counting signature is simply the total size of the
partition from~(\ref{partition}) in which the events have at least
one lepton and 4~or less jets. This was the only observable taken on
this data set that made our list of the most effective observables.
The next two signatures are related to ``spoiler'' modes for the
trilepton signal. Note that the trilepton signal itself did {\em
not} make the list: this is a wonderful discovery mode for
supersymmetry, but the event rates between a model with $\alpha =0$
and one with non-vanishing $\alpha$ were always very similar (and
low). This made the trilepton counting signature ineffective at
distinguishing between models. By contrast, counting the number of
b-jet pairs (a proxy for counting on-shell Higgs bosons) or the
number of opposite-sign electron or muon pairs whose invariant mass
was within 5~GeV of the Z-mass (a proxy for counting on-shell
Z-bosons) {\em were} excellent signatures for separating models from
time to time. This was especially true when the two models in
question had very different values of $\alpha$ such that the mass
differences between $\wtd{N}_2$ and $\wtd{N}_1$ were quite different
in the two cases. We will give specific examples of such outcomes in
Section~\ref{results}.

The following three sections of Table~\ref{tbl:sigC} involve some of
the same types of observables as in the previous tables, with a few
notable changes and surprises. First note that several of the
observables in Table~\ref{tbl:sigC} involve some sort of
normalization. In particular numbers~8, 10~and 11. Our estimate of
the correlations among signatures found that the fluctuations of
these normalized signatures tended to be less correlated with other
observables for that partition than the un-normalized quantities.
However, normalizing signatures in this way also tended to reduce
their ability to distinguish models. Signature \#8 is defined as the
following ratio
\begin{equation} r_{\rm jet} \equiv \frac{p_T^{\rm jet 3} + p_T^{\rm
jet 4}}{p_T^{\rm jet 1} + p_T^{\rm jet 2}} \label{rjet}
\end{equation}
where $p_T^{\rm jet\,i}$ is the transverse momentum of the $i$-th
hardest jet in the event. For this signature we require that there
be at least three jets with $p_T > 200 \GeV$. This signature, like
the $p_T$ of the hardest lepton or the $p_T$ of the 4th~hardest jet,
was effective at capturing the increasing softness of the products
of cascade decays as the value of $\alpha$ was increased away from
$\alpha = 0$.

Let us note that Lists~A, B~and~C are not mutually disjoint. For
example, signatures~4, 5~and~12 of Table~\ref{tbl:sigC} also appear
in Table~\ref{tbl:sigB}. The signature mix is determined by
attempting to minimize $L_{\rm min}$ via the formula
in~(\ref{resistor}) while attempting to keep the correlations
between any pair of signatures below the targets set above in the
text. As mentioned earlier, larger lists are not always better --
the more signatures one adds, the larger the likelihood that some
pair will be correlated with one another to an unsatisfactory
amount. Furthermore, when signatures are added which are only
occasionally useful, the resolving power of the ensemble can
actually be degraded since the statistical threshold defined by
$\lambda_{\rm min}$ in Table~\ref{tgammatable} grows with the number
of signatures.

We will see some examples of this perverse effect in the next
section in which we will examine the effectiveness of these three
lists. We will do this first against our benchmark models from
Section~\ref{theory} and then against a large ensemble of random
MSSM model lines. Before doing so let us note that by fixing a
particular set of $n$ signatures in every instance -- and indeed,
with the fixed integration ranges indicated in the Tables -- we are
very likely to often be far from the {\em optimal} signature mix and
integration ranges. That is, we should not expect to achieve the
absolute $L_{\rm min}$ value of Figure~\ref{fig:ri_plot} for any
particular pair or points along a model line. If we have chosen our
signature list well, however, then we can hope that the result of
adding the contributions of all $n$~signatures
using~(\ref{resistor}) will be close to the optimal $L_{\rm min}$
value over a large array of model pairs.

\section{Analysis Results} \label{results} 

In this section we will examine how well our signature lists in
Tables~\ref{tbl:sigA}, \ref{tbl:sigB}~and~\ref{tbl:sigC} perform in
measuring the value of the parameter $\alpha$ which appears
in~(\ref{mirage_ratios}). Recall that our specific goal is to
distinguish between a model with $\alpha = 0$ and another with {\em
all other soft terms held equal}, but with $\alpha \neq 0$. We would
like to do this with the least amount of data (or integrated
luminosity) as possible for the smallest values of $\alpha$
possible. We will first demonstrate how the lists perform on our
benchmark cases before turning to an analysis of their performance
on a large ensemble of randomly-generated supersymmetric models.

\subsection{Benchmark Models Analysis}
\label{sec:bm analysis}

We begin with the theory-motivated benchmark models briefly
mentioned at the end of Section~\ref{theory} and discussed at length
in the Appendix. The input values for the soft
supersymmetry-breaking parameters are listed in
Table~\ref{tbl:inputs} at the very end of Section~\ref{theory}. To
remind the reader, model~A is an example of a heterotic string
compactification with K\"{a}hler stabilization of the dilaton while
model~B is an example of a Type~IIB string model with flux
compactification. Each of these examples predicts a particular value
of $\alpha$ as a function of other parameters in the theory;
specifically, model~A predicts $\alpha\simeq0.3$, while model~B
predicts $\alpha\simeq1$.
Further details can be found in the Appendix (and references
therein), but these details are not relevant for our purposes in
this section.

\begin{table}[t]
{\begin{center}
\begin{tabular}{|l|c|c|c|l|c|c|}\cline{1-7}
Parameter & Point A & Point B & & Parameter & Point A & Point B \\
\cline{1-7} $m_{\wtd{N}_1}$ & 85.5 & 338.7 &
& $m_{\tilde{t}_{1}}$ & 844.7 & 379.9 \\
$m_{\wtd{N}_{2}}$ & 147.9 & 440.2 &
& $m_{\tilde{t}_{2}}$ & 1232 & 739.1 \\
$m_{\wtd{N}_{3}}$ & 485.3 & 622.8 &
& $m_{\tilde{c}_{L}}$, $m_{\tilde{u}_{L}}$ & 1518 & 811.7 \\
$m_{\wtd{N}_{4}}$ & 494.0 & 634.3 & & $m_{\tilde{c}_{R}}$,
$m_{\tilde{u}_{R}}$ & 1520 & 793.3 \\ \cline{1-7}
$m_{\wtd{C}_{1}^{\pm}}$ & 147.7 & 440.1 & & $m_{\tilde{b}_{1}}$ & 1224 & 676.8
\\
$m_{\wtd{C}_{2}^{\pm}}$ & 494.9 & 635.0 & & $m_{\tilde{b}_{2}}$ & 1507 & 782.4
\\
$m_{\tilde{g}}$ & 510.0 & 818.0 &
& $m_{\tilde{s}_{L}}$, $m_{\tilde{d}_{L}}$ & 1520 & 815.4 \\
$\mu$ & 476.1 & 625.2 & & $m_{\tilde{s}_{R}}$, $m_{\tilde{d}_{R}}$ &
1520 & 793.5 \\ \cline{1-7}
$m_{h}$ & 115.2 & 119.5 & & $m_{\tilde{\tau}_{1}}$ & 1487 & 500.4 \\
$m_{A}$ & 1557 & 807.4 & & $m_{\tilde{\tau}_{2}}$ & 1495 & 540.4 \\
$m_{H^0}$ & 1557 & 806.8 & & $m_{\tilde{\mu}_{L}}$, $m_{\tilde{e}_{L}}$ & 1500 &
545.1 \\
$m_{H^\pm}$ & 1559 & 811.1 & & $m_{\tilde{\mu}_{R}}$, $m_{\tilde{e}_{R}}$ & 1501
& 514.6 \\
\cline{1-7}
\end{tabular}
\end{center}}
{\caption{\label{tbl:bm inputs phys masses}\footnotesize{\bf Low
energy physical masses for benchmark points}. Low energy physical
masses (in units of GeV) are given at the scale~1~TeV. All points
are taken to have $\mu > 0$. The actual value of $\tan\beta$ is
fixed in the electroweak symmetry-breaking conditions.}}
\end{table}

The input values of Table~\ref{tbl:inputs} were evolved from the
input scale $\Lambda_{\UV} = 2 \times 10^{16} \GeV$ to the
electroweak scale of~1~TeV by solving the renormalization group
equations. For this we use the computer package {\tt
SuSpect}~\cite{Djouadi:2002ze}, utilizing two-loop running for all
parameters {\em except} for the gaugino masses. For these we use
one-loop RGEs only in order to maintain the parametrization for the
gaugino soft parameters in terms of $\alpha$ given
by~(\ref{mirage_ratios}). Once run to the low scale the physical
spectra and mixings of the models were computed by {\tt SuSpect}.
The result of this process for our two benchmark models is given in
Table~\ref{tbl:bm inputs phys masses}.

From here we performed a simulation using the combined package of
{\tt PYTHIA} + {\tt PGS4} as described in Section~\ref{sec:sigs}.
For each of these two models a model-line was generated by varying
the parameter $\alpha$ from $\alpha=0$ to $\alpha = 1$, in
increments of 0.05, while keeping all other soft parameters fixed.
Along these model lines the gluino soft mass $M_3$ was held constant
to set the overall scale, and the two parameters $M_1$ and $M_2$
were varied according to the ratios in~(\ref{mirage_ratios}). For
each point 500,000 events were generated using the L1 trigger
options in {\tt PGS4}. After applying further initial cuts as
described in Section~\ref{sec:sigs} the signatures associated with
each of the three lists in Tables~\ref{tbl:sigA},
\ref{tbl:sigB}~and~\ref{tbl:sigC} were constructed. We then used the
criterion for distinguishability described in
Section~\ref{sec:separate} to determine the minimum luminosity
$L_\mathrm{min}$ needed to separate $\alpha = 0$ from all other
points along the line.

\begin{figure}[t]
\begin{center}
    {
      \includegraphics[scale=0.5]{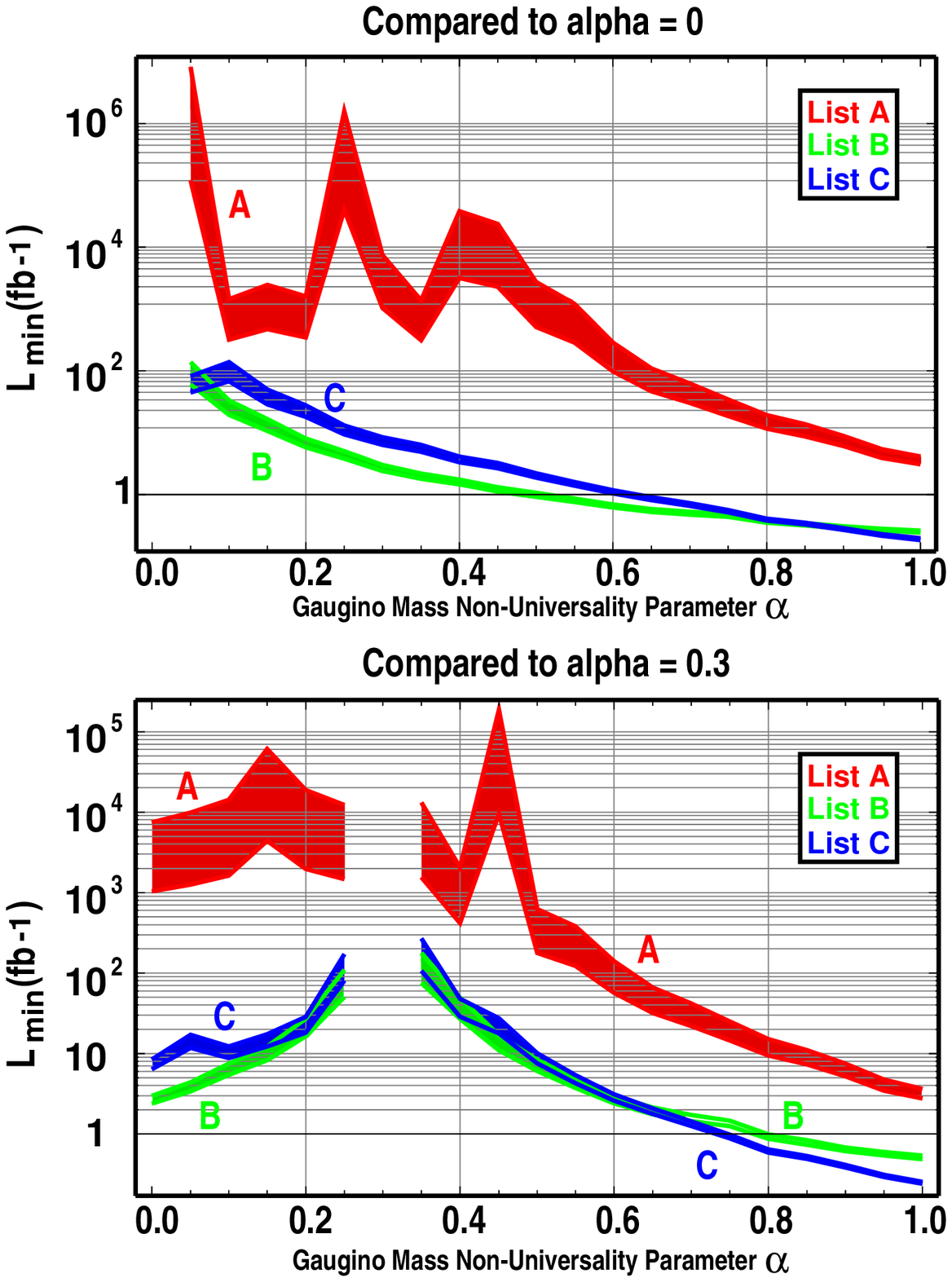}
    }
    {
      \includegraphics[scale=0.5]{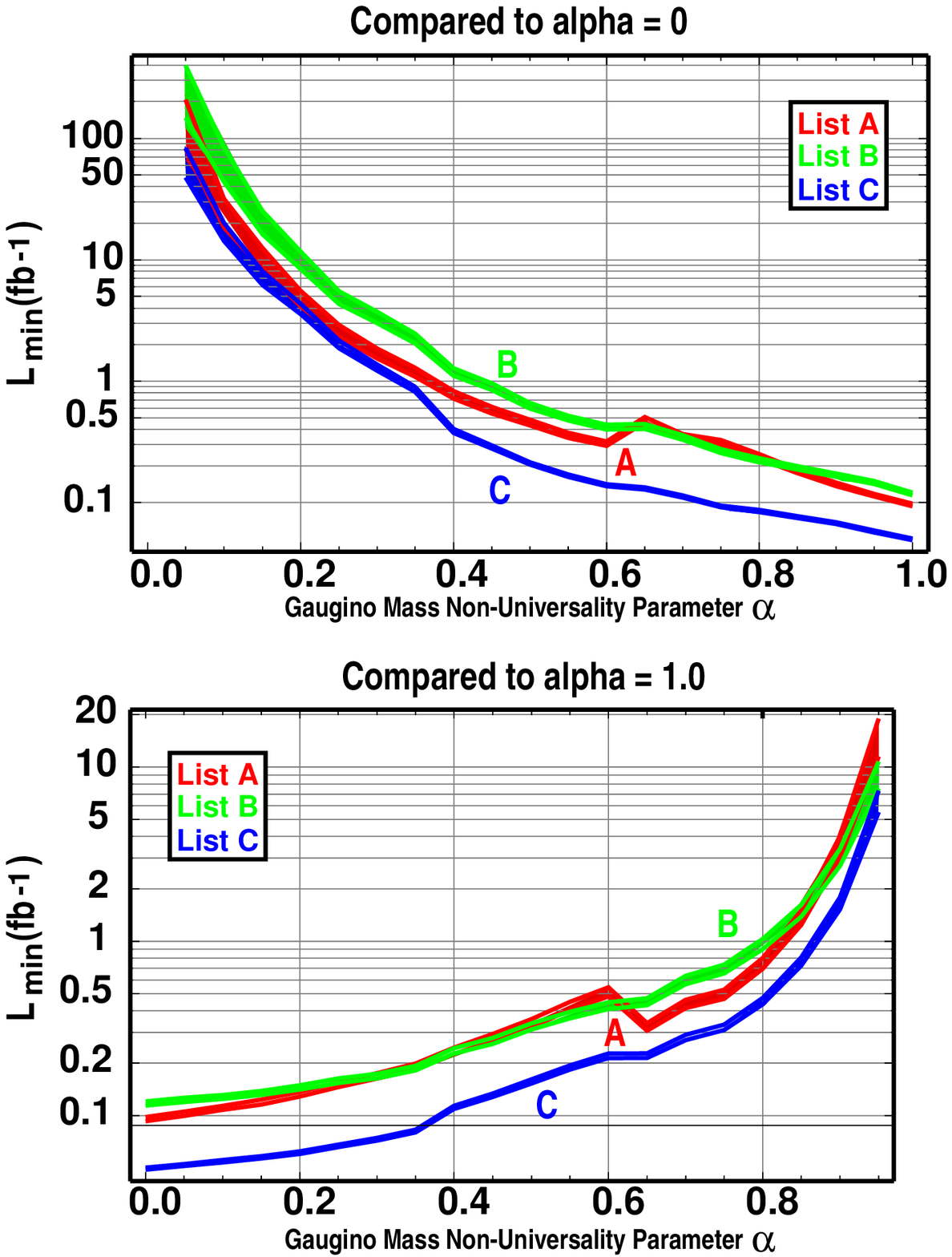}
    }
\caption{\label{plot:bm models Lmin}\footnotesize{\textbf{
$L_\mathrm{min}$ as a function of $\alpha$ for the two benchmark
models.} The three shaded regions correspond to the three signature
lists as indicated by the legend. The lower bound of each shaded
region indicates the minimum integrated luminosity $L_\mathrm{min}$
needed to separate the model with the specified $\alpha$ from
$\alpha = 0$ (top panels) or the predicted value of $\alpha$ (lower
panels). The upper bound of the shaded region represents an estimate
of the 1~sigma upper bound on the calculated value of $L_{\rm min}$
caused by statistical fluctuations.}}
\end{center}
\end{figure}

The results of this analysis are presented in the top panels of
Figure~\ref{plot:bm models Lmin}. The plot on the left corresponds
to benchmark model~A while the one on the right corresponds to
benchmark model~B. The vertical axis shows the minimum luminosity
needed to separate a given $\alpha \neq 0$ scenario from the unified
case of $\alpha = 0$. The three shaded regions represent the three
model lists we used to analyze the data. At the lower edge of each
region is the value of $L_{\rm min}$ as calculated using the
relations in~(\ref{Lmindef}) and~(\ref{resistor}). The upper edge of
each region represents an estimate of the 1~sigma upper bound on the
calculated value of $L_{\rm min}$ caused by statistical fluctuations
({\em i.e.} the fact that the cross-sections extracted from the data
or simulation are not the true cross-sections for each signature).
The lower panels in Figure~\ref{plot:bm models Lmin} represent the
same analysis, but now each of the two models are compared to their
predicted values: $\alpha = 0.3$ for model~A and $\alpha = 1.0$ for
model~B. With the exception of the straw-man List~A in the case of
benchmark model~A, all the lists do an adequate job of
distinguishing points along these alpha-lines with moderate amounts
of integrated luminosity. Naturally, as the two points being
compared approach one another the signature difference between them
become smaller and the needed $L_{\rm min}$ increases. It is
instructive to consider the case of model~A to understand why some
approaches to extracting the parameter $\alpha$ succeed and others
fail.


%
Model~A has nearly universal scalar masses at a rather high scale of
approximately 1.5 TeV, yet the light gluino makes this the model
with the higher overall cross-section. All supersymmetric
observables in this benchmark model are therefore dominated by
gluino pair production and their eventual cascade decays through
highly off-shell squarks. In the analysis the gluino mass is kept
constant along an alpha-line, so the cross-section for the dominant
process $gg \to \tilde{g}\tilde{g}$ is fixed at $\sigma(gg \to
\tilde{g}\tilde{g}) = 13.4\,{\rm pb}$ for this alpha-line. Any
signatures related to this variable will depend on $\alpha$ only via
the change in the gluino branching fractions, which are nearly
constant as a function of the parameter $\alpha$.\footnote{Only the
highly suppressed three-body decay $\tilde{g} \to \tilde{C}_1 q
\bar{q}'$ with $q$ and $\bar{q}'$ representing third-generation
quarks shows any significant dependence on the value of the
parameter $\alpha$ for this benchmark model.} Blunt signatures like
the total $M_{\rm eff}$ variable of~(\ref{sigA}) indicate roughly
the total production cross-section and crude mass scale of the
superpartner being predominantly produced. This is an example in
which the most inclusive possible observable is simply too inclusive
to detect the change in gaugino mass ratios. For this one must
consider processes that produce electroweak gauginos, which are
subdominant by as much as a factor of ten in the case of benchmark
model~A.

\begin{figure}[t]
\begin{center}
\includegraphics[scale=0.65]{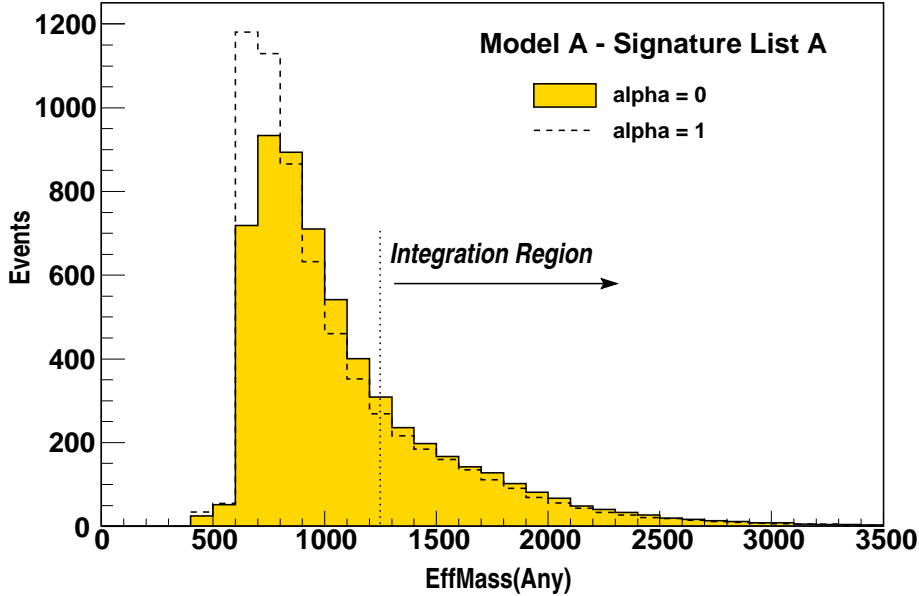}
\caption{\label{plot:Meff_histoA}\footnotesize{\textbf{Distribution
of the variable $M_{\rm eff}^{\rm any}$ from signature List~A for
benchmark model~A.}} Solid filled histogram is the case for
$\alpha=0$, dotted histogram is the case for $\alpha=1$. The lower
bound for the integration region is indicated by the dotted line at
1250~GeV. The sharp lower bound in the distribution is an artefact
of the event-level cuts imposed on the data as described in
Section~\ref{sec:sigs}. In this case the failure of List~A to
separate the two cases is apparent: the difference between the two
histograms is negligible above the value $M_{\rm eff}^{\rm any} =
1250\GeV$. The resolving power would improve dramatically if this
lower bound was relaxed to $M_{\rm eff}^{\rm any} = 500\GeV$, as
demonstrated in Figure~\ref{plot:relax_cut}.}
\end{center}
\end{figure}

\begin{figure}[t]
\begin{center}
\includegraphics[scale=0.85]{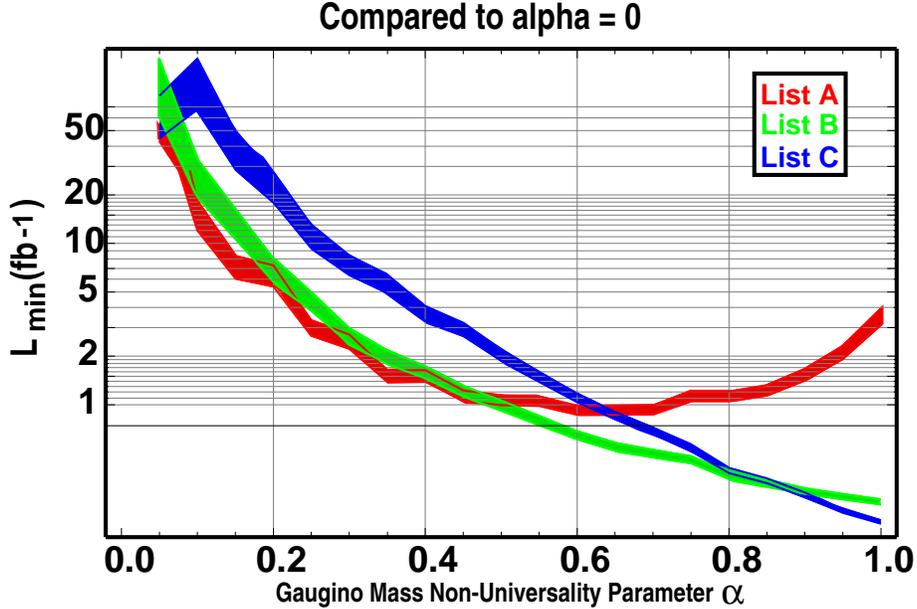}
\caption{\label{plot:relax_cut}\footnotesize{\textbf{
$L_\mathrm{min}$ as a function of $\alpha$ for benchmark model~A
with relaxed lower bound on $M_{\rm eff}^{\rm any}$.}} The three
shaded regions correspond to the three signature lists as in the
upper left panel of Figure~\ref{plot:bm models Lmin}. In this case
the lower bound of the integration range for the single observable
of List~A has been relaxed to 500~GeV.}
\end{center}
\end{figure}

Further compounding the problems for the inclusive signature of
List~A is the fact that the count rate for this particular final
state is varying only very slowly with $\alpha$. Despite the fact
that this count rate can be quite large in this model, the resulting
value of $L_{\rm min}$ is high because the $\Delta S_{AB}$ value for
this particular signature is very near zero. As a result, small
statistical fluctuations in the data or the simulation result in
large fluctuations in the resulting value of $L_{\rm min}$ needed to
truly separate different values of the parameter $\alpha$. This
reflects itself in both the width of the shaded region in the left
panels of Figure~\ref{plot:bm models Lmin} and in the volatility of
the extracted value itself. In Figure~\ref{plot:Meff_histoA} we plot
the distribution of the List~A variable~(\ref{sigA}) in benchmark
model~A for the case of $\alpha=0$ (solid line) and $\alpha =1$
(dashed line). Above our integration cut of 1250~GeV there is very
little difference between the distributions, even for this extreme
case. However, it is clear that some discrimination power is
available had we chosen a different lower bound for integration.
When the lower bound on this particular variable is relaxed to
500~GeV the inclusive $M_{\rm eff}$ variable becomes competitive
with the other signature lists, as shown in
Figure~\ref{plot:relax_cut}.

Benchmark model~A therefore provides us with an example where the
procedure of optimizing the signature list over a wide ensemble of
models has produced a prescription that is most definitely {\em not}
optimal for this particular case. Once a particular model framework
is established it will of course be possible to tailor analysis
techniques to optimize the statistical power of any given signature.
But for our quasi-model-independent analysis we must forgo
optimization in favor of generality. Nevertheless, we gain resolving
power by simply expanding the list of signatures to include those
which are more sensitive to the changes in the lower-mass
electroweak gaugino spectrum. Returning to the left panels of
Figure~\ref{plot:bm models Lmin} it is clear that Lists~B and~C do
far better at measuring the parameter $\alpha$ than the single
$M_{\rm eff}$ variable alone. For example, the jet invariant mass
variables in both lists, as well as the normalized $\met$ signatures
and $p_T({\rm Jet}_4)$ observable of List~C are much more sensitive
to changes in $\alpha$ for this benchmark model than the observable
in~(\ref{sigA}).

But note the {\em reduction} in resolving power that occurs when we
choose the largest signature list. As discussed in
Section~\ref{sec:sigs}, it is clear that the largest possible
signature list is not always the most effective at separating two
theories. In this particular example many of the additional
observables in List~C are not at all helpful in separating different
$\alpha$ values -- particularly the counting variables for which the
total rates are low and the differences across the alpha-line are
small. These additional variables were designed to be most effective
when the mass hierarchies in the superpartner spectrum change as the
value of $\alpha$ is modified, so that dramatic changes in
production rates and/or branching ratios occur. Such threshold
effects do not occur over the $\alpha$ range probed in benchmark
model~A, but do in fact occur for benchmark model~B. This is clearly
evident in the right panels of Figure~\ref{plot:bm models Lmin},
where additional resolving power is obtained when using the expanded
signature List~C.

\begin{figure}[t]
\begin{center}
\includegraphics[scale=0.75]{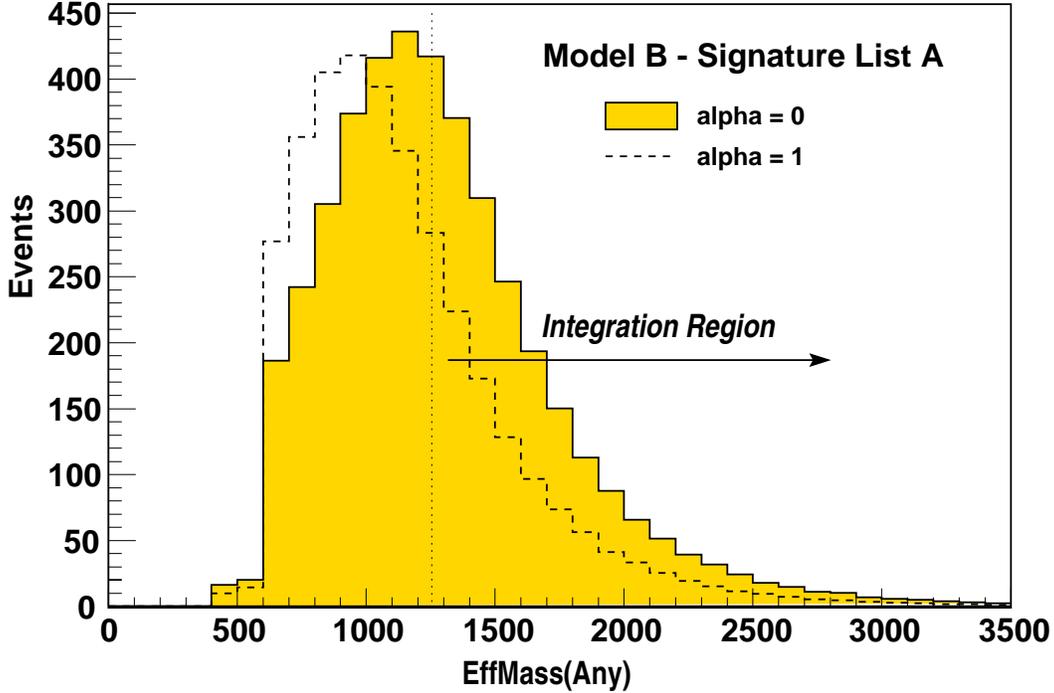}
\caption{\label{plot:Meff_histoB}\footnotesize{\textbf{Distribution
of the variable $M_{\rm eff}^{\rm any}$ from signature List~A for
benchmark model~B.}} Solid filled histogram is the case for
$\alpha=0$, dotted histogram is the case for $\alpha=1$. The lower
bound for the integration region is indicated by the dotted line at
1250~GeV. The sharp lower bound in the distribution is an artefact
of the event-level cuts imposed on the data as described in
Section~\ref{sec:sigs}.}
\end{center}
\end{figure}

We note that the single inclusive variable of~(\ref{sigA}) is much
more effective in benchmark model~B in part because the production
cross-sections for all $SU(3)$-charged superpartners are roughly
equal in magnitude. The inclusive $M_{\rm eff}$ variable no longer
tracks the mass and decay products of a single heavy state so
variations with the parameter $\alpha$ are now more prominent. This
is shown in Figure~\ref{plot:Meff_histoB}, which should be compared
to the case of model~A in Figure~\ref{plot:Meff_histoA}. Note that
the total area under the two curves in Figure~\ref{plot:Meff_histoB}
is nearly identical, highlighting the need to choose a wise value of
the lower bound on the integration region to achieve a high degree
of differentiation. Model~B is similar to the randomly-generated
models we used to design our signature lists and thus the chosen
value of 1250~GeV for this particular observable is close to what
would be the optimal choice for this particular model comparison.

\begin{figure}[t]
\begin{center}
      \includegraphics[scale=0.7]{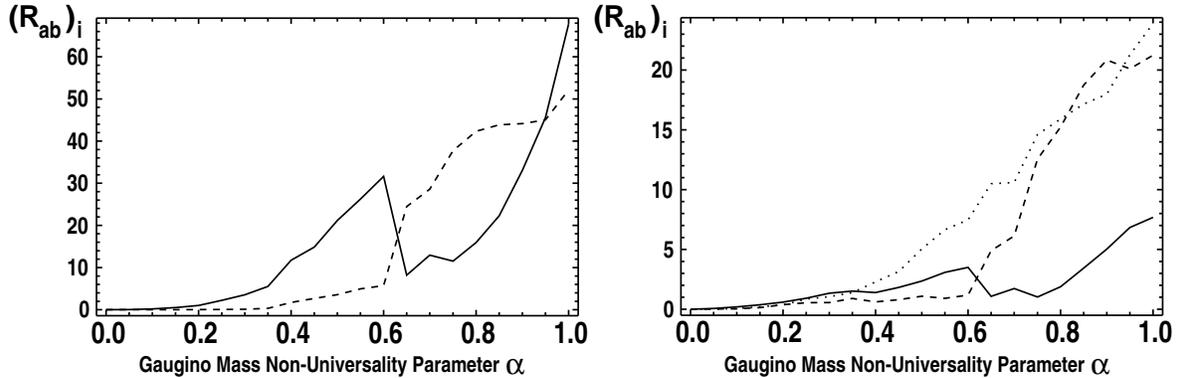}
\caption{\label{plot:B_ListB_Rs}\footnotesize{\textbf{Values of
$(R_{AB})_i$ for the five signatures of List~B as a function of
$\alpha$ for benchmark model~B.}} The ability of each individual
signature from List~B to resolve the case $\alpha =0$ from the
indicated value of $\alpha$ is given by the height of the curve
$(R_{AB})_i$ in the above plots. In the left panel we display
signature~1 (solid curve) and signature~5 (dashed curve). In the
right panel we display signature~2 (solid curve), signature~3
(dashed curve) and signature~4 (dotted curve).}
\end{center}
\end{figure}

Despite the lower overall cross-section for the supersymmetric
signal in benchmark model~B, the three signature lists succeed in
distinguishing the case $\alpha = 0$ from non-vanishing cases with
far less integrated luminosity. In large part this is due to the
richness of the particle spectrum for this model. The superpartner
masses given in Table~\ref{tbl:bm inputs phys masses} are for the
case $\alpha=1$. As $\alpha$ approaches zero the masses of the
lighter neutralinos and lightest chargino fall relative to that of
the gluinos and squarks (which remain constant). Along this
alpha-line several important thresholds are crossed, resulting in
dramatic changes in the relevant branching fractions for the heavier
states. The mix of signatures in List~B and List~C that contribute
most strongly to the resolving power of the overall list changes as
we move along the alpha-line. For example, consider the $(R_{AB})_i$
values of~(\ref{Rdef}) for the five signatures of List~B. We plot
these values in Figure~\ref{plot:B_ListB_Rs} for model~A
corresponding to $\alpha = 0$ and model~B corresponding to the
indicated value of $\alpha \neq 0$.

\begin{figure}[t]
\begin{center}
      \includegraphics[scale=0.5]{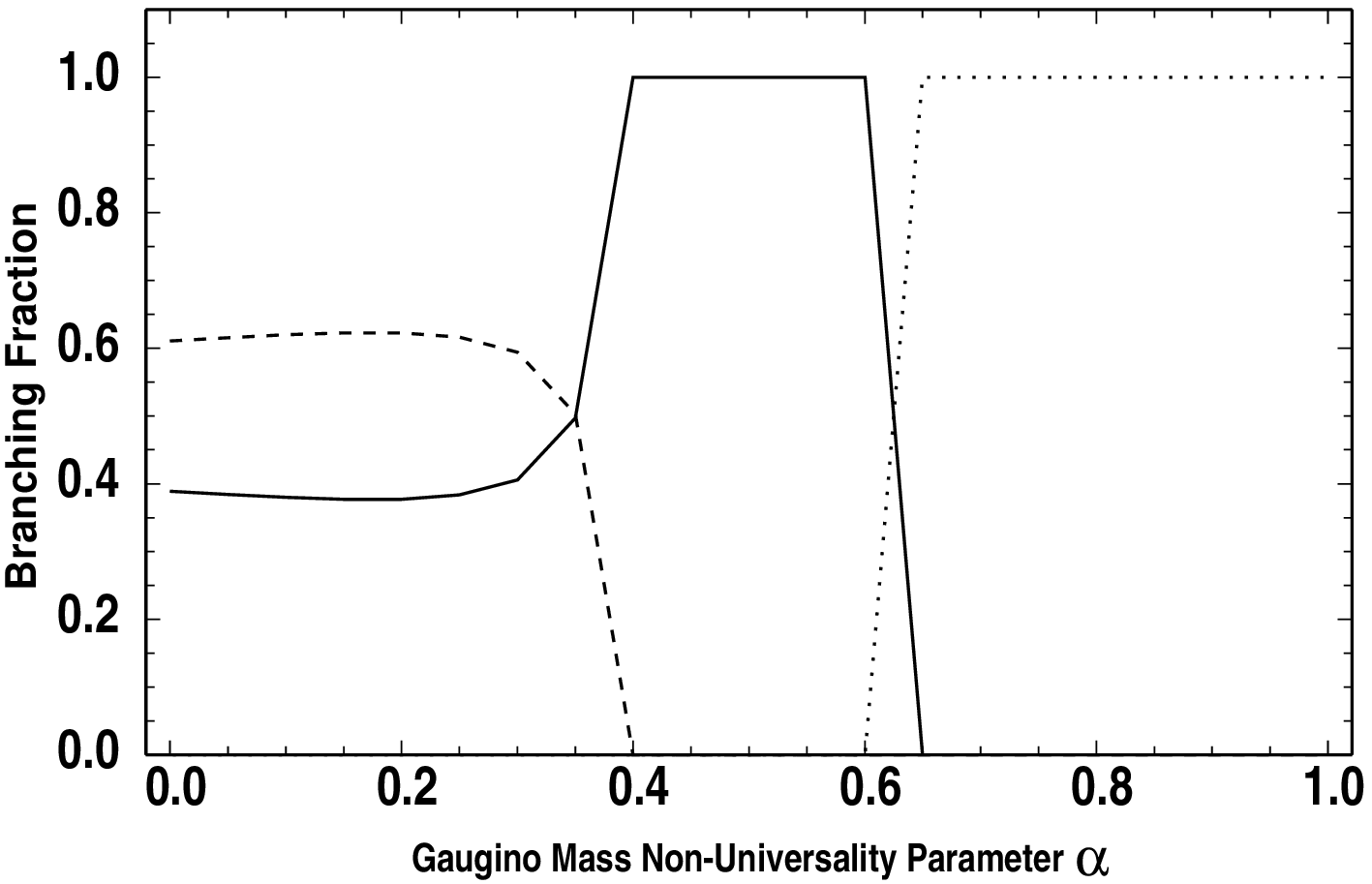}
      \includegraphics[scale=0.5]{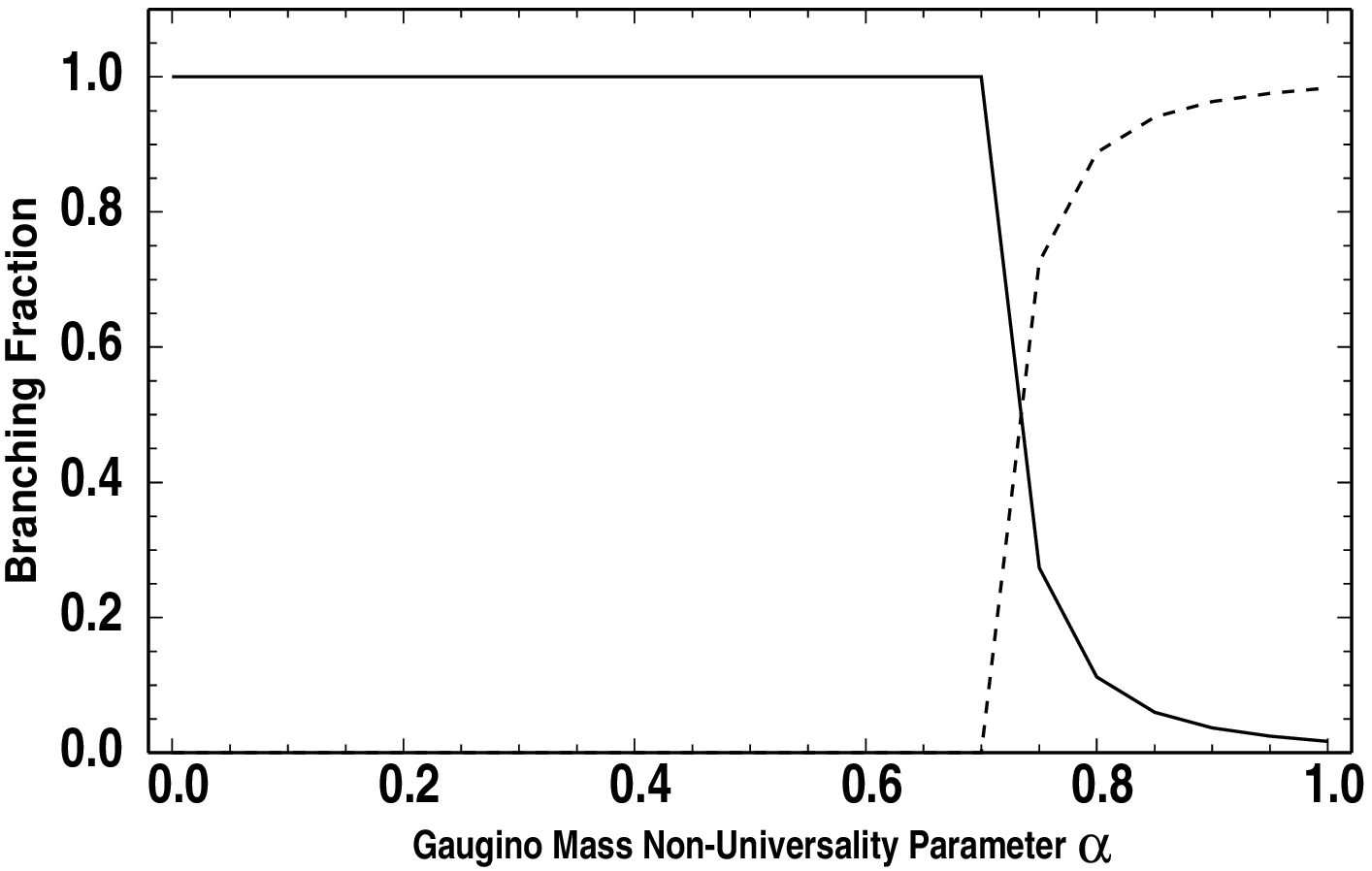}
\caption{\label{plot:stopBR}\footnotesize{\textbf{Branching
fractions for principal decay modes of lightest stop (left panel)
and lightest chargino (right panel) as a function of $\alpha$ for
benchmark model~B.}} In the left panel the decay modes are
$\tilde{t}_1 \to \tilde{N}_1 t$ (dashed curve), $\tilde{t}_1 \to
\tilde{C}_1 b$ (solid curve), and $\tilde{t}_1 \to \tilde{N}_1 c$
(dashed curve). In the right panel the decay modes are $\tilde{C}_1
\to \tilde{N}_1 W$ (solid curve) and $\tilde{C}_1 \to \tilde{t}_1
\bar{b}$ (dashed curve).}
\end{center}
\end{figure}

To understand these curves, we first note that the dominant SUSY
production processes in benchmark model~B are the pair production of
stops and associated production of light squarks with a gluino. The
branching fraction for three of the more important decay modes of
the stop are plotted versus the parameter $\alpha$ in the left panel
of Figure~\ref{plot:stopBR}. For values of $\alpha \lappeq 0.35$,
when both the chargino $\tilde{C}_1$ and the LSP $\tilde{N}_1$ are
sufficiently light, the direct two-body decay into the LSP and a top
quark is dominant. About 50\% of the time the W-bosons from the top
decays on both sides of the events will decay hadronically and the
event will be captured by the first observable in List~B. For the
intermediate region $0.35 \lappeq \alpha \lappeq 0.6$ the stop
decays predominantly via $\tilde{t}_1 \to \tilde{C}_1 b$ and the
final state topology is determined by the subsequent decay of the
chargino. The branching fractions for the primary decay channels of
the chargino $\tilde{C}_1$ are given in the right panel of
Figure~\ref{plot:stopBR}. In this intermediate $\alpha$ region the
chargino is decaying primarily to a W-boson, populating all of the
signatures in List~B.

For larger values of $\alpha \gappeq 0.6$ the chargino $\tilde{C}_1$
and the LSP $\tilde{N}_1$ are now massive enough that the only decay
channel available for the stops is the process $\tilde{t}_1 \to
\tilde{N}_1 c$, producing $\met$ and two jets only. These events are
captured by the second and (especially) fifth observables in List~B,
as evidenced by their rapid growth in significance. For $\alpha
\gappeq 0.7$ charginos that are directly produced (or produced
through cascade decays of heavier squarks) will now decay {\em into}
stops via $\tilde{C}_1 \to \tilde{t}_1 b \to \tilde{N}_1 c b$. This
boosts the resolving power of the signatures with lepton vetoes
relative to the other signatures in List~B.

\begin{figure}[t]
\begin{center}
      \includegraphics[scale=0.5]{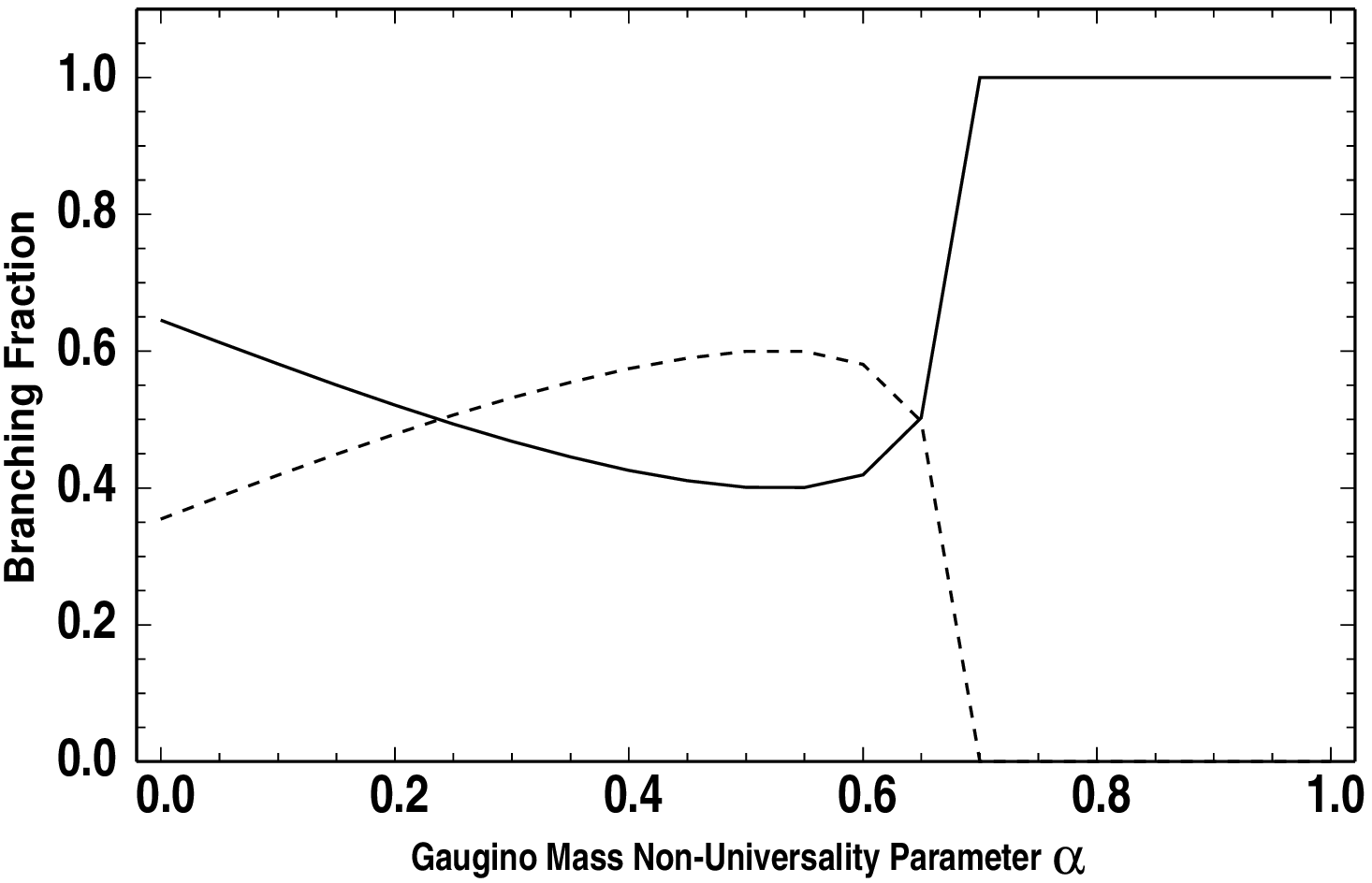}
      \includegraphics[scale=0.7]{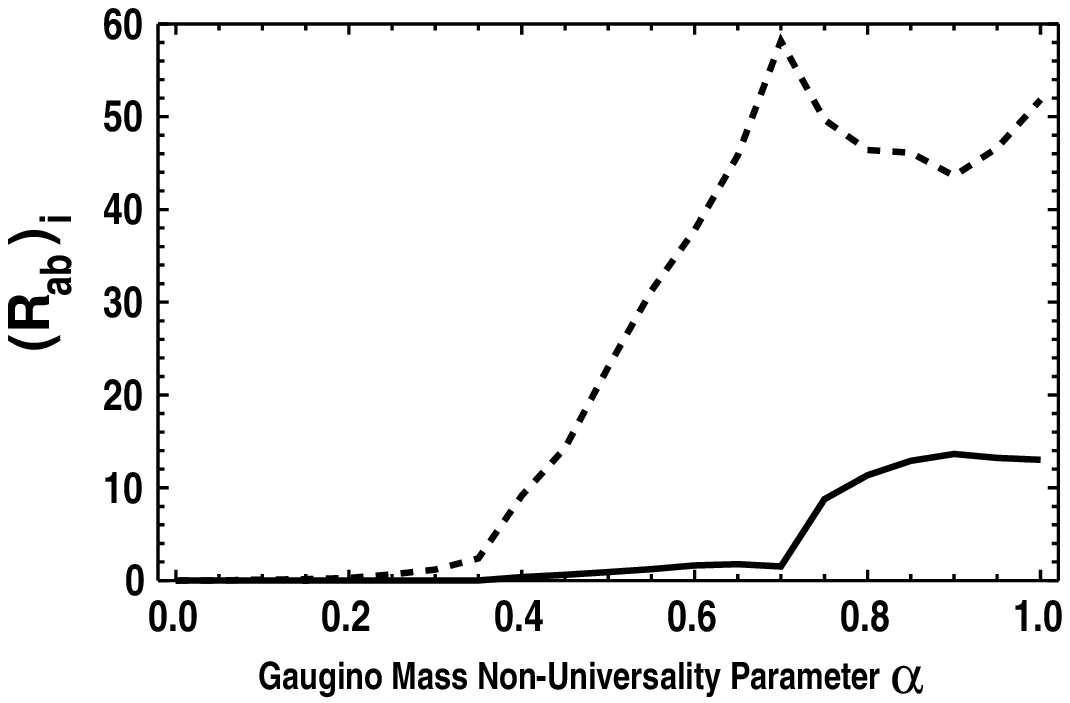}
\caption{\label{plot:gauginoBR}\footnotesize{\textbf{Branching
fraction for next-to-lightest neutralino (left) and $(R_{AB})_i$
values for key counting signatures from List~C (right).}} The
branching fraction of the next-to-lightest neutralino $\tilde{N}_2$
for benchmark model~B is plotted as a function of $\alpha$ in the
left panel. The decay modes are $\tilde{N}_2 \to \tilde{N}_1\, h$
(dashed curve) and $\tilde{N}_2 \to \tilde{N}_1\, Z$ (solid curve).
In the right panel the $(R_{AB})_i$ values for the inclusive
leptonic counting signature (signature~1 -- solid curve) and the
inclusive B-jet counting signature (signature~3 -- dashed curve) are
plotted as function of $\alpha$.}
\end{center}
\end{figure}

Similar arguments explain the behavior of the expanded list of
observables in List~C. Here we will only take a moment to mention
the counting signatures which make their first appearance in our
analysis. Generally speaking, counting signatures are sensitive only
to the total cross-section for the final state being counted.
Changes in the $p_T$ of Standard Model particles produced in
cascades are washed out, making them less useful for comparing
different gaugino mass hierarchies. Counting signatures are
therefore only effective when the two $\alpha$ values being compared
correspond to different decay patterns altogether. This happens in
several instances in benchmark model~B, as we indicated above. The
counting signatures in List~C are specifically designed to consider
changes in the decay table for the next-to-lightest neutralino
$\tilde{N}_2$ -- particularly the appearance of the so-called
``spoiler'' modes for the classic trilepton signal. In the left
panel of Figure~\ref{plot:gauginoBR} the primary decay modes of the
next-to-lightest neutralino $\tilde{N}_2$ are given. We observe that
both of the on-shell decays $\tilde{N}_2 \to \tilde{N}_1 h$ and
$\tilde{N}_2 \to \tilde{N}_1 Z$ are available for this state when
$\alpha \lappeq 0.7$, with the Higgs mode peaking around $\alpha
\simeq 0.6$ before becoming kinematically inaccessible. This
changeover is reflected in the $R_i$ values for the leptonic
counting signature and the B-jet counting signature of List~C, as
shown in the right panel of Figure~\ref{plot:gauginoBR}. Note that
the light stop in benchmark model~B makes this a very B-jet rich
point. In fact, this particular counting signature is one of the
most effective observables in List~C along the alpha-line for this
point.

\subsection{Analysis of a Large Set of Model Variations}
\label{sec:phill}

\begin{table}[t]
{\begin{center}
\begin{tabular}{|c|c|} \cline{1-2}
Input Parameter Range & Variation \\  \cline{1-2}
$400\GeV\ge M_{3} \ge 800\GeV$ & 5 steps\\
$400\GeV\ge \mu \ge 1000\GeV$ & 5 steps\\
$300\GeV\ge (m_{\tilde{e}_{L,R}}, m_{\tilde{\tau}_{L,R}}) \ge 700\GeV$ & 5 steps\\
$500\GeV\ge (m_{\tilde{Q}_L}, m_{\tilde{q}_L}, m_{\tilde{t}_{L,R}},
m_{\tilde{b}L,R}) \ge 1000\GeV$ & 5 steps \\ \cline{1-2}
$\tan\beta=10$ & Fixed \\
$m_A = 1000 \GeV$ & Fixed \\
$A_{\tau}$, $A_t$, $A_b$, $A_e$, $A_u$, $A_d = 0$ & Fixed \\
\cline{1-2}
\end{tabular}
\end{center}}
{\caption{\label{tbl:large_sample_inputs}\footnotesize{\bf MSSM soft
parameters ranges and variation steps used to generate controlled
sample.} These values are given at the electroweak scale. For each
choice of MSSM input, the gaugino unification parameter $\alpha$ was
varied in four steps, from $\alpha=0$ to $\alpha=1.0$}}
\end{table}

We next examine the efficacy of our method by testing it on a large
sample of varying model points.
We will do this in two steps: first
on a controlled sample of models and subsequently on a random
collection of model lines.
Ranges for the MSSM input parameters and variation steps used for
our controlled sample are given in
Table~\ref{tbl:large_sample_inputs}. Only $M_{3}$, $m_{\tilde{l}}$,
$m_{\tilde{q}}$, and $\mu$ were allowed to vary across 5 uniform
steps.  All other soft parameters were held constant. The gaugino
universality parameter $\alpha$ was also varied in 4 steps from
$\alpha=0$, to 0.33, 0.66, and 1.0. These choices discretize the
range of parameter space into 2500 individual model points. Note
that the parameters of Table~\ref{tbl:large_sample_inputs} are given
at the low-energy electroweak scale.
%
%
We emphasize the fact that in this first step we have chosen to
sample the parameter space on a discrete grid rather than sampling
it randomly. While a truly random sampling is necessary for
ultimately testing our method, we here wish to study the performance
of our signature sets as key parameters are varied. Our discrete
grid is designed to keep the overall supersymmetric production rate
roughly fixed, allowing for a more straightforward comparison of
$L_{\rm min}$ values. This course sampling also allows a large
degree of model variation while keeping computation time to a
minimum. Our analysis of a random collection of models will appear
at the end of this subsection.

Simulated data for the model points was generated with the following
procedure. For each model, the {\tt SuSpect} partner code {\tt
SusyHIT}~\cite{Djouadi:2006bz} was used to compute the low-scale
spectrum from the input MSSM soft terms. No renormalization group
evolution was necessary because the input parameters were given at
the electroweak scale. As before {\tt PYTHIA} + {\tt PGS4} was used
to simulate the detector response for each point. A check was
performed to ensure that each model point had a neutralino LSP, and
also that each $\alpha \ge 0$ model point simulated had an
associated $\alpha = 0$ counterpart, so that the minimum luminosity
required to distinguish between the two models, $L_{\rm min}$, could
be computed. Only models satisfying these requirements were retained
for analysis. Exactly 1449~model pairs ($\alpha=0$ and $\alpha\ne0$)
were retained after applying this selection procedure.


\begin{table}[t]
{\begin{center}
\begin{tabular}{|c|c|c|c|c|}
\multicolumn{5}{c}{Largest Production Channel} \\ \hline
Mode & $\alpha=0$ & $\alpha=0.33$ & $\alpha=0.66$ & $\alpha=1.0$\\
\hline \hline
$gg\rightarrow \tilde{g}\tilde{g}$ & 44.6\% & 45.2\% & 42.9\% & 44.8\%\\
$fg\rightarrow \tilde{q}_{R}\tilde{g}$ & 31.1\% & 30.2\% & 33.1\% &
35.7\%\\
$fg\rightarrow \tilde{q}_{L}\tilde{g}$ & 24.3\% & 25.5\% & 23.9\% &
19.4\%\\
\hline \hline \multicolumn{5}{c}{Second Largest Production Channel}
\\ \hline
Mode & $\alpha=0$ & $\alpha=0.33$ & $\alpha=0.66$ & $\alpha=1.0$\\
\hline \hline
$gg\rightarrow \tilde{g}\tilde{g}$ & 2.7\% & 2.1\% & 2.8\% & 1.4\%\\
$fg\rightarrow \tilde{q}_{R}\tilde{g}$ & 42.0\% & 48.8\% & 47.5\% &
45.2\%\\
$fg\rightarrow \tilde{q}_{L}\tilde{g}$ & 42.0\% & 47.1\% & 49.6\% &
53.3\%\\
$f_{i}f_{j}\rightarrow \tilde{\chi}_{2}^{0} \tilde{\chi}_1^{\pm}$ &
13.2\% & 1.9\% & - & -\\
\hline
\end{tabular}
\end{center}}
{\caption{\label{tbl:large_sample_xsec}\footnotesize{\bf Dominant
production modes across all model variations}. At a given $\alpha$
choice, the upper table indicates the percentage of models for which
these modes had the largest cross section, while the lower table
indicates the percentage for which the modes had the second-largest
cross-section. All models exhibit predominantly gluino pair
production, or gluino-quark associated production. A small fraction
of $\alpha=0$ models exhibit neutralino-chargino pair production.
This mode 'switches off' as $\alpha$ is increased from zero, as the
gaugino masses increase.}}
\end{table}

Table~\ref{tbl:large_sample_xsec} gives the dominant production
modes across the entire set of model variations.  The upper table
indicates the mode and percentage of models, for a given~$\alpha$
choice, that occur with the largest cross-section. The lower table
gives the same information for the modes that occur with the
second-largest cross-section.  The majority of models exhibit
squark-gluino associated production, or gluino pair production as
the dominant production mechanism.  Approximately 13\% of $\alpha=0$
models, and  about 2\% of $\alpha=0.33$ models have
neutralino-chargino production as the second most dominant mode.

The particle decay behavior varies throughout the range of model
simulations. However, gluino decays are largely insensitive to
changes in $\alpha$. For the case $\alpha=0$, approximately 68\% of
models have $\tilde{g} \rightarrow \tilde{\chi}_{1}^{\pm}+\bar{q}q'$
as the primary decay channel (the channel having the largest
branching fraction), while 31\% of models have instead
$\tilde{g}\rightarrow \tilde{b}_1 + b$ as the primary channel.  The
$\alpha=0.33$ and 0.66 models exhibit similar ratios. The
$\alpha=1.0$ models show a slight variation, with the distribution
shifting to 70\% and $\sim$30\% respectively. For all $\alpha$
values, approximately 68\% of model variations also exhibit
$\tilde{g} \rightarrow \tilde{\chi}_{1}^{\pm}+\bar{q}q'$ as the
dominant secondary channel (having the second-largest branching
fraction), while 30\% have decays to an on-shell second-generation
squark + quark as the secondary channel.

The first- and second-generation squark decays are equally
insensitive to variations in~$\alpha$.  For all~$\alpha$,
approximately 50\% of models indicate $\tilde{q}_L \rightarrow
\tilde{g} + q$ is the primary decay channel, while the other 50\%
have $\tilde{\chi}_{1}^{\pm} + q'$ as the primary channel.  This is
also the dominant secondary channel in 48\% of the models. Another
40\% have $\tilde{\chi}_{2}^{0} + q$ as the secondary channel. The
$\tilde{q}_R$ are slightly different, with approximately 62\% of
models indicating $\tilde{q}_R \rightarrow \tilde{g} + q$ as the
primary channel, and another 37\% $\tilde{q}_R \rightarrow
\tilde{\chi}_{1}^{0} + q$.  This is also the dominant secondary
channel in 63\% of models, with $\tilde{\chi}_{2}^{0} + q$ the
secondary channel for another 32\%, and the remaining 5\% having
$\tilde{q}_R \rightarrow \tilde{g} + q$.

Due to dependence on the gaugino mass parameters, the chargino
decays are significantly more sensitive to variations of~$\alpha$.
For the $\alpha=0$ case, approximately 74\% of models have
$\tilde{\chi}_{1}^{\pm}\rightarrow W^{\pm}+\tilde{\chi}_{1}^{0}$ as
the primary decay channel. Another 25\% have
$\tilde{\chi}^{\pm}\rightarrow \tilde{\chi}_{1}^{0} + \bar{q}+q'$ as
the primary channel (here the quarks are from the first or second
generation), while the remaining 1\% have instead
$\tilde{\chi}^{\pm}\rightarrow \tilde{\chi}_{1}^{0} + \tau +
\nu_{\tau}$.  As~$\alpha$ increases these three decay channels
persist, however their distribution across each set of models begins
to change, and additional channels begin to appear. For the
$\alpha=0.33$ case, the above channels occur in 65\%, 31\%, and 1\%
of models, respectively. However, now the remaining 3\% of models
have $\tilde{\chi}_{1}^{\pm}\rightarrow
\tilde{\tau}_{1}^{\pm}+\nu_{\tau}$ as the primary channel.  The
$\tilde{\chi}^{\pm}\rightarrow \tilde{\chi}_{1}^{0} + \bar{q}+q'$
channel is the dominant secondary channel for all~$\alpha$
variations.

The $\tilde{\chi}_{2}^{0}$ decay behavior is similarly diverse.  For
case $\alpha=0$, approximately 39\% of models have
$\tilde{\chi}_{2}^{0}\rightarrow \tilde{\chi}_{1}^{0} + q\bar{q}$ as
the primary decay channel, while 23\% have
$\tilde{\chi}_{2}^{0}\rightarrow \tilde{\chi}_{1}^{0} + Z^0$, 28\%
have $\tilde{\chi}_{2}^{0}\rightarrow \tilde{\chi}_{1}^{0} + h^0$,
and another 10\% have $\tilde{\chi}_{2}^{0}\rightarrow
\tilde{\chi}_{1}^{0} + \tau^{+}\tau^{-}$ as the primary channel.
This distribution shifts slightly for $\alpha=0.33$ to 40\%, 26\%,
18\%, and 13\%, respectively.  Here, another 3\% of models have
$\tilde{\chi}_{2}^{0}\rightarrow \tilde{\tau}_{1}^{\pm} + \tau$ as
the dominant channel. For $\alpha=0.66$ it is shifted further to
46\%, 18\%, 15\%, 15\%, where here the remaining 3.4\% of models now
having $\tilde{\chi}_{2}^{0}\rightarrow \tilde{\nu}_{eL} + \nu_{e}$
as the primary channel.  For $\alpha=1.0$, the $Z^{0}$ and $h^{0}$
decays occur less frequently, with only 8\% and 5\% of models having
these as the primary channel.  The $\tilde{\chi}_{1}^{0} +
q\bar{q}$, $\tilde{\chi}_{1}^{0} + \tau^{+}\tau^{-}$, and
$\tilde{\tau}_{1}^{\pm} + \tau$ channels appear with the largest
branching fraction in 56\%, 19\%, and 9\% of models, respectively.


\begin{table}[t]
{\begin{center}
\begin{tabular}{|l||c|c|c||c|c|c||c|c|c|}
\multicolumn{1}{c}{} & \multicolumn{3}{c}{$\alpha = 0.33$} &
\multicolumn{3}{c}{$\alpha = 0.66$} & \multicolumn{3}{c}{$\alpha =
1.0$} \\ \hline
$L_{\rm min}$ value & List A & List B & List C & List A & List B & List C
 & List A & List B & List C \\ \hline
$\leq 1\,{\rm fb}^{-1}$ & 115 & 206 & 282 & 271 & 417 & 474 & 410 & 475 & 484 \\
$\leq 2\,{\rm fb}^{-1}$ & 35 & 93 & 86 & 52 & 36 & 10 & 38 & 9 & 0 \\
$\leq 4\,{\rm fb}^{-1}$ & 49 & 57 & 42 & 52 & 35 & 2 & 24 & 0 & 0 \\
$\leq 10\,{\rm fb}^{-1}$ & 42 & 73 & 50 & 48 & 8 & 0 & 10 & 0 & 0 \\
$\leq 100\,{\rm fb}^{-1}$ & 130 & 40 & 8 & 72 & 0 & 0 & 2 & 0 & 0 \\
$> 100\,{\rm fb}^{-1}$ & 98 & 0 & 0 & 1 & 0 & 0 & 0 & 0 & 0 \\
\cline{1-10}
\end{tabular}
\end{center}}
{\caption{\label{tbl:Phill_Lmin}\footnotesize{\bf Minimum integrated
luminosity $L_{\rm min}$ to separate $\alpha = 0$ from $\alpha \neq
0$ in controlled model sample.} Distribution of $L_{\rm min}$ values
for the three signature sets of Tables~\ref{tbl:sigA},
\ref{tbl:sigB}~and~\ref{tbl:sigC}. In each case we are comparing the
indicated value of $\alpha$ with the case $\alpha = 0$ for the same
set of background model parameters.}}
\end{table}

As with the benchmark models, we analyze the 1449~model pairs using
the three signature sets given in Tables~\ref{tbl:sigA},
\ref{tbl:sigB}~and~\ref{tbl:sigC}.  Due to the large number of model
points we present results statistically in the form of the observed
distribution of $L_{\rm min}$. Table~\ref{tbl:Phill_Lmin} shows the
minimum luminosity required to distinguish between models with
$\alpha=0$ and those with $\alpha\neq 0$ when using, respectively,
signature Lists~A, B~and~C.
%
%
Considering the case of $\alpha = 0.33$ first, signature List~A is
able to successfully resolve a large number of model pairs with
fairly low luminosity. However, only 241 out of the 469 model
variations analyzed for this value of $\alpha$ can be resolved with
less than $10\;{\rm fb}^{-1}$. Signature Lists~B and~C exhibit
significantly stronger resolving power, with List~B able to
distinguish 429 variations, and List~C 461 out of the 469 total
model variations considered. Both Lists~B and~C allow the majority
of model variations to be distinguished with $\le 4\;{\rm fb}^{-1}$
integrated luminosity, however List~C exhibits the best performance
overall, as it is able to distinguish the models with a consistently
lower luminosity requirement.
%
%
For the $\alpha=0.66$ models, all three signature sets allow the
majority of model points to be distinguished from $\alpha=0$ with
less than $4\;{\rm fb}^{-1}$ integrated luminosity. Only List~A was
unable to resolve all model variations with less than $10\;{\rm
fb}^{-1}$, as 73 out of 496 models required higher luminosity.
Signature List~C exhibits the best performance, allowing nearly all
model variations to be resolved with $\le 2\;{\rm fb}^{-1}$.  The
$\alpha=1.0$ models are sufficiently different from the $\alpha=0$
case that all three of the signature sets are able to distinguish
the two cases with exceptionally low luminosity. Signature List~C
again exhibits the best performance, allowing all models to be
distinguished with less than $1\;{\rm fb}^{-1}$ of data.


\begin{figure}[t]
\begin{center}
\includegraphics[scale=0.7]{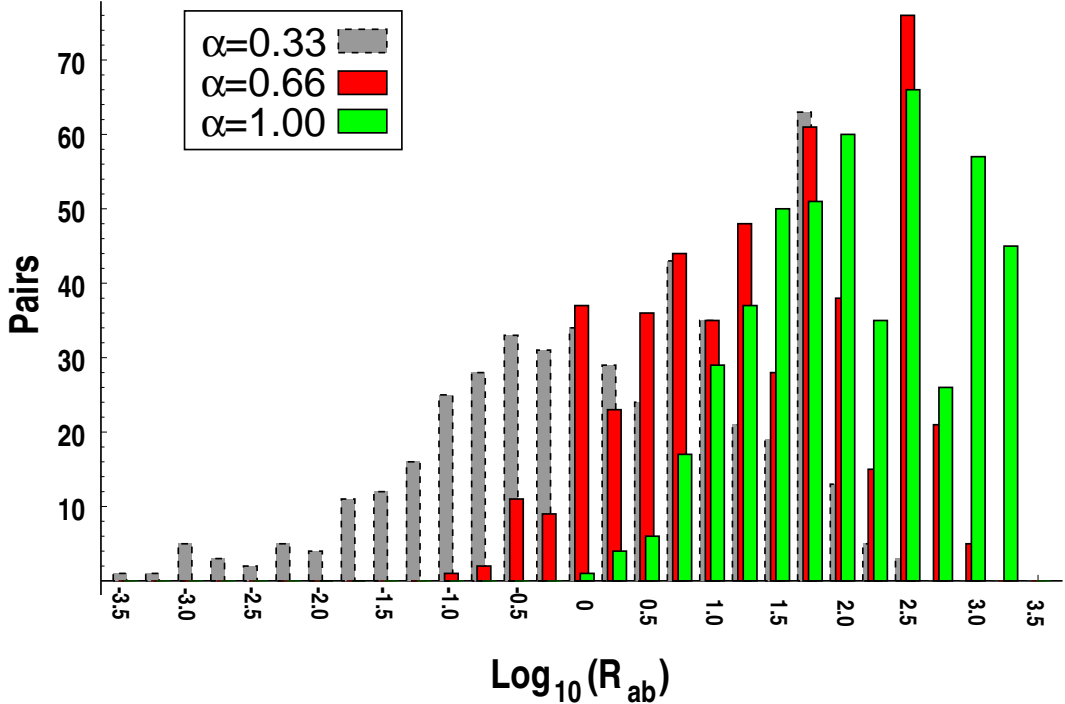}
{\caption{\label{fig:rab_distA}\footnotesize{\bf Distribution of
$(R_{AB})$ values for signature List~A.} The distribution of
$R_{AB}$ values for the single signature of List~A is given for the
parameter sets $\alpha=0.33$, $\alpha=0.66$, and $\alpha=1.0$. In
each case we are comparing the indicated value of $\alpha$ with the
case $\alpha = 0$ for the same set of background model parameters.
Note that larger values of $R_{AB}$ imply lower values of $L_{\rm
min}$.}}
\end{center}
\end{figure}

We can understand these results by examining the individual
$(R_{AB})_i$ response of each signature.  From
equation~(\ref{Lmin}), the minimum luminosity required to
distinguish two models, A~and~B, is inversely proportional to
$R_{AB}$, which is the sum of the individual $(R_{AB})_i$ values of
each signature.  Because $(R_{AB})_i$ reflects the sensitivity of
the $i$-th signature to changes between models~A and~B (a larger
$(R_{AB})_i$ value being more sensitive), signatures that have high
sensitivity to physical changes associated with $\alpha$ provide a
greater contribution to the total $R_{AB}$, and thus reduce the
$L_{\rm min}$ requirement.

The distribution of $R_{AB}$ values for the single signature of
List~A is shown in Figure~\ref{fig:rab_distA}. For the $\alpha=0.33$
case the distribution is localized to relatively low values of
$R_{AB}$. For the $\alpha=0.66$ and $\alpha=1.0$ cases the
distribution begins to spread out, with many models having
significantly larger $R_{AB}$ values. This indicates the signature
is becoming increasingly more sensitive to the differences brought
on by changes in $\alpha$ as this parameter is increased. However,
with only a single signature it is not possible to guarantee that it
will be as effective for other models as it is in this example.  In
order for this approach to work across a broad range of potential
physics scenarios it is advantageous to adopt a combination of
signatures, where each may be sensitive to one or more aspects of a
particular class of models.

\begin{figure}[t]
\begin{center}
\includegraphics[scale=0.6]{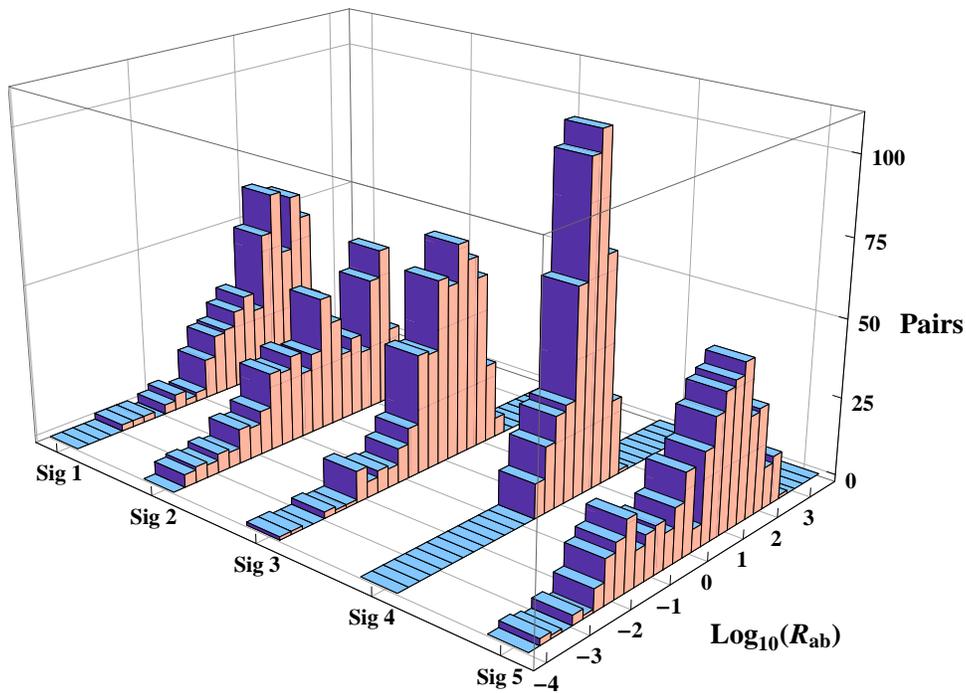}
{\caption{\label{fig:rab_distB1}\footnotesize{\textbf{Distribution
of $(R_{AB})_i$ values for signature List~B [$\alpha = 0.33$ versus
$\alpha = 0$].} The distribution of $(R_{AB})_i$ values for the five
signatures of List~B is given for the case of comparing $\alpha = 0$
with $\alpha = 0.33$. For the definition of the five signatures, see
Table~\ref{tbl:sigB}.}}}
\end{center}
\end{figure}

\begin{figure}[t]
\begin{center}
\includegraphics[scale=0.6]{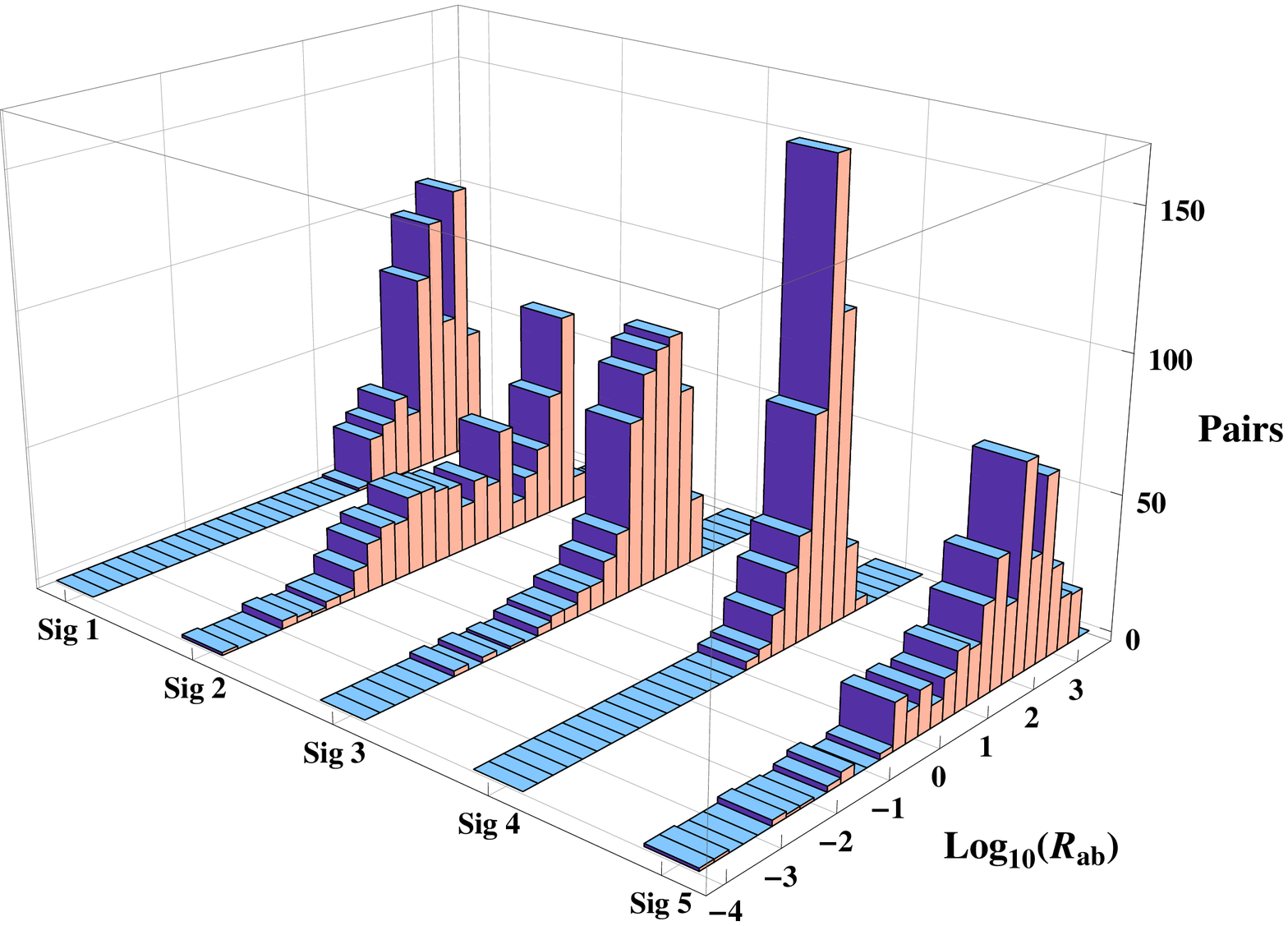}
\caption{\label{fig:rab_distB2}\footnotesize{\textbf{Distribution of
$(R_{AB})_i$ values for signature List~B [$\alpha = 1$ versus
$\alpha = 0$].} The distribution of $(R_{AB})_i$ values for the five
signatures of List~B is given for the case of comparing $\alpha = 0$
with $\alpha = 1$. For the definition of the five signatures, see
Table~\ref{tbl:sigB}.}}
\end{center}
\end{figure}

Figures~\ref{fig:rab_distB1} and~\ref{fig:rab_distB2} show the
distributions of $(R_{AB})_i$ obtained for the five signatures of
signature List~B.  Each figure represents five histograms where the
variable being considered is $\log_{10}[(R_{ab})_i]$, with the
comparison being between $\alpha = 0$ and $\alpha = 0.33$ in
Figure~\ref{fig:rab_distB1} and between $\alpha = 0$ and $\alpha =
1$ in Figure~\ref{fig:rab_distB2}. In a similar fashion to the
single signature of List~A, the distributions are in general
clustered at low $(R_{AB})_i$ for $\alpha=0.33$, and begin to spread
out considerably, taking on much larger values as $\alpha$
increases. Comparing the individual distributions to those in the
single signature of List~A, the overall spread of values is not
significantly different.  However, recall that $R_{AB}$ is the sum
of the individual $(R_{AB})_i$ values.  Therefore we gain a
significant enhancement by simply including additional signatures. A
similar effect occurs with the larger set of signatures in List~C.
As we saw in Section~\ref{method}, however, there is ultimately a
point of negative returns and a maximum efficacy is obtained.

\begin{figure}[p]
\begin{center}
      \includegraphics[scale=0.65]{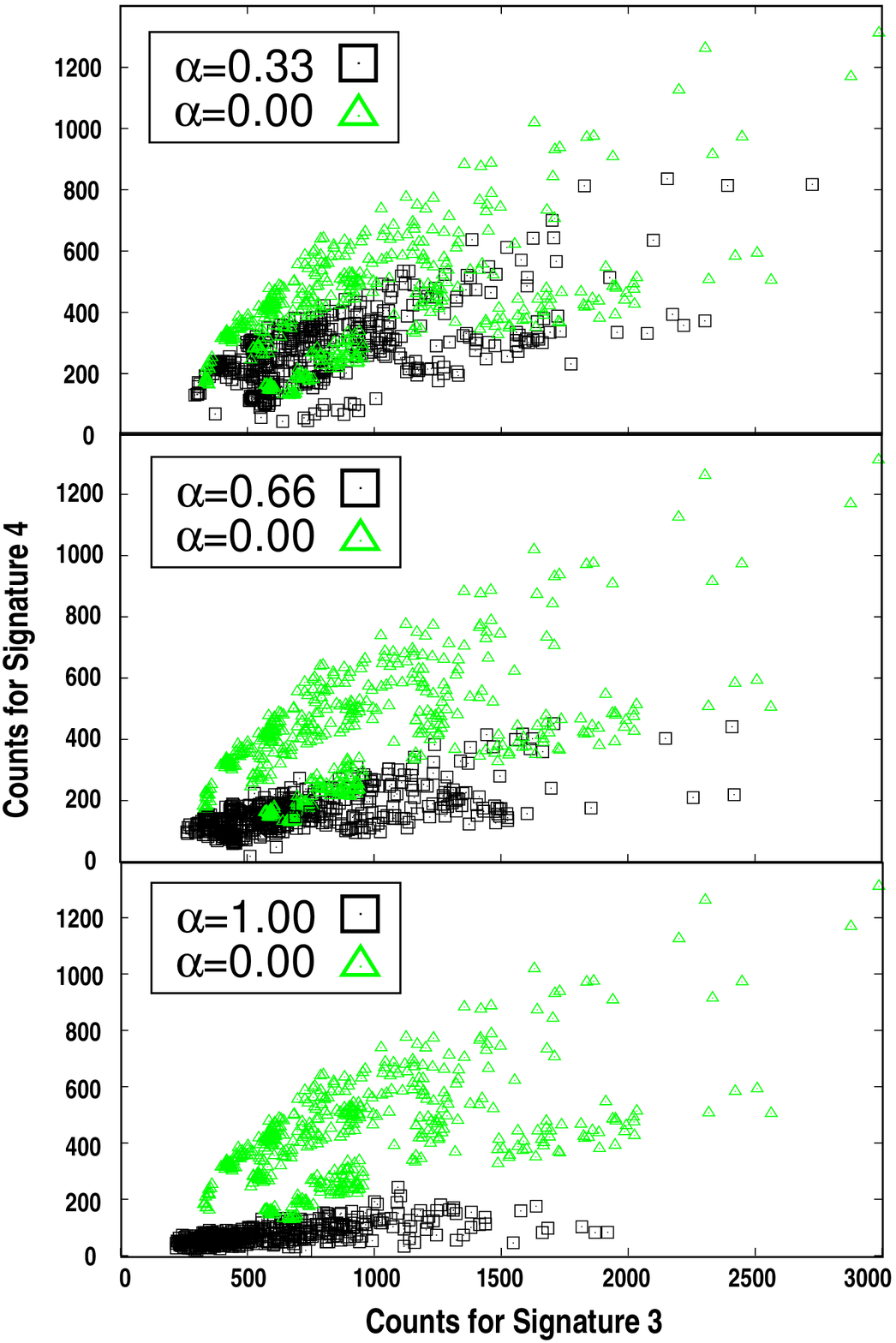}
\caption{\label{fig:scatterB}\footnotesize{\textbf{Footprint-style
plot for a pair of signatures from List~B.} Total counts for
signature \#3 versus signature \#4 of List~B is given for the case
$\alpha = 0$ (green triangles) $\alpha \neq 0$ (black squares). The
cases shown are for $\alpha = 0$ versus $\alpha = 0.33$ (top panel),
$\alpha = 0.66$ (middle panel) and $\alpha = 1$ (bottom panel). The
axes measure the number of events for which the kinematic quantity
was in the range given in Table~\ref{tbl:sigB}. Larger values of the
non-universality parameter $\alpha$ correspond to a greater degree
of separation between the two model ``footprints.''}}
\end{center}
\end{figure}
\begin{figure}[p]
\begin{center}
      \includegraphics[scale=0.65]{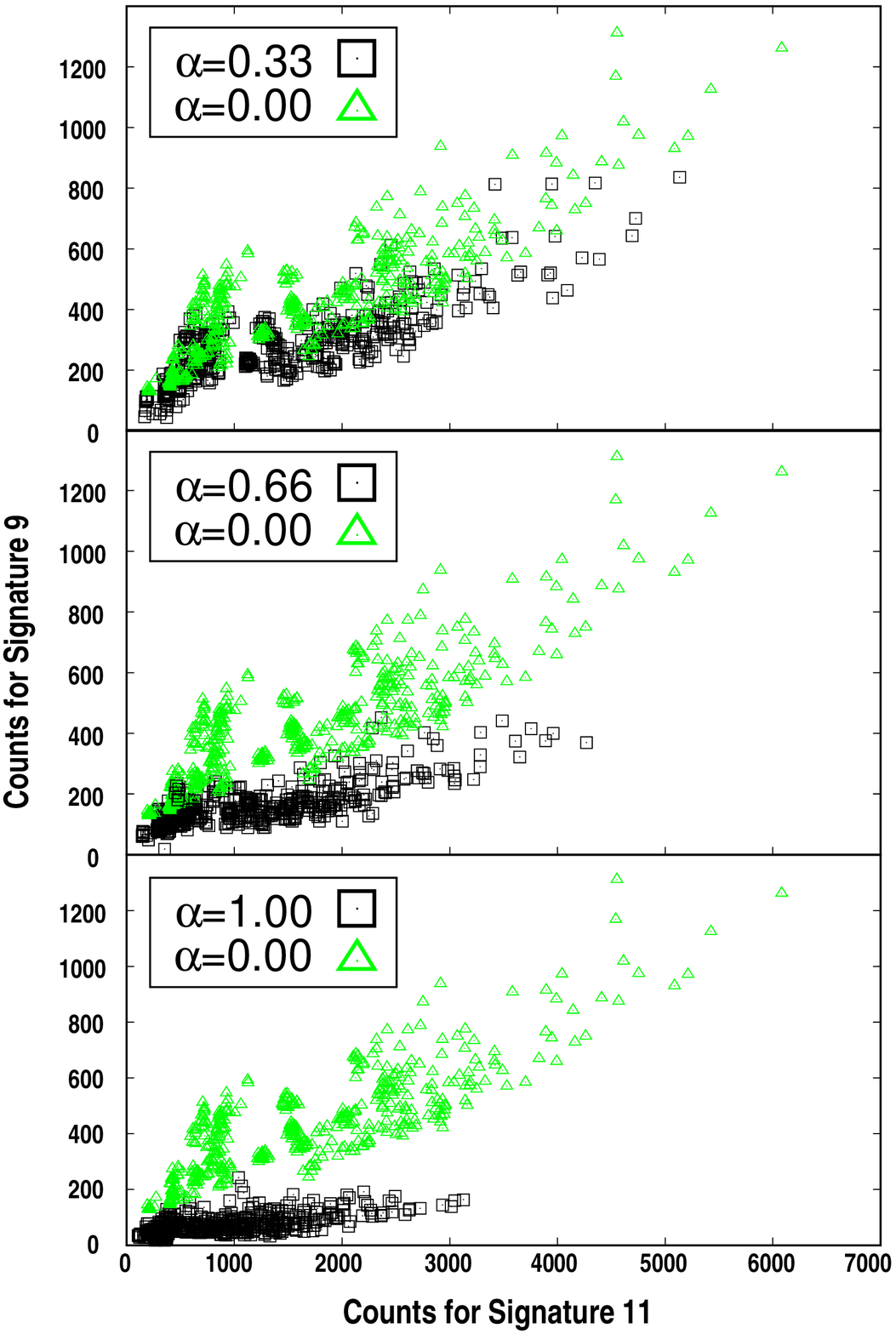}
\caption{\label{fig:scatterC}\footnotesize{\textbf{Footprint-style
plot for a pair of signatures from List~C.} Total counts for
signature \#11 versus signature \#13 of List~C is given for the case
$\alpha = 0$ (green triangles) $\alpha \neq 0$ (black squares). The
cases shown are for $\alpha = 0$ versus $\alpha = 0.33$ (top panel),
$\alpha = 0.66$ (middle panel) and $\alpha = 1$ (bottom panel). The
axes measure the number of events for which the kinematic quantity
was in the range given in Table~\ref{tbl:sigC}. Larger values of the
non-universality parameter $\alpha$ correspond to a greater degree
of separation between the two model ``footprints.''}}
\end{center}
\end{figure}

Thus far we have presented the results of our approach in terms of
the minimum integrated luminosity required to resolve two model
classes ($\alpha=0$ and $\alpha \neq 0$) using our set of optimized
signatures. To understand why this approach works, it is useful to
examine the signature results themselves. Figures~\ref{fig:scatterB}
and~\ref{fig:scatterC} show examples of two-dimensional slices of
the signature space ``footprint'' for our large set of model
variations. In these figures the results have been normalized to
$5\;{\rm fb}^{-1}$ of data.


Figure~\ref{fig:scatterB} compares the count rates for the third and
fourth signatures of List~B for the case $\alpha = 0$ versus $\alpha
= 0.66$ (left panel) and $\alpha = 1$ (right panel).
Figure~\ref{fig:scatterC} compares the count rates for signatures
\#11 and \#13 of List~C for the case $\alpha = 0$ versus $\alpha =
0.66$ (left panel) and $\alpha = 1$ (right panel). In this case the
two signatures are both taken from the set of events containing at
least one lepton and five or more jets (see Table~\ref{tbl:sigC}).
We have chosen this pair for the dramatic separation that can be
achieved, though similar results can be obtained with other pairs of
signatures.

The power of our inclusive signature list approach lies in the
choice of signatures and their ability to remain highly sensitive to
changes in the physical behavior of each model. This feature is
reflected qualitatively in the visual clustering of the data points,
which become progressively more distinct as the parameter~$\alpha$
is increased. As the regions separate it becomes increasingly less
likely that a model from one class can be confused with a model from
the other class, even when considering statistical fluctuations. In
our approach this manifests itself when one computes $R_{AB}$, which
reflects the ``distance'' in signature space between the two models
under comparison, and which becomes large when the models are
sufficiently different from one another.


The idea of using repeated pairings of targeted observables in order
to separate model classes was studied in previous
``footprint-style''
analyses~\cite{Bourjaily:2005ja,Kane:2006yi,Kane:2007pp}. If we
consider the universal gaugino mass scenario ({\em i.e.} $\alpha =
0$) as a ``model,'' and the case of non-universal gaugino masses as
a separate model, then a set of signatures will be truly targeted at
this particular model feature if the set of all such two-dimensional
planes implies complete separation between the models. With this in
mind it is interesting to examine distinguishability between the two
values of $\alpha$ from a somewhat different perspective. Adopting
the approach of~\cite{Kane:2007pp} we can ask how many degeneracies
exist between the two classes of models, where by degeneracy we mean
two models that exist at different points in the microscopic
parameter space, but occupy the same point in signature space (up to
statistical fluctuations). If it is possible, through application of
one or more signatures, to ensure that no degeneracies exist we can
claim to that it is possible to completely discriminate between the
two classes.

As an example of how this idea can be applied, we can consider the
analysis performed in~\cite{Kane:2007pp}. Let one particular value
of the parameter $\alpha$ (such as $\alpha = 0$) be ``model A'' and
let some other value of the parameter $\alpha$ be ``model B.''
Choose any pair of signatures in one of the signature lists. From
our controlled sample we can choose an individual case $B_j \in B$
and compute the quantity $(\Delta S_{A_i B_j})^2$ between that
particular point and all the points $A_i \in A$ for this pair of
signatures. If the value for all such  $(\Delta S_{A_i B_j})^2$ is
always greater than the two-signature threshold given by
$\gamma_2(0.95)$ in Table~\ref{gammatable} we will claim the point
$B_j$ has been separated from the entire footprint of model~A. We
can then repeat this exercise over all cases of model~B. The number
of cases of model~B that have {\em not} been separated from the
entire footprint of model~A we will denote as $N_{BA}$. This is a
type of degeneracy count for model~B with respect to model~A.
Clearly the process can be performed for model~A with respect to
model~B, producing a degeneracy count $N_{AB}$. In general we expect
these two numbers to be roughly equivalent in magnitude, but not
necessarily precisely equal.

\begin{figure}[t]
\begin{center}
      \includegraphics[scale=0.4]{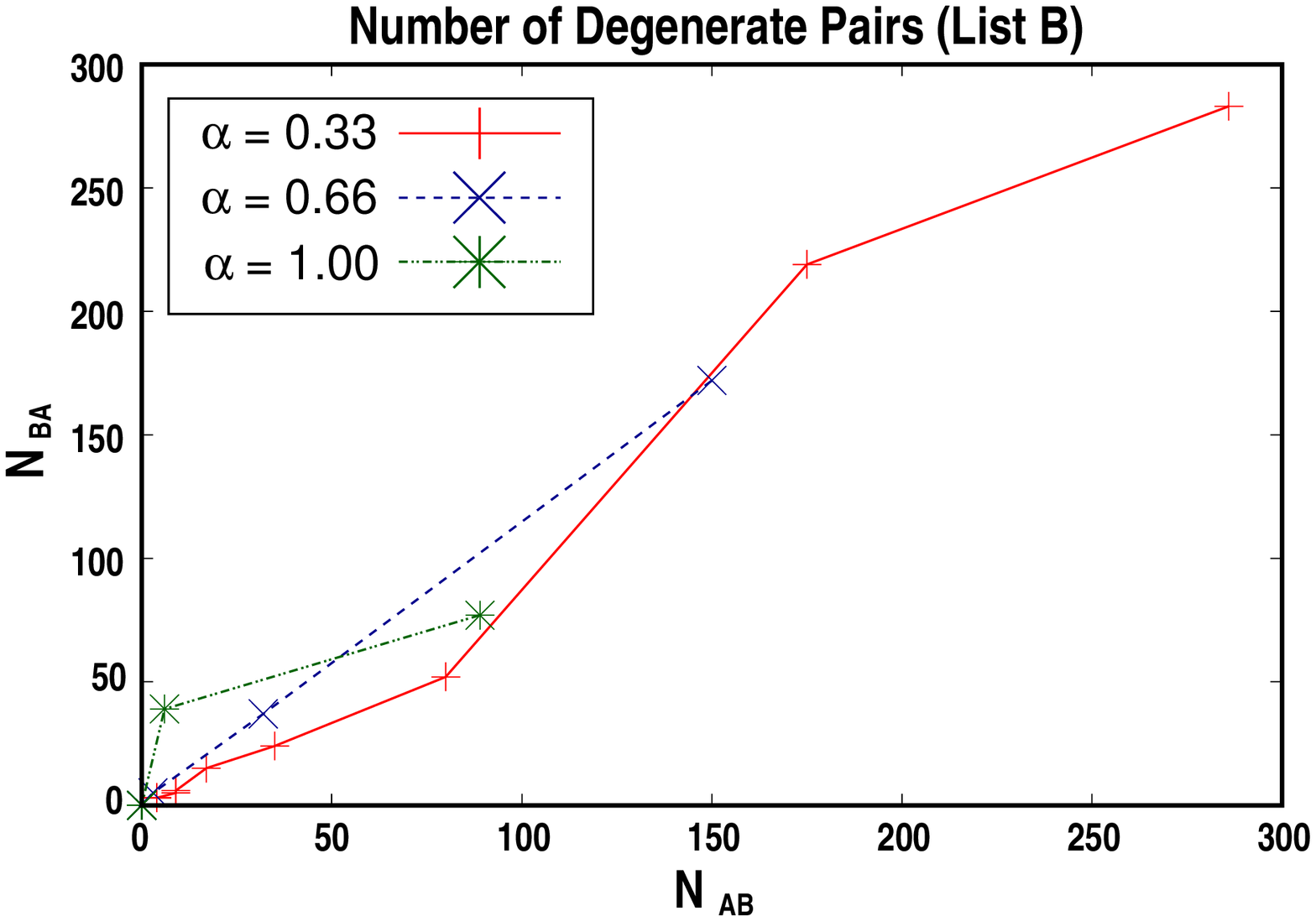}
      \includegraphics[scale=0.4]{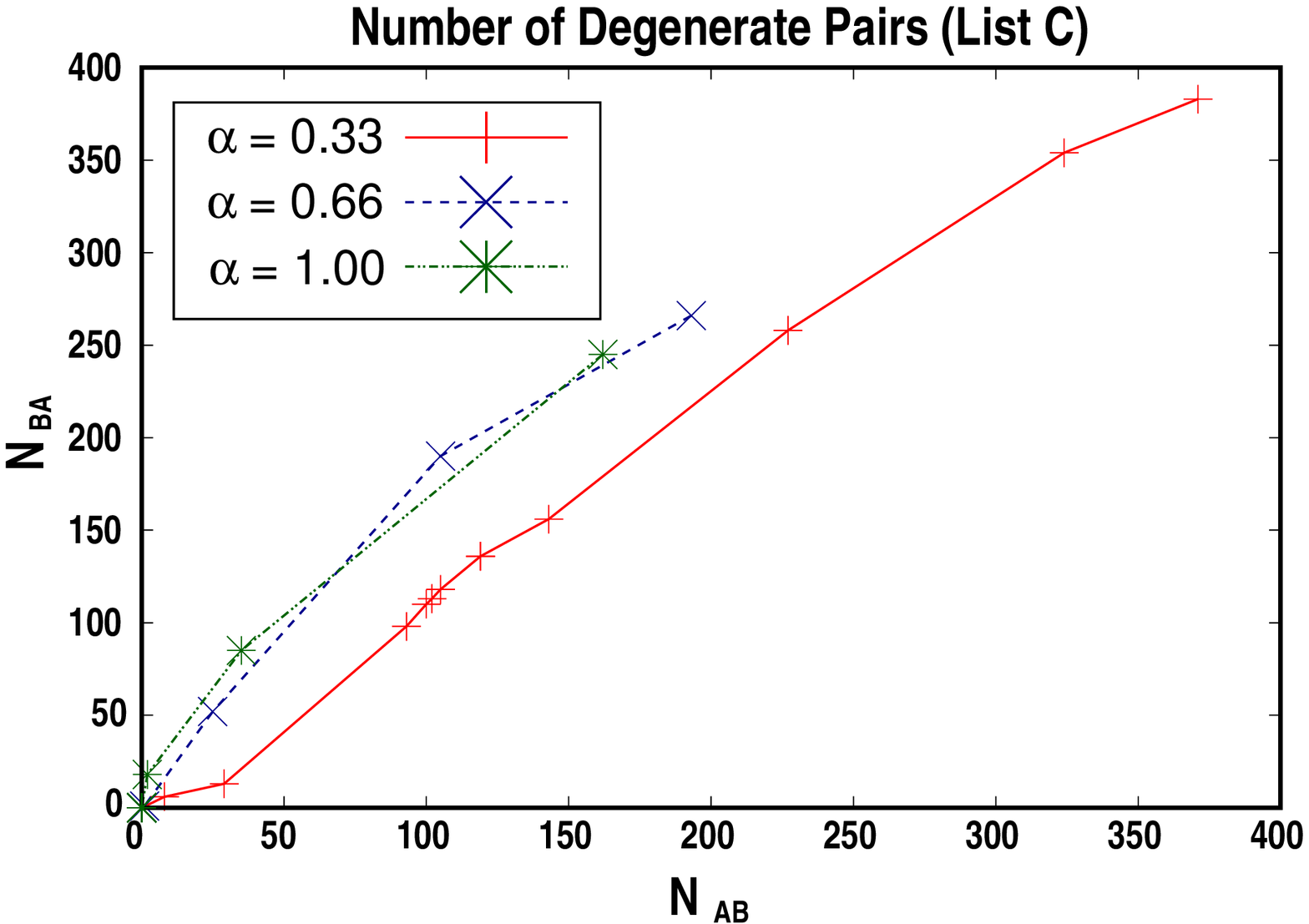}
\caption{\label{fig:degen}\footnotesize{\textbf{Degeneracy counts
for List~B (left panel) and List~C (right panel).} The relative
degeneracy counts $N_{AB}$ and $N_{BA}$ that result from successive
application of pairs of signatures from List~B and List~C are
plotted for our controlled model sample. In each case model~A is the
case with $\alpha = 0$ while model~B is the case with the indicated
value of $\alpha \neq 0$. Once all model pairs have been applied the
total degeneracy count vanishes for both lists and for all values of
$\alpha \neq 0$.}}
\end{center}
\end{figure}

If either $N_{AB}$ or $N_{BA}$ are non-vanishing then the two
footprints are not yet disjoint in the multi-dimensional signature
space. We can then choose any other pair of signatures and repeat
the procedure, this time restricting $A_i$ and $B_j$ to run only
over the degenerate cases. If we have chosen a good set of
signatures the quantities $N_{AB}$ and $N_{BA}$ should rapidly
converge to zero as the algorithm is successively applied. The
results of performing this exercise on the controlled model sample
generated by the parameters of Table~\ref{tbl:large_sample_inputs}
is shown in Figure~\ref{fig:degen}. In the left panel we show the
successive values of $N_{AB}$ and $N_{BA}$ as pairs of signatures
from List~B are used to compute the separability parameter $(\Delta
S_{AB})^2$, while the right panel uses pairs of signatures from
List~C. In both cases ``model~A'' represents the set of models with
$\alpha = 0$, while ``model~B'' represents the case with the
indicated value of $\alpha =0.33$, 0.66 and 1.0. For all three
values of the parameter $\alpha$ the lists do an excellent job of
converging towards $N_{AB} = N_{BA} = 0$ after only a few pairings
are considered. This suggests that the signature lists of
Tables~\ref{tbl:sigB} and~\ref{tbl:sigC} should be able to reveal
the departure of the gaugino soft masses from the universal ratios
on a truly {\em general} supersymmetric model with a high degree of
reliability and in a small amount of integrated luminosity.


\begin{figure}[t]
\begin{center}
\includegraphics[scale=0.65,angle=0]{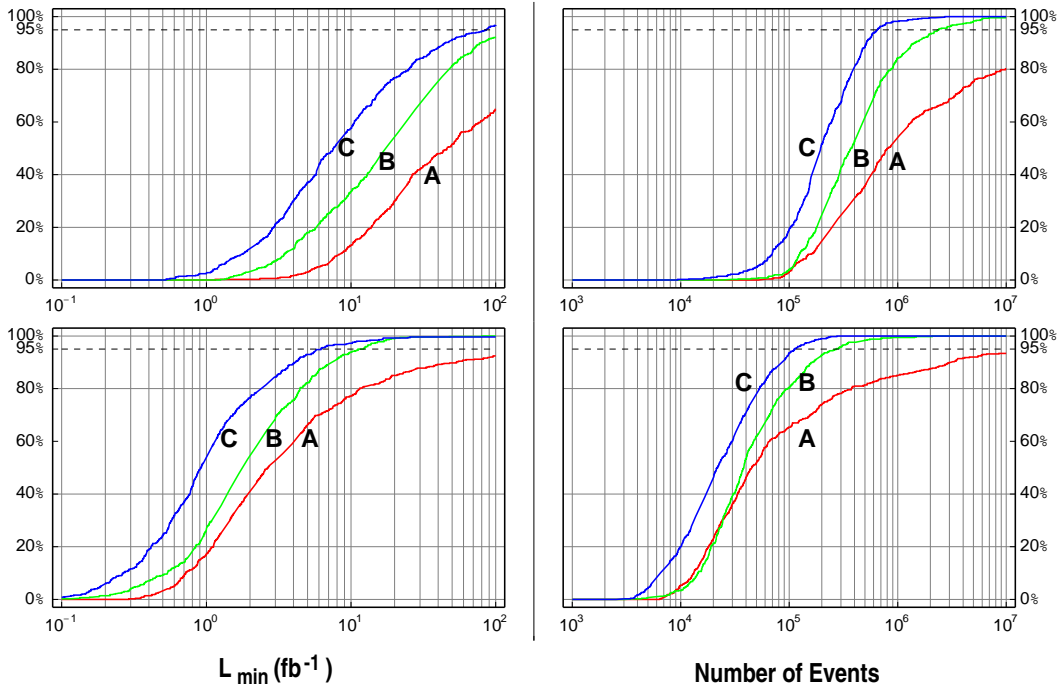}
\caption{\footnotesize \textbf{Efficiencies of the three signature
lists.} The ability of the three signature lists to separate the
case $\alpha = 0.1$ from $\alpha = 0$ is indicated in the top pair
of plots and the simpler case $\alpha = 0.3$ from $\alpha = 0$ in
the bottom pair of plots. On the left, the percentage of cases that
could be distinguished using each of the three signature lists of
Tables~\ref{tbl:sigA}, \ref{tbl:sigB}~and~\ref{tbl:sigC} is given as
a function of integrated luminosity in units of ${\rm fb}^{-1}$. On
the right the same percentage is shown as a function of the number
of supersymmetric events. The 95\% separability threshold is
indicated by the dashed horizontal line.} \label{fig:results1}
\end{center}
\end{figure}

To honestly confirm this hypothesis we must generate a more random
set of models. After all, the signature lists of
Tables~\ref{tbl:sigA}, \ref{tbl:sigB}~and~\ref{tbl:sigC} were
constructed precisely with the sorts of models of our controlled
sample in mind. But as we saw in Section~\ref{sec:bm analysis},
models such as benchmark model~A can prove more challenging for our
analysis algorithm. To allow for the possibility of more perverse
cases than those of our controlled sample, an additional set of 500
models were generated with six points on the $\alpha$-lines ranging
from 0 to 0.5. In this case a~16-dimensional parameter space defined
by the quantities in~(\ref{paramset}) was considered. Specifically,
slepton and squark masses were allowed to vary in the range 300~GeV
to 1200~GeV with the masses of the first and second generation
scalars kept equal. The gaugino mass scale given by $M_3$ and the
$\mu$-parameter were also allowed to vary in this range. The
pseudoscalar Higgs mass $m_A$ was fixed to be 850 GeV and the value
of $\tan\beta$ was allowed to vary from 2~to~50. If all points along
the $\alpha$-line satisfied all experimental constraints on the
superpartner mass spectrum, then 100,000 events were generated for
each of the six points along the $\alpha$-line in the manner
described in Section~\ref{method}. Using this data the value of
$L_{\rm min}$ was computing using~(\ref{Lmindef})
and~(\ref{resistor}) for each of our three signature sets.

\begin{figure}[t]
\begin{center}
      \includegraphics[scale=0.4]{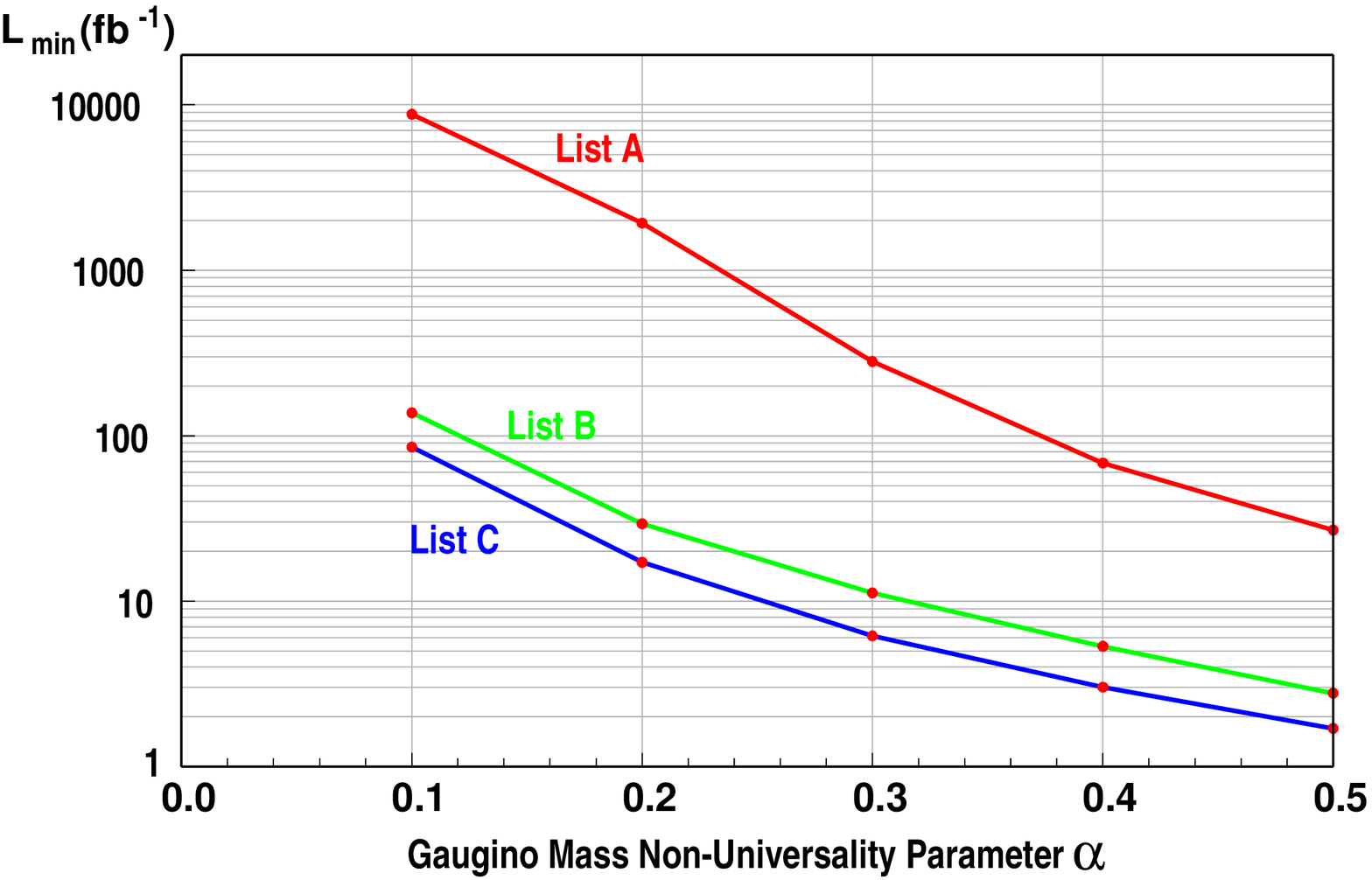}
      \includegraphics[scale=0.4]{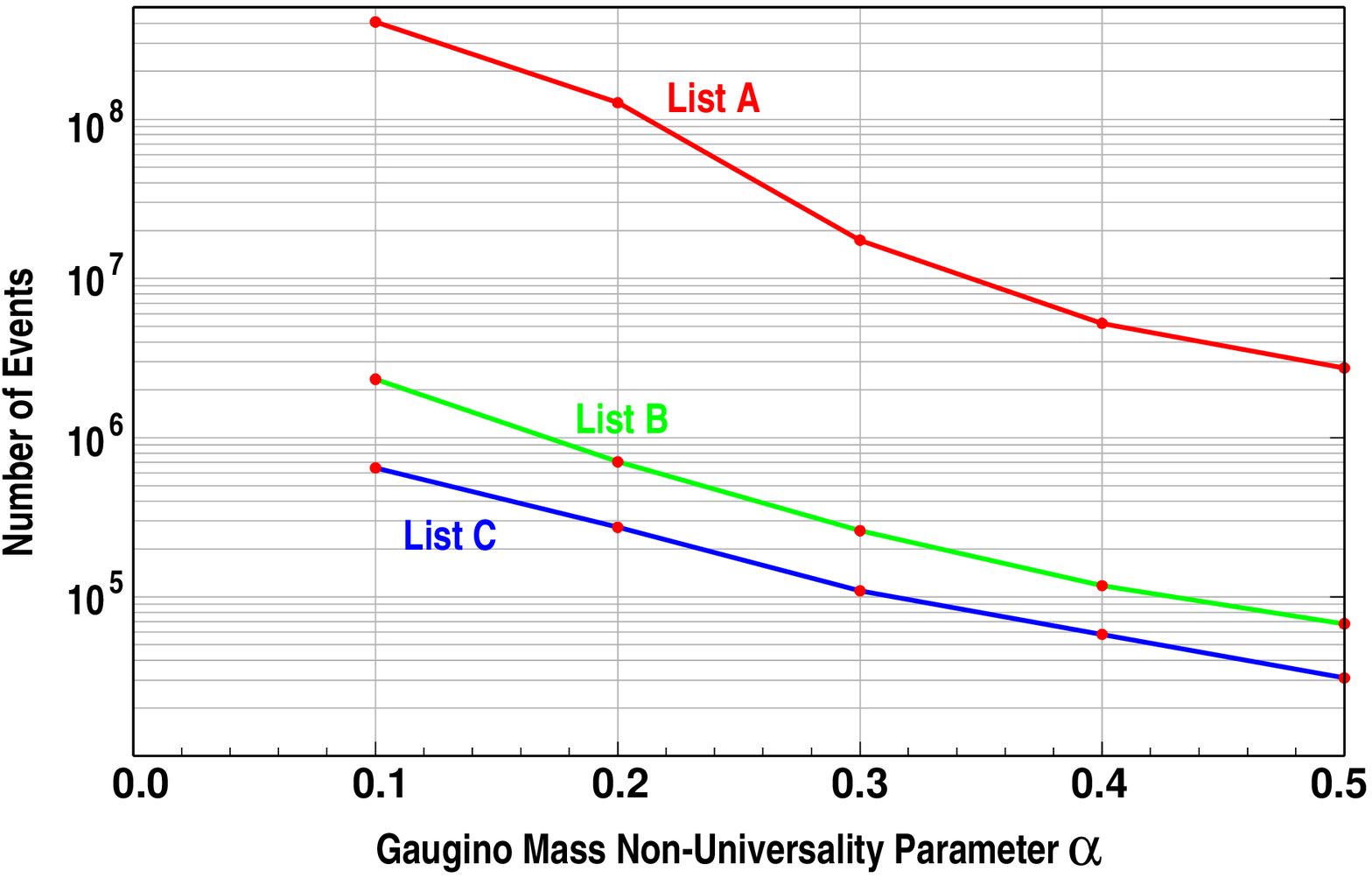}
\caption{\label{fig:results2}\footnotesize{\textbf{$L_{\rm min}$ and
$N_{\rm min}$ required to detect $\alpha \neq 0$ for 95\% of the
random models.} }}
\end{center}
\end{figure}

The results of this analysis are given in Figures~\ref{fig:results1}
and~\ref{fig:results2}. Figure~\ref{fig:results1} considers the
ability of our signature lists to separate the case $\alpha = 0.1$
from $\alpha = 0$ (top pair of plots) and the simpler case $\alpha =
0.3$ from $\alpha = 0$ (bottom pair of plots). On the left, the
percentage of cases that could be distinguished using each of the
three signature lists of Tables~\ref{tbl:sigA},
\ref{tbl:sigB}~and~\ref{tbl:sigC} is given as a function of
integrated luminosity in units of ${\rm fb}^{-1}$. Since the random
model sample includes examples with very different superpartner mass
scales, the overall supersymmetric production cross-section varies
much more across this sample than in the controlled model sample
described above. We therefore take this into account by plotting the
same percentage in terms of the number of supersymmetric events on
the right side of Figure~\ref{fig:results1}. The 95\% separability
threshold is indicated by the dashed horizontal line. Even using our
best set of signatures (List~C) it will require nearly 100~${\rm
fb}^{-1}$ to be able to detect non-universality at the level of
$\alpha \simeq 0.1$ for an arbitrary supersymmetric model. Yet for
the vast majority of models the departure from universality should
become apparent after just 10-20~${\rm fb}^{-1}$. Departures from
universality at the level of $\alpha \simeq 0.3$ should be apparent
using this method for most supersymmetric models after just a
few~${\rm fb}^{-1}$. In Figure~\ref{fig:results2} the integrated
luminosity (or number of supersymmetric events) needed to detect
$\alpha \neq 0$ for 95\% of our random models is given as a function
of the five non-vanishing $\alpha$ values simulated.

\section{Conclusions}

If supersymmetry is discovered at the LHC the high energy community
will be blessed with a large number of new superpartners whose
masses and interactions will need to be measured. At the same time
the community will be cursed by a large model space with many
Lagrangian parameters which cannot themselves be directly measured
experimentally. Undoubtedly performing global fits of the many
observables to the parameter space of certain privileged and
well-defined benchmark models will be of great help in making sense
of this embarrassment of richness. But recent work suggests that
unless these models are determined by very few parameters it is
likely (if not perhaps inevitable) that multiple points in the
parameter space will fit the data well. It then becomes an
interesting question to ask whether it is possible to fit to certain
model {\em characteristics} rather than to any particular model
itself.

In our opinion one of the most important such characteristic is the
pattern of soft supersymmetry-breaking gaugino masses. No other
property of the low-energy soft Lagrangian is more easily linked to
underlying high-scale physics, particularly if that high-scale
physics is of a string-theoretic origin. Only the related issue of
the wavefunction of the LSP is of more importance to low-energy
physics and cosmology. We are thus interested in asking whether we
can identify the presence on non-universalities in the gaugino
sector independent of all other properties of the superpartner
spectrum. The manner by which any such undertaking can be tackled is
by no means clear -- though neither is it clear that such an
undertaking is inherently impossible. In the present work we have
decided to begin this process with a simple parametrization of the
gaugino masses determined by a single parameter which can be thought
of as the ratio of bulk gravity and anomaly contributions to gaugino
masses. We developed model ``lines'' in the spirit of previous
benchmark studies such as the Snowmass Points \& Slopes in which
only the single non-universality parameter is varied. By
understanding how the observable physics at the LHC is affected by
this parameter -- and then repeating the analysis many times with
the other supersymmetric parameters varied -- we can learn which LHC
signatures are most directly ``targeted'' at this important
underlying characteristic.

Our procedure depends on certain analytic results that improve on
the methods first introduced concretely by Arkani-Hamed et al. These
analytic results in turn depend on the assumption that the
signatures considered have fluctuations which are largely
uncorrelated with one another. This severely limits the type of
signature ensembles one might construct. Yet this restriction does
not imply a loss of resolving power, as the ``optimal'' signature
list is rarely the largest possible list one can imagine. Our
analysis has suggested two signatures ensembles which perform
remarkable well at the task of measuring the value of the
non-universality parameter we introduce. Broadly speaking, we find
that a non-universality at the 10\% level can be measured with 10-20
fb$^{-1}$ of integrated luminosity over approximately 80\% of the
supersymmetric parameter space relevant for LHC observables. If we
are interested in measurements at only the 30\% level these numbers
change to 5-10 fb$^{-1}$ over approximately 95\% of the relevant
parameter space.

This is remarkable progress, but the task we set out for ourselves
is admittedly still somewhat artificial. There are two independent
mass ratios that can be constructed from the three soft
supersymmetry breaking gaugino masses -- our parametrization is
therefore not fully general. It would be of great interest to study
more general departures from non-universality to see if the optimal
signature lists change substantially. Of greater import is the need
to perform a Monte Carlo simulation in order to compare a candidate
model to the ``data'' at the LHC. To perform such a comparison we
must assume knowledge of all input parameters {\em apart} from the
one we are attempting to measure. While this is a common practice in
benchmark studies at colliders, it is far from the reality that
theorists and experimentalists will encounter in the early stages of
the LHC era. Our study demonstrated the efficacy of certain targeted
observables in extracting the non-universality parameter $\alpha$
while keeping all other parameters fixed for the two models. This is
quite a strong assumption and future work should relax this
constraint. In other words, one would like to distinguish between
two models (with different values of $\alpha$) even if the other
parameters for the two models are not the same. There are many
directions by which this may be pursued. For example, in the current
analysis we have not allowed ourselves any knowledge of the mass
spectrum, though analysis of kinematic end-points will certainly
provide some information in this regard early on in LHC data-taking.
In addition, techniques such as the use of on-shell effective
theories~\cite{ArkaniHamed:2007fw} might provide sufficient
information about the dominant production and decay modes for new
mass eigenstates to allow an approximation to our analysis to be
performed before the full mass spectrum is reconstructed. We hope to
pursue both avenues for introducing greater realism in future work.

\section*{Acknowledgments}

The authors would like to thank Piyush Kumar for important
discussions at the beginning of this project. B.A. and B.D.N. would
like to thank the Michigan Center for Theoretical Physics for
hospitality during the early stages of this work. B.A., M.H.~and
B.D.N. are supported by National Science Foundation Grant
PHY-0653587. G.K.~and~P.G. are supported in part by the Department
of Energy.

\section*{Appendix: Some Specific Examples} \label{models}
\setcounter{equation}{0} \setcounter{section}{1}
\renewcommand\thesection{\Alph{section}}
\renewcommand{\theequation}{\Alph{section}.\arabic{equation}}


The low energy limit of four-dimensional string constructions can be
studied as a supergravity theory defined by three functions of the
massless chiral superfields: the K\"ahler potential $K$, the
superpotential $W$ and the gauge kinetic function $f_a$. As in the
previous subsection the label $a$ refers to a particular gauge group
$\mathcal{G}_a$. This last function is naturally of most import for
the soft supersymmetry breaking gaugino masses, but all three
functions play a role in determining the nature of supersymmetry
breaking in the observable sector. In string models the gauge
kinetic function is typically determined by gauge-singlet chiral
superfields which we will simply refer to as moduli. The K\"ahler
potential and superpotential are generally functions of both moduli
and gauge-charged matter superfields.

Let us assume, as is so often the case at tree-level in string
theory models, that the gauge kinetic function is linear in the
moduli\footnote{Unless explicitly written otherwise, in this
appendix we are always using Planck units in which $M_{\PL} = 1$.}
\begin{equation} f_a = k_a X_a \, . \label{f1} \end{equation}
Here $X_a$ represents a generic modulus field and we have allowed
for the possibility that each gauge group can have its own modulus
dependence. The proportionality constant $k_a$ can be thought of as
the affine level at which the gauge group $\mathcal{G}_a$ is
realized in the underlying conformal field theory. We will hereafter
always set this constant to unity. Note that the real part of the
lowest (scalar) component of $X_a$ must acquire a vacuum expectation
value (VEV) in order to determine the size of the corresponding
gauge coupling
\begin{equation}
<\re \; x_a>\; = \frac{<x_a + \bar{x}_a>}{2}\; =
\frac{1}{g_{a}^{2}}, \label{xvev}
\end{equation}
where $x_a=X_a\lowest$. If we wish to entertain the notion of gauge
coupling unification then we must either arrange these VEVs to be
equal, or there must be a single universal modulus $X$ to couple to
all gauge groups with equal strength. The latter is the case for the
two models we will consider below so we will assume
\begin{equation}
f_a = X; \qquad <\re \; x>\; = 1/g_{\STR}^{2} , \label{f2}
\end{equation}
where $g_{\STR}$ is the universal gauge coupling at the string
scale. The highest component of the chiral superfield $X$ is the
auxiliary field $F^X$. A non-zero expectation value for this field,
or indeed of any other such auxiliary field, is an indication of
spontaneous breaking of supersymmetry. The manifestation of this
supersymmetry breaking in the form of gaugino masses is given (at
tree level) by the expression
\begin{equation} M_a =  \frac{g_a^2}{2}  F^N \partial_N f_a , \label{Ma}
\end{equation}
where the repeated index $n$ sums over all chiral superfields
present in the function $f_a$ and the expression on the left is
understood to be evaluated in the vacuum. For the case of~(\ref{f2})
this implies
\begin{eqnarray}
M_{a} &=& \frac{g_{\STR}^{2}}{2}F^{X}\nonumber \\
 &=& \lang \frac{F^X}{x+\bar{x}} \rang \, , \label{Ma1}
\end{eqnarray}
where in the last line we have made explicit the vacuum evaluation
at the string scale. Now we have arrived at a universal contribution
to the three gaugino masses of the Standard Model, which gives rise
to the term $M_a^1(\Lambda_{\UV})$ of~(\ref{piece1}).

Additional contributions to~(\ref{Ma}) appear at the loop level. The
structure of these terms have been computed
elsewhere~\cite{Gaillard:1999yb,Bagger:1999rd,Binetruy:2000md} and
they generally depend on details of the complete theory beyond the
form of the tree level gauge kinetic function. A subset of these
terms can be derived completely from the superconformal anomaly, the
most important of which is universal for any supergravity theory
\begin{equation} M_a|_{\rm an} =
-\frac{g_{a}^{2}(\Lambda_{\UV})}{2} \frac{b_a}{16\pi^2}
\frac{2\oline{M}}{3} \, . \label{Man}
\end{equation}
The coefficient $b_a$ is as defined in~(\ref{ba}) and the field $M$
is the auxiliary field of the supergravity multiplet whose
expectation value determines the gravitino mass $m_{3/2} =
-\frac{1}{3}\lang \oline{M}\rang$. In the limit where this is the
{\em only} significant one loop correction to the gaugino masses we
recover the expression in~(\ref{piece2}) where $M_g \equiv m_{3/2}$.

We now have our two components to the mirage gaugino mass pattern.
Our next task is to ask how the magnitudes of $M_u = \lang
F^X/(x+\bar{x})\rang$ and $M_g = -\lang \oline{M}\rang/3$ might be
related to one another. As both $F^X$ and $M$ are auxiliary fields
their equations of motion are easy to obtain, relating these
quantities to the K\"ahler potential and superpotential via
\begin{equation}
F^M = - e^{K/2} K^{M\oline{N}} \left(\oline{W}_{\oline{N}} +
K_{\oline{N}} \oline{W} \right), \; \; \oline{M} = -3e^{K/2}
\oline{W} \label{EQM}
\end{equation}
with $W_{\oline{N}} = \partial W / \partial \oline{Z}^{\bar{N}}$,
$K_{\oline{N}} = \partial K / \partial \oline{Z}^{\bar{N}}$ and
$K^{M \bar{N}}$ being the inverse of the K\"ahler metric $K_{M
\oline{N}}=
\partial^2 K / \partial Z^M \partial \oline{Z}^{\bar{N}}$. Here
$Z^N$ represents any chiral superfield, including our particular
modulus $X$ from the gauge kinetic functions. Given a specific model
of supersymmetry breaking -- such as gaugino condensation -- the
modulus dependence on the non-perturbatively generated
superpotential terms can be computed and~(\ref{EQM}) can be used to
explicitly relate the size of the gravitino mass to the size of
$\lang F^X \rang$. However, if we make the assumption that the
scalar potential has vanishing vacuum expectation value in the
ground state of the theory then we can bypass this complication and
use this assumed constraint directly~\cite{Brignole:1993dj}. The
scalar potential is given by
\begin{equation}
V= K_{M\oline{N}} F^M  \oline{F}^{\bar{N}} - \frac{1}{3} M \oline{M}
\, \label{pot}
\end{equation}
where repeated indices are again summed. The condition $\lang V
\rang = 0$ immediately relates the auxiliary $F$-terms to the
gravitino mass. In particular, if $F^X$ is the only non-vanishing
$F$-term component in the theory then we have
\begin{equation}
\lang F^{X} \rang = \sqrt{3} m_{3/2} \lang (K_{x\bar{x}})^{-1/2}
\rang \label{FX} \, ,
\end{equation}
up to a possible phase. For the moduli we will consider in this
paper the tree level K\"ahler potential is typically $K_{\rm
tree}(X,\oline{X}) = -\ln (X+\oline{X})$ and thus the imposition of
vanishing scalar potential in the vacuum implies
\begin{equation} M_u = \lang \frac{F^X}{(x+ \bar{x})} \rang = \sqrt{3}
m_{3/2} = \sqrt{3} M_g \, . \label{MuvsMg} \end{equation}
Clearly such a situation will {\em not} result in the ratio $r =
M_g/M_u \sim \order(10-100)$ and therefore if~(\ref{MuvsMg}) holds
the contribution from~(\ref{Man}) will be only a small perturbation
on the universal contribution from~(\ref{Ma1}).

But this is where a thorny ``problem'' for string phenomenology
becomes an opportunity. The problem is that the vast majority of
simple, explicit models of dynamical supersymmetry breaking (such as
the gaugino condensation mentioned above) do not produce vanishing
vacuum energy. In other words, when the values of $m_{3/2}$ and
$\lang F^X \rang$ are computed from first principles via~(\ref{EQM})
the relation in~(\ref{FX}) typically fails to be true. This is often
considered an embarrassment for string models and much effort in
string phenomenology is devoted to stabilizing moduli and breaking
supersymmetry while simultaneously achieving $\lang V \rang = 0$.
While many solutions have been postulated through the years, we can
group them here into two broad classes. In the first class the
simple structure of the scalar potential in~(\ref{pot}) is retained,
with a single modulus carrying non-vanishing auxiliary VEV, but the
K\"ahler potential is assumed to differ from the tree level form
$K_{\rm tree}(X,\oline{X}) = -\ln (X+\oline{X})$ so
that~(\ref{MuvsMg}) is modified and $\lang V \rang = 0$ is obtained.
For the second class a new sector is brought into the theory to
produce a new contribution to the scalar potential $\delta V$ of
approximate magnitude $\delta V \simeq m_{3/2}^2$. If this new
sector does not interact with the observable sector then its sole
impact is to approximately cancel the large (negative) vacuum energy
associated with the second term in~(\ref{pot}), leaving $\lang F^X
\rang$ essentially disconnected from the size of the gravitino mass.
As we will see below, explicit examples of both classes of solutions
have the remarkable property of giving rise to the same general
pattern of gaugino masses as in~(\ref{Malowfull3}).

\subsection{Class 1: K\"ahler Stabilization Models}

As an example of the first class of models we will consider the
weakly coupled heterotic string models studied by Binetruy, Gaillard
and Wu (BGW)~\cite{Binetruy:1996xja,Binetruy:1996nx} and reviewed
in~\cite{Gaillard:2007jr}. The presentation here will follow that
of~\cite{Kane:2002qp} from which we will take our benchmark
scenario.

For the heterotic string gauge coupling unification is a result of a
single modulus, the dilaton $S$, appearing universally in all gauge
kinetic functions. The BGW construction postulates the existence of
some non-perturbative correction to the action for the dilaton
field, along the lines of that originally suggested by
Shenker~\cite{Shenker:1990uf}, which results in a modification of
the K\"ahler metric for the dilaton scalar. Borrowing the notation
of~(\ref{FX}) with $X \to S$ it is sufficient for our purposes to
parameterize this modification as follows
\begin{equation}
F^{S} = \sqrt{3} m_{3/2} (K_{s\bar{s}})^{-1/2} =
\sqrt{3}m_{3/2}a_{\rm np}(K_{s\bar{s}}^{\rm tree})^{-1/2} ,
\label{FS}
\end{equation}
where we have introduced the parameter
\begin{equation}
a_{\rm np} \equiv \(\frac{K_{s\bar{s}}^{\rm tree}}{K_{s\bar{s}}^{\rm
true}}\)^{1/2} \label{acond}
\end{equation}
designed to measure the departure of the dilaton K\"ahler potential
from its tree level value. Recall that $\lang (K_{s\bar{s}}^{\rm
tree})^{1/2} \rang = \lang 1/(s+\bar{s}) \rang = g_{\STR}^{2}/2
\simeq 1/4$ and $s = S\lowest$.

In order to be more concrete we must build a model for supersymmetry
breaking in which $a_{\rm np}$ is calculable. Here we will take an
indirect approach. Consider the field-theoretic non-perturbative
phenomenon of gaugino condensation. Using the relation between the
dilaton and the gauge coupling it is easy to see that the effective
superpotential generated by the gaugino condensate will have the
form $W(S) \propto (e^{-(8\pi^2/b_{a})S})^3$ were $b_{a}$ is the
beta-function coefficient of a condensing gauge group ${\cal G}_{a}$
of the hidden sector. Let us simplify things by assuming a single
condensing gauge group, which we will denote by ${\cal G}_{+}$, with
beta-function coefficient $b_+ = b_a/16\pi^2$. The values of $b_+$
can be quite a bit larger than analogous values for the Standard
Model groups, but a limiting case for the weakly coupled heterotic
string is that of a single $E_8$ gauge group condensing in the
hidden sector, so that ${\cal G}_{+} = {\cal G}_{E_8}$ and $b_+ =
90/16\pi^2 = 0.57$. Clearly we must insist $b_+ > 0$ in order for
gaugino condensation to happen at all.

However, if we do not insist on the tree level dilaton K\"ahler
potential then the vanishing of the vacuum energy
implies~\cite{Casas:1996zi}
\begin{equation}
(K_{s\bar{s}})^{-1}\left|K_s - \frac{3}{2b_{+}} \right|^{2} =3 \; \;
\to (K_{s\bar{s}})^{-1/2} = \sqrt{3}
\frac{\frac{2}{3}b_{+}}{1-\frac{2}{3}b_{+}K_{s}}\, , \label{Ktrue}
\end{equation}
where we have used the equations of motion~(\ref{EQM}) for $F^S$ and
$W(S) = e^{-3S/2b_{+}}$. So provided $K_s \sim \order(1)$ so that
$K_s b_{+} \ll 1$ we can immediately see that a K\"ahler potential
which stabilizes the dilaton while simultaneously providing zero
vacuum energy will necessarily imply a suppressed dilaton
contribution to soft supersymmetry breaking. Indeed,
from~(\ref{acond})
\begin{equation}
a_{\rm np}=\sqrt{3}\frac{\frac{2}{3}\frac{g_{\STR}^{2}}{2}b_{+}}{1-
\frac{2}{3}K_{s}b_{+}} \ll 1 , \label{aBGW}
\end{equation}
and
\begin{equation} r = M_g/M_u = m_{3/2} \lang
\frac{(s+\bar{s})}{F^S} \rang = \frac{1}{\sqrt{3}a_{\rm np}} \gg 1
\, . \label{rvalA} \end{equation}
It is not hard to construct explicit examples which achieve the
outcome in~(\ref{Ktrue})
and~(\ref{aBGW})~\cite{Binetruy:1997vr,Gaillard:1999et}.

\begin{figure}[tb]
\begin{center}
\includegraphics[scale=0.55,angle=0]{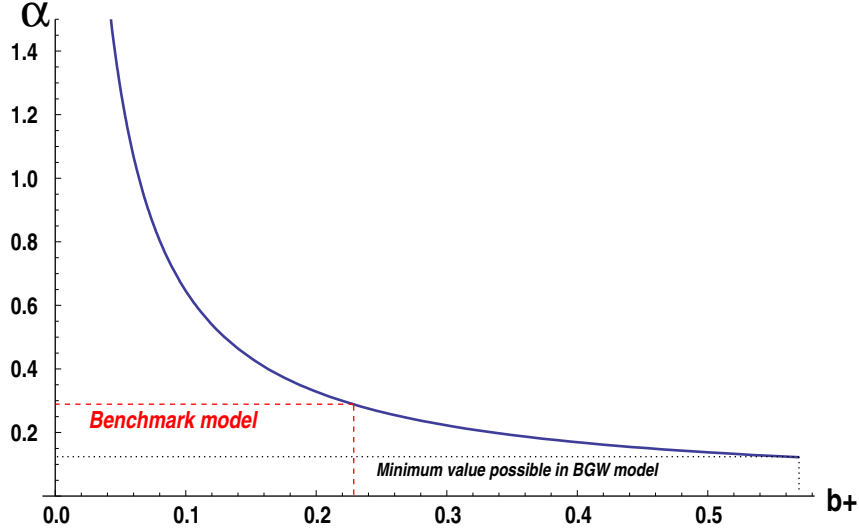}
\caption{\footnotesize \textbf{Effective value of $\alpha$ as a
function of $b_+$ in the BGW class of models.} The parameter $b_+$
represents the beta-function coefficient of the largest gauge group
which experiences gaugino condensation in the hidden sector. This
parameter controls the effective value of $\alpha$ for the gaugino
masses at the electroweak scale. Since the largest possible
confining group would be $E_8$ there is a minimal size to the
effective $\alpha$ parameter in this class of theories. The
benchmark point considered in the text corresponds to $b_+ =
36/16\pi^2$.} \label{fig:BGWalpha}
\end{center}
\end{figure}

The value of the parameter $\alpha$ associated with~(\ref{rvalA})
can be readily computed from~(\ref{alpha})
\begin{equation} \alpha = \frac{1}{\sqrt{3}\ln\(M_{\PL}/m_{3/2}\)a_{\rm
np}} \label{alphaBGW}\, . \end{equation}
Using~(\ref{aBGW}) and the assumption that $\lang K_s \rang =
-g_{\STR}^{2}/2$ we can plot the predicted value of $\alpha$ as a
function of condensing group beta-function coefficient $b_+$. The
result is shown in Figure~\ref{fig:BGWalpha}. Note that the largest
possible value of $b_+$ ($b_+ = b_{E_8}= 90/16\pi^2 = 0.57$)
corresponds to the smallest possible $\alpha$ value. We immediately
see that if this class of models is realized in Nature then the
$\alpha \to 0$ limit cannot be obtained and {\em departures from the
mSUGRA gaugino mass regime are a prediction of the theory}. Our
benchmark point will take $b_+ = 36/16\pi^2$ which corresponds to
$a_{\rm np} = 1/15.77$. Such a value for $b_+$ could arise from a
condensation of a sector consisting only of $E_6$ Yang-Mills fields
and no matter charged under the $E_6$ group. This benchmark point
was studied in~\cite{Kane:2002qp} and we give the explicit soft
supersymmetry breaking mass parameters for the point in
Table~\ref{tbl:inputs} at the end of this Appendix.

The corresponding effective $\alpha$ value at the scale
$\Lambda_{\EW} = 1 \TeV$ is $\alpha = 0.28$. We note that extraction
of the value of $\alpha$ from low-scale gaugino soft masses depends
on the renormalization group scale, as is apparent from expressions
such as~(\ref{Malowfull2}) and~(\ref{Malowfull3}). In particular,
the value of $\alpha = 0.28$ can be extracted using the ratios
in~(\ref{mirage_ratios}) provided the gaugino masses are evaluated
at the scale $\Lambda_{\EW} = 1 \TeV$. The value of $\alpha$ at
other scales can be found by using the more general
formula~(\ref{Malowfull2}).

\subsection{Class 2: Type~IIB Models with Flux Compactifications}

The second type of solution to the vacuum energy problem --
introducing a new sector whose purpose is to cancel negative
contributions to the vacuum energy arising from the last term
of~(\ref{pot}) -- is realized in certain constructions of Type~IIB
string theory compactified on Calabi-Yau
orientifolds~\cite{Kachru:2003aw}. In this class of theories NS~and
RR~three-form fluxes are introduced to stabilize many of the moduli
upon compactification. The presence of this flux warps the bulk
geometry of the Calabi-Yau, resulting in a ``throat'' of the
Klebanov-Strassler type~\cite{Klebanov:2000hb}. At the infrared end
of this throat a hidden sector gaugino condensate exists on a set of
$D_7$-branes and is thus ``sequestered'' from the observable sector,
in the language of Randall and Sundrum~\cite{Randall:1998uk}. For
gauge theories living on $D_7$ branes the gauge coupling is
determined by the K\"ahler modulus $T$, as opposed to the dilaton
$S$ of the heterotic example presented previously. But apart from
this small notational change much of the phenomenology is strikingly
similar to the example in the previous subsection.

The K\"ahler potential for the modulus $T$ is again taken to be $K =
K_{\rm tree}(T,\oline{T}) = -\ln (T+\oline{T})$ and we assume that
the Standard Model exists on a second collection of $D_7$ branes
such that the (tree level) gauge kinetic functions for the Standard
Model gauge groups are universal and of the form $f_a = T$. In the
effective supergravity theory just below the string compactification
scale, the presence of the three-form fluxes is represented by a
constant $W_0$ in the effective superpotential. Combined with the
effect of gaugino condensation in the hidden sector the total
effective superpotential is then
\begin{equation}
W = W_0 + \sum_i A_i e^{-a_i T}\, . \label{WKKLT} \end{equation}
For simplicity, let us assume a single condensate from the gauge
group $\mathcal{G}_+$ with coefficients $A_+=1$ and $a=a_+$. To make
contact with the notation of the previous section we need merely
identify\footnote{We have changed notation so as to ease comparisons
with the original literature.}
\begin{equation} a_+ \to \frac{3}{2b_+} \, .\label{a2b} \end{equation}
Minimizing the resulting scalar potential $V(t,\bar{t})$ with $t=
T\lowest$ generates a non-vanishing value for $\lang t + \bar{t}
\rang$ at which the auxiliary field $F^T$
vanishes~\cite{Choi:2004sx}. Restoring the Planck units to the
second term in~(\ref{pot}) we see that the vacuum must therefore
have an energy density given by $\lang V \rang = -3 m_{3/2}^2
M_{\PL}^2$. The size of the VEV for ${\rm Re}\,t$, as well as the
size of the gravitino mass $m_{3/2}$ are determined by the size of
the constant term $W_0$ in~(\ref{WKKLT}). In particular we
have~\cite{Choi:2004sx}
\begin{eqnarray} \lang a_+ {\rm Re}\,t \rang &\simeq& \ln(1/W_0) \nonumber \\
m_{3/2} & \simeq & M_{\PL} \frac{W_0}{(2\lang{\rm
Re}\,t\rang)^{3/2}}\, . \label{relations} \end{eqnarray}
An acceptable phenomenology requires that the constant $W_0$ be
finely-tuned to a value $W_0 \sim \order(10^{-13})$ in Planck units.
That such a fine-tuning is possible at all is a particular feature
of Type~IIB compactifications with three-form fluxes. Combining the
two relations in~(\ref{relations}) we see that the model will assume
an appropriate value of $W_0$ such that
\begin{equation} \lang a_+ {\rm Re}\,t \rang \simeq
\ln(M_{\PL}/m_{3/2}) \, . \end{equation}

The remaining component to the model is the sector that resolves the
issue of the large negative vacuum energy. Here it is postulated
that at the far tip of the Klebanov-Strassler throat there is an
additional source of supersymmetry breaking. In this case we assume
the presence of $\oline{D}_3$-branes which break supersymmetry {\em
explicitly}. Being at the end of the warped throat the effect of
this hard supersymmetry breaking is presumed to be mild on the
observable sector $D_7$-branes. The vacuum stabilization for the
K\"ahler modulus $t = T\lowest$ is thus largely unaffected. Being an
explicit breaking of supersymmetry it is not possible to perfectly
capture the effects of the $\oline{D}_3$-branes in the form of
corrections to the supergravity effective Lagrangian in superspace.
However, it can be approximated~\cite{Choi:2005ge,Choi:2005uz} by
assuming a correction to the pure-supergravity part of the action
\begin{equation} \Lag \ni -2\int \diff^4 \theta E  \to
-2 \int \diff^4 \theta \[ E + P(T,\oline{T})\] \label{Lag}
\end{equation}
which gives rise to a new contribution to the scalar potential for
the modulus $T$. When the modulus-dependence of $P(T,\oline{T})$ is
trivial and $P(T,\oline{T})=C$ then the resulting scalar potential
contribution is simply
\begin{equation} V_{\rm lift} = \frac{C}{(t+\bar{t})^2} \, ,
\label{Vlift1} \end{equation}
and the more general case of $P(T,\oline{T})=C(T+\oline{T})^{n}$
gives rise to
\begin{equation} V_{\rm lift} = \frac{C}{(t+\bar{t})^{(2-n)}}\, .
\label{Vlift2} \end{equation}
Under these conditions the equations of motion for the auxiliary
field for the K\"ahler modulus has the approximate solution
\begin{equation} M_u = \lang \frac{F^T}{t+\bar{t}} \rang \simeq
m_{3/2} \frac{2-n}{a_+ \lang t + \bar{t}\rang}\, . \label{MuKKLT}
\end{equation}

To see how this generates a mirage pattern of masses, we look again
at the ratio $r=M_g/M_u$
\begin{equation} r  = m_{3/2} \lang
\frac{(t+\bar{t})}{F^T} \rang = \frac{a_+ \lang t +
\bar{t}\rang}{2-n} \simeq \ln(M_{\PL}/m_{3/2}) \gg 1 \, .
\label{rvalB} \end{equation}
Provided the VEVs in~(\ref{relations}) can be arranged, the mirage
pattern of gaugino masses necessarily follows. The implied value of
$\alpha$ follows from the definition in~(\ref{alpha})
\begin{equation} \alpha = \frac{2}{2-n} +
\order\(\ln(m_{3/2}/M_{\PL})\) \label{alphaKKLT}\, . \end{equation}
In the minimal case with $n=0$ we therefore have the prediction that
$\alpha \simeq 1$ for this class of theories.

We note that in the case $n=0$ we can rewrite the quantity $r$
in~(\ref{rvalB}) in the following way
\begin{equation} r = a_+ \lang {\rm Re}\, t\rang
 =  a_+ \frac{1}{g_{\STR}^2} \quad \to \quad \frac{3}{2b_+ g_{\STR}^2}
 =  \frac{\sqrt{3}}{2a_{\rm np}}\, . \label{connection}
\end{equation}
Our two classes of theories are very different, yet they both result
in a mirage pattern of gaugino masses in which the relative sizes of
the contributions to soft supersymmetry breaking depend on the
hidden sector gaugino condensation in a similar manner, as seen by
their functional dependence on the parameters $a_+$ and/or $b_+$.
Should we find this surprising? Perhaps not, since both aim to solve
the same problem (namely, large negative vacuum energy) using the
dynamics of a single real scalar field. And both methods ultimately
involve adding a correction to the action for this real scalar of
the form~(\ref{Lag}).\footnote{Furthermore, in both constructions
there are elements of this addition~(\ref{Lag}) that are not under
full calculational control.} The {\em requirement} that $\lang
V_{\rm total} \rang = 0$ in the ground state then  {\em dictates}
the necessary values for the parameters such that the ratio
$r=M_g/M_u$ dependence on the gaugino condensate is as
in~(\ref{connection}). We hasten to add, however, that the two
models are indeed quite distinct in other regards. In particular
they make quite different predictions for the other soft
supersymmetry breaking parameters. For the case of the flux
compactifications of Type~IIB we refer the reader to the relevant
literature~\cite{Choi:2005uz,Falkowski:2005ck} for more details on
how these additional terms are computed. We have chosen as a
benchmark point a scenario studied in~\cite{Choi:2005uz} in which
$n=0$ so that $\alpha =1$. The overall scale was treated as a free
parameter in~\cite{Choi:2005uz} and we here take that scale to be
$m_{3/2} = 16.3 \TeV$. The precise values of the soft supersymmetry
breaking parameters for both benchmark models are collected in
Table~\ref{tbl:inputs} in Section~\ref{theory} of the main text.


\clearpage


\end{document}